\documentclass[aps,prd,nofootinbib,tightenlines,notitlepage,floatfix,superscriptaddress,showkeys,twocolumn]{revtex4-1}

\usepackage[utf8]{inputenc}          
\usepackage{graphicx}         
\usepackage[usenames,dvipsnames]{xcolor}
\usepackage{array,dcolumn,longtable} 
\usepackage{gensymb}

\usepackage{enumerate}
\usepackage{multirow}
\newcommand{\sr}{\rule[-0.45cm]{0pt}{0.9cm}}

\usepackage{amsmath,amssymb,amsfonts,slashed} 

\usepackage{hyperref}
\hypersetup{linktocpage,breaklinks}

\usepackage{txfonts}
\usepackage{bm}
\usepackage{stmaryrd}
\usepackage[utf8]{inputenc}

\usepackage{epsfig}
\usepackage{epstopdf}

\usepackage{natbib}
\usepackage{cleveref}

\usepackage{makecell}

\definecolor{mred}{RGB}{127,0,25}
\definecolor{mdgr}{RGB}{51,51,51}
\definecolor{mag}{RGB}{211, 54, 130}
\definecolor{verm}{RGB}{164, 25, 0}

\hypersetup{colorlinks=true,
            citecolor=NavyBlue,
            linkcolor=NavyBlue,
            urlcolor=NavyBlue}
            
\usepackage{xcolor}

\newcommand{\beq}{\begin{equation}}
\newcommand{\eeq}{\end{equation}}
\newcommand{\bea}{\begin{eqnarray}}
\newcommand{\eea}{\end{eqnarray}}

\newcommand{\MeV}{\, {\rm MeV}}
\newcommand{\CEFT}{$\chi$EFT }
\newcommand{\nsat}{$n_{\rm sat}$ }
\newcommand{\Msol}{M$_\odot$ }

\begin{document}

\preprint{APS/123-QED}

\title{Nontrivial features in the speed of sound inside neutron stars}
\author{D. Mroczek}
\affiliation{Illinois Center for Advanced Studies of the Universe,\\ Department of Physics, 
University of Illinois at Urbana-Champaign, 1110 W. Green St., Urbana IL 61801-3080, USA}
\author{M. C. Miller}
\affiliation{Department of Astronomy and Joint Space-Science Institute, University of Maryland, College Park, MD 20742-2421 USA}
\author{J. Noronha-Hostler}
\affiliation{Illinois Center for Advanced Studies of the Universe,\\ Department of Physics, 
University of Illinois at Urbana-Champaign, 1110 W. Green St., Urbana IL 61801-3080, USA}
\author{N. Yunes}
\affiliation{Illinois Center for Advanced Studies of the Universe,\\ Department of Physics, 
University of Illinois at Urbana-Champaign, 1110 W. Green St., Urbana IL 61801-3080, USA}

\begin{abstract}

Measurements of neutron star masses, radii, and tidal deformability have direct connections to nuclear physics via the equation of state (EoS), which for the cold, catalyzed matter in neutron star cores is commonly represented as the pressure as a function of energy density. Microscopic models with exotic degrees of freedom display nontrivial structure in the speed of sound ($c_s$) in the form of first-order phase transitions and bumps, oscillations, and plateaus in the case of crossovers and higher-order phase transitions. 
We present a procedure based on Gaussian processes to generate an ensemble of EoS that include nontrivial features.
Using a Bayesian analysis incorporating measurements from X-ray sources, gravitational wave observations, and perturbative QCD results, we show that these features are compatible with current constraints. We investigate the possibility of a global maximum in $c_s$ that occurs within the densities realized in neutron stars -- implying a softening of the EoS and possibly an exotic phase in the core of massive stars -- and find that such a global maximum is consistent with, but not required by, current constraints. 

\end{abstract} 
\maketitle
   
\section{Introduction}

One of the main goals of modern nuclear physics is to determine the phase structure of Quantum Chromodynamics (QCD). The cold, catalyzed nuclear matter in neutron stars probes the zero-temperature, isospin asymmetric regime of QCD at baryon number densities ($n_{B}$) ranging from sub-nuclear to several times nuclear saturation density ($n_{\rm sat} \equiv 0.16 \textrm{ fm}^{-3}$) in the core \cite{Baym:2017whm}. In this regime, first-principle QCD calculations are not yet feasible because of the  fermion sign problem \cite{Borsanyi:2021sxv,Borsanyi:2022qlh}, and effective models and parameterizations of the equation of state (EoS) are used instead. 

A variety of models have been developed to compare against astronomical observations, all of which have different regimes of validity, advantages and disadvantages. One such model arises from chiral effective field theory ($\chi$EFT), which breaks down at densities around 2 $n_{\rm sat}$ \cite{Holt:2016pjb,Drischler:2021kxf}. In this effective theory, one prescribes a general Lagrangian that respects the symmetries of low-energy QCD (with nucleons and pions as degrees of freedom). This Lagrangian is then expanded order-by-order in two- and multi-nucleon interactions. The low-density crust ($n_B \lesssim$ 0.5 $n_{\rm sat}$) and the high-density inner core  ($n_B \gtrsim$ 1.1-1.5 $n_{\rm sat}$) \cite{Huth:2021bsp,Raaijmakers:2021uju} of neutron stars, however, require additional modeling and assumptions beyond \CEFT about the underlying degrees of freedom and relevant interactions. 

Another class of models relies on mean-field approximations of an effective Lagrangian with nucleon, electron, and muon degrees of freedom. These models lead to a squared speed of sound $c_s^2 = {\rm d} p/{\rm d} \varepsilon$ (where $p$ is the pressure and $\varepsilon$ is the energy density) that increases monotonically with the density (see, e.g., Refs.~\cite{Douchin:2001sv,Gulminelli:2015csa,Vinas:2021vmv}). Such behavior leads to acausal sound speeds in non-relativistic models at densities only a few times that of nuclear saturation and therefore cannot be the correct description of nuclear matter at those densities. Relativistic hadronic frameworks also break down at high densities ($n_B \gtrsim$ 6 $n_{\rm sat}$), when nucleons start to overlap \cite{Alford:2022bpp}. 

Yet another set of results are available from perturbative QCD (pQCD) calculations, in which the QCD field equations are solved perturbatively in a small-coupling expansion. These calculations have found that at very high densities ($n_B \gtrsim$ $40 n_{\rm sat}$), $c_s^2 \rightarrow 1/3$ (in units where the speed of light $c=1$) from below and high-density quark matter is approximately mass-scale-invariant, or ``conformal'' \cite{Kurkela:2009gj,Kurkela:2014vha,Gorda:2021znl}. Astronomical observations, however, strongly suggest that $c_s^2 > 1/3$ in the core of neutron stars \cite{Legred:2021hdx}, at densities in the range of $2\lesssim n_B/n_{\rm sat} \lesssim 6$. This result indicates that $c_s^2$ must display non-monotonic behavior with increasing density\footnote{While exact conformal symmetry implies $c_s^2 = 1/3$, and, thus, that other EoS-related quantities must take on specific values, the reverse is not a sufficient condition to establish conformal symmetry. Indeed, it is possible for $c_s^2$ to pass through 1/3 a number of times before eventually approaching it from below at high densities.}, which has motivated searches for evidence that deconfinement into approximately conformal quark matter occurs within densities realized in neutron stars \cite{Annala:2023cwx,Annala:2021gom,Annala:2019puf,Marczenko:2022jhl}. 

The onset of conformal quark matter is, however, not the only question relevant to constraining the cold nuclear EoS. Models that include heavy resonances and exotic hadronic phases and/or strange and quark degrees of freedom predict, respectively, higher-order phase transitions and crossovers and/or first-order phase transitions \cite{Buballa:2014jta,Ferreira:2020kvu, Jakobus:2020nxw,Dutra:2015hxa,McLerran:2018hbz, Li:2019fqe,Alford:2017qgh,Zacchi:2015oma,Alvarez-Castillo:2018pve,Zacchi:2019ayh,Zhao:2020dvu,Minamikawa:2020jfj,Duarte:2020xsp,Sen:2020qcd,Minamikawa:2020jfj,Pisarski:2021aoz,Stone:2019blq,Kapusta:2021ney,Somasundaram:2021ljr,Motornenko:2019arp,Baym:2019iky,Baym:2017whm,Malfatti:2020onm}. An $N$th order phase transition occurs when the $N$th susceptibility of the pressure (i.e.,~the $N$th partial derivative of the pressure with respect to the chemical potential) presents non-analytic behavior (such as a discontinuous jump or a divergence). A crossover occurs when there is no phase separation, and the change in degrees of freedom happens gradually over some range in density (i.e., all derivatives of the pressure are continuous). These different types of phase transitions and crossovers do not necessarily predict an approach of $c_s^2$ to 1/3 within neutron star densities (though it may happen at much higher $n_{\rm B}$, well beyond the densities at which the star would collapse to a black hole). 

Different physical processes (i.e.,~phase transitions of different order or crossovers) lead to unique and nontrivial structure in $c_s^2$ as a function of $n_{\rm B}$  (see \cite{Tan:2021ahl} for details and extensive examples from microphysical models). 
Generally, a first-order phase transition is associated with the onset of new degrees of freedom. In neutron stars, a first-order transition could separate a hadronic phase from a quark phase, for example. When a first-order phase transition takes place, the presence of latent heat leads to a range in the energy density $\varepsilon$ where the pressure $p$ is constant, which appears as a \textit{plateau} in $c_s^2(\varepsilon)$ over which $c_s^2 = 0$ for a system in equilibrium. 
On the other hand, a crossover (or phase transitions of higher order) is typically associated with the emergence of a new state, new degrees of freedom/particles, or new interactions, that occur gradually across a range of $n_B$. 
These new particles or interactions lead to a \textit{bump} in $c_s^2$, which may be wide (like a \textit{positive plateau}) or narrow (like a \textit{positive spike}), depending on whether the crossover occurs over a wide or narrow region in baryon density (see examples of quarkyonic matter \cite{McLerran:2018hbz,Duarte:2023cki,Duarte:2021tsx,Duarte:2020kvi} or percolation approaches \cite{Kojo:2015fua}).
Second-order phase transitions are associated with critical points, or, at vanishing temperatures, quantum critical points. In this case, $c_s^2$ displays a \textit{negative spike} approaching zero (for an example at finite temperatures see Fig 2 from \cite{Dore:2022qyz}). Higher-order phase transitions are also possible and may occur due to exotic baryon states or new types of interactions that could lead to a kink in $c_s^2$ \cite{CMF_MUSES}. An EoS can display one, or a combination of such features depending on the assumptions made about the relevant degrees of freedom and interactions. 

Recently, astronomical observations across the electromagnetic and gravitational-wave spectra have placed constraints on the macroscopic properties of neutron stars, such as the mass ($M$), radius ($R$), and tidal deformability ($\Lambda$). These measurements have also made it possible to  indirectly infer the allowed EoS via model-to-data Bayesian comparisons, since the EoS determines $M$, $R$, and $\Lambda$ as a function of central number density $n_B^{\rm max}$. Analyses typically include binary tidal deformability ($\tilde{\Lambda}$) posteriors from the LIGO gravitational wave observations of events GW170817 \cite{LIGOScientific:2017vwq,De:2018uhw,LIGOScientific:2018cki} and GW190425 \cite{LIGOScientific:2020aai},
the existence of heavy pulsars \cite{arzoumanian2018nanograv, antoniadis2013massive,cromartie2020relativistic} and NASA's Neutron Star Interior Composition Explorer (NICER) joint $M-R$ posteriors from PSR J0030-0451 \cite{Miller:2019cac} and PSR J0740+6620 \cite{Miller:2021qha} (see, respectively, \cite{Riley:2019yda} and \cite{Riley:2021pdl} for independent analyses of these two pulsars from a separate group within the NICER collaboration). 

Other constraints are also available from the measured properties of nuclei at $n_{\rm sat}$. These properties include the symmetry energy ($S = E_{\rm SNM} -E_{\rm PNM}$), defined as the difference in the binding energy per nucleon between symmetric nuclear matter (SNM) and pure neutron matter (PNM) as a function of density, and the slope parameter (L), which determines how the symmetry energy changes with density \cite{Tsang:2012se,Li:2019xxz,Drischler:2020hwi,Drischler:2020yad,PREX:2021umo,Reed:2021nqk,Yue:2021yfx}. 

From the theory perspective, it recently became possible to consistently extrapolate pQCD results to densities as low as $\sim$ 2.5 \nsat \cite{Rhoades:1974fn, Gorda:2022jvk}. These constraints are based on the mechanical stability and causality of the EoS ($0 \leq c_s^2 \leq 1 $) and the consistency of the underlying thermodynamic potential that connects the low-density regime of the EoS to the high-density regime ($\gtrsim 40 \ n_{\rm sat}$) constrained by pQCD. These constraints offer information at each $n_{\rm B}$ about the region in $p-\varepsilon$ that can be connected to the high-density perturbative results via a stable and causal EoS and a consistent thermodynamic potential. For a given EoS, it is possible to check its compatibility with stability, causality, and consistency constraints at any density between $\sim 2.5 - 40$ $n_{\rm sat}$. We will refer to these constraints collectively as the pQCD constraints from here on. 

The lack of first-principle approaches for the $\beta$-equilibrated, zero-temperature nuclear EoS between $\sim 1.1$ $n_{\rm sat}$ up to $n^{\rm max}_{B}$ realized in neutron stars means that astronomical observations are the only direct probe of the EoS in this regime. Thus, model-to-data Bayesian comparisons of generic functional forms of the EoS are the state-of-the-art for obtaining posterior distributions for the EoS. Nonetheless, microscopic models are vital in providing guidance for the behavior of functional forms of the EoS, especially so that specific features associated with the onset of new degrees of freedom and interactions can be correctly identified. 

The posterior distribution that is extracted from a Bayesian analysis is sensitive to how data and theoretical input are incorporated \cite{Miller:2019nzo}, as well as prior-imposed assumptions about the EoS (e.g.,~correlations across density scales) \cite{Greif:2018njt, Raaijmakers:2018bln, Riley:2018ekf,Legred:2022pyp}. Parametric descriptions, such as spectral expansions \cite{Lindblom:2010bb,Lindblom:2018rfr} or piecewise polytropes, provide a framework to represent the EoS without relying on micro-physics models. Spectral representations of the EoS assume the adiabatic index $\Gamma (p)$ as a function of pressure can be expanded in terms of a set of spectral basis functions and coefficients, which uniquely determine the EoS \cite{Lindblom:2010bb,Lindblom:2018rfr,Lindblom:2022mkr}. Piecewise polytropes divide the EoS into a number of segments and represent the pressure for each segment as a polytrope, $p = \kappa \rho^\Gamma$, where $\kappa$ and $\Gamma$ are the fixed polytropic constant and the adiabatic index, respectively \cite{Read:2008iy}. Parametric representations have been widely used, since they do not rely on as many assumptions as physics-based models (see, e.g., Refs. \cite{Raithel:2017ity,Annala:2023cwx,Annala:2021gom, Annala:2019puf,Annala:2017llu,Marczenko:2022cwq,Hebeler:2013nza,Miller:2019cac,Miller:2021qha,LIGOScientific:2018cki,Read:2008iy,Margalit:2017dij,Ozel:2015fia}, though this list is far from comprehensive). However, the question of whether these parameterizations are flexible enough to capture all relevant physics has recently been raised in the literature \cite{Greif:2018njt,Tan:2020ics,Legred:2022pyp}. Specifically, it has been shown that parameterized EoS can introduce undesired correlations across density scales \cite{Greif:2018njt,Legred:2022pyp} and are unable to capture behavior consistent with state-of-the-art nuclear physics models with exotic degrees of freedom \cite{Tan:2020ics}. 

Physics-agnostic frameworks based on Gaussian processes (GPs) offer more flexibility in the modeling of the EoS at the cost of increased functional complexity \cite{Landry:2018prl,Han:2021kjx,Legred:2021hdx,Landry:2020vaw,Essick:2019ldf}. Generally, a GP models the speed of sound as a continuous function over a specified domain. The properties of the probability density for the speed of sound at each point of the domain are determined by a mean vector and covariance matrix. The covariance matrix is calculated using a specific kernel function that requires hyperparameters that can be fixed or sampled from a hyperprior. The hyperprior may be model-agnostic or conditioned to more closely reproduce a set nuclear physics models \cite{Essick:2019ldf}. 
In principle, a GP can be tailored to resemble any continuous function across some domain. So far, GPs have been implemented with a fixed set of hyperparameters for an individual EoS \cite{Landry:2018prl,Han:2021kjx,Legred:2021hdx,Landry:2020vaw,Essick:2019ldf,Miller:2021qha,Riley:2021pdl,Gorda:2022jvk,Annala:2023cwx}, though priors may contain samples drawn from mixture of multiple stationary kernels, probing a wide range of potential correlation properties \cite{Legred:2021hdx,Essick:2019ldf}. In contrast with the assumption of uniform correlations across density scales, many state-of-the-art nuclear physics models with exotic degrees of freedom display multiscale correlations across various densities \cite{Tan:2020ics}. Furthermore, the features that emerge in the speed of sound as a result are known to be important for understanding heavy, ultra-heavy (neutron stars with masses above $2.5\ M_\odot$), and twin stars \cite{Tan:2021ahl,Tan:2021nat,Tan:2020ics}. 

With that motivation, we introduce modified GPs (mGPs) as a framework for modeling EoS with nontrivial features that possess long, medium, and short-range correlations across densities. First, we produce a family of EoS from a benchmark model of GP EoS that contain only long and fixed-range correlations in $c_s^2$. We then generate a family of EoS from mGPs, which introduce multiscale correlations in the form of nontrivial features in $c_s^2$. With these two families of EoS, we carry out Bayesian parameter estimation analysis against  observational and experimental data and input from pQCD. The results of this analysis allow us to compare the marginalized posteriors of the mass and radius curve and the speed of sound and number density curve when we use the benchmark GP model and the modified GP to represent the EoS. We find that neither EoS family is favored over the other by current data. We do find, nonetheless, that the marginalized posterior for the speed of sound at densities $\sim 1.5-2$ $n_{\rm sat}$ is not identical for EoS from GPs compared to mGPs, although the data are not informative enough yet to discriminate between these posteriors. 

The remainder of this manuscript presents the details of the analysis summarized above. In Sec.~\ref{sec:EOSgen}, we discuss GPs as a model-agnostic framework for generating the EoS and how we introduce multiscale correlations to the EoS with mGPs. Section \ref{sec:priors} outlines how we generate EoS priors from the GP and mGP frameworks. Statistical methods are discussed in detail in Sec.~\ref{sec:statmethods}, followed by results in Sec.\ \ref{sec:results} and conclusions and discussion in Sec.\ \ref{sec:conclu}. Throughout this manuscript, we use $c = 1$ and the Einstein summation convention when necessary. Thermodynamic quantities are in cgs units (unless otherwise stated), with the exception of $c_s^2$, which we always normalize by $c^2$.

\section{Generating the Equation of State}\label{sec:EOSgen}

Both the benchmark EoS model that is a standard GP and the modified GP EoS are built from model-agnostic GPs, which approximate functional forms of $c_s^2$ as a function of the pressure over a fixed domain (for a more comprehensive overview of Gaussian processes, we recommend Refs. \cite{Landry:2018prl,williams2006gaussian}.) We now discuss the details and motivation for the construction of both models. 

A GP provides the joint probability density for a continuous function $f(x)$ over a domain of interest, which we represent here by a sample of discrete values labeled $x_i$. This probability density is assumed to be a multivariate Gaussian distribution (no summation over $i$ implied)
\begin{equation}\label{eq:phidist}
    f(x_i) = \mathcal{N}\left[\mu_i(x_i),\Sigma_{ij}(x_i)\right],
\end{equation}
where ${\cal{N}}(\cdot,\cdot)$ is the normal distribution function at $x_i$, with a mean $\mu_i$ that varies with $x_i$ and a covariance matrix $\Sigma_{ij}$, which gives the correlation between the values of $f$ at $x_i$ and $x_j$, where $i$ can equal $j$.

In the context of extracting the properties of neutron stars from data using an ensemble of synthetic EoS, GPs have been used to approximate the EoS from samples of functional forms of $c_s^2$ \cite{Essick:2019ldf,Komoltsev:2021jzg,Landry:2020vaw,Landry:2018prl,Essick:2021kjb,Essick:2020flb,Legred:2021hdx}. Because the range of Gaussian distributions is infinite,  
while $c_s^2$ is bounded by stability and causality ($0 \leq c_s^2 \leq 1$), it is common to use the GP to approximate an auxiliary variable that compactifies the range of a Gaussian distribution (infinite in both the positive and negative directions) to the range of $c_s^2$. Let us call this variable $\phi$ \cite{Lindblom:2010bb,Landry:2018prl} and define it via
\begin{equation}\label{eq:phitocs2}
\phi \equiv \ln(d\varepsilon/dp - 1) = \ln(1/c_s^2 - 1),
\end{equation}
where $p$ is the pressure and $\varepsilon$ is the energy density. 
This auxiliary variable has the desired range for a GP, but when mapped to $c_s^2$ using the definition above, $\phi \rightarrow +\infty$ corresponds to $c_s^2 \to 0$, while $\phi \rightarrow -\infty$ corresponds to $c_s^2 \to 1$. 
It is common in the literature to model $\phi$ as a function of $\log_{10} p$ in cgs units \cite{Miller:2021qha,Essick:2019ldf,Landry:2020vaw,Landry:2018prl,Essick:2021kjb,Essick:2020flb,Legred:2021hdx}. More explicitly, Eq. (\ref{eq:phidist}) becomes
\begin{equation}\label{eq: GPform}
    \phi(\log_{10} p_i) = \mathcal{N}\left[\mu_i(\log_{10} p_i), \Sigma_{ij}\right].
\end{equation}
Other units and thermodynamic variables can be used instead (e.g.~Ref.~\cite{Komoltsev:2021jzg} uses baryon density in units of $n_{\rm sat}$), but we will use the parameterization presented above. 

A more computationally practical implementation of a GP is to decompose it into a mean and a scatter via 
\begin{equation}\label{eq:Cholesky}
    \phi(\log_{10} p_i) = \mu_i(\log_{10} p_i) + L_{ij} u_j,
\end{equation}
where $L_{ij}$ is the Cholesky decomposition of the covariance matrix plus a white-noise kernel contribution, i.e.~$ L_{ik}L_{kj}^{\rm T} := \Sigma_{ij} + \sigma_{\rm wn}^2\delta(x_i -x_j)$, with $u_j = \mathcal{N}(0,1)$ and $\sigma_{\rm wn}$ a constant white-noise variance. The white-noise kernel (the second term on the right-hand side of the Cholesky decomposition) is added for numerical stability, since the determinant of $\Sigma_{ij}$ can be nearly singular. The effect of the white-noise kernel is to slightly smear the GP by adding noise to the diagonal elements of $\Sigma_{ij}$. A small $\sigma_{\rm wn}$ is sufficient to dramatically increase the stability of the calculation without changing the overall properties of the final sample. We use $\sigma^2_{\rm wn} = 0.0003$, but any other value of the same order of magnitude would produce similar results.

\begin{figure*}
\begin{minipage}{0.49\linewidth}
  \centering
  \includegraphics[width=\linewidth]{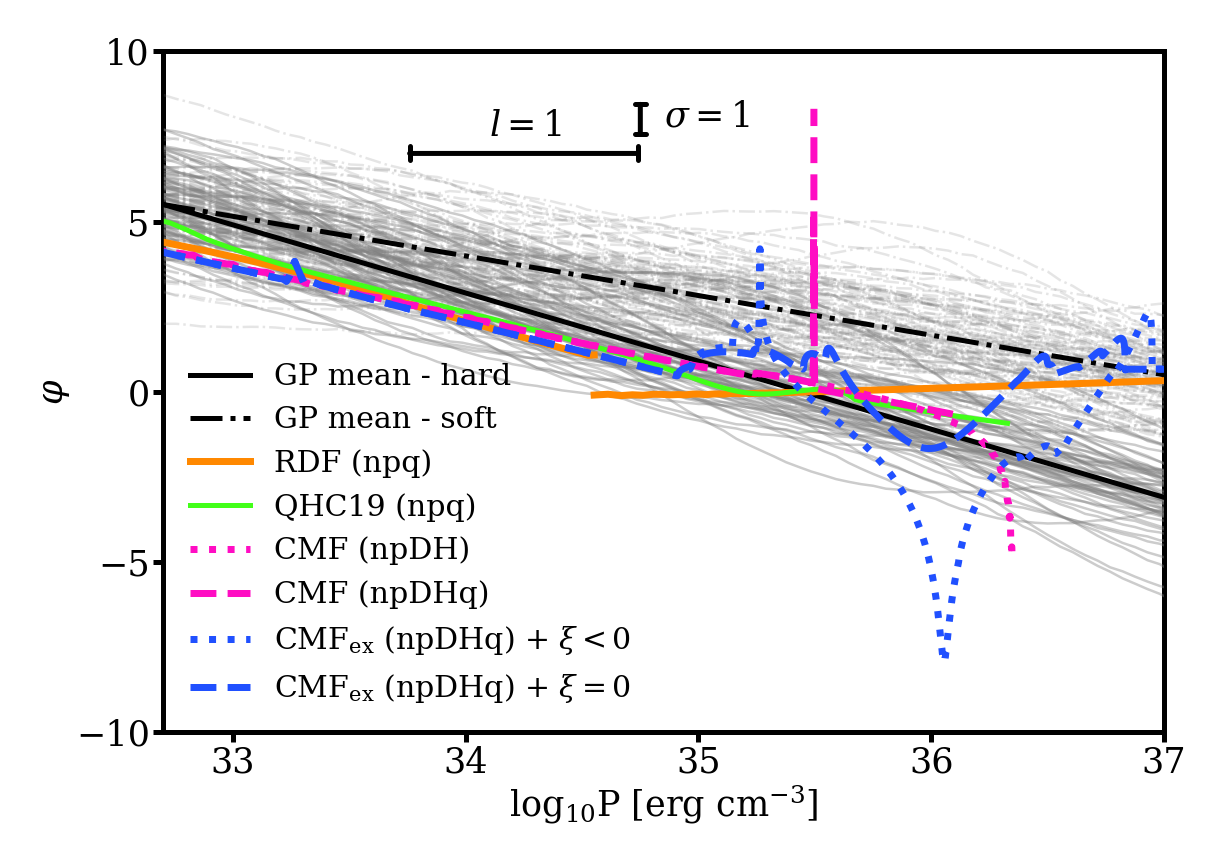} 
\end{minipage}
\hfill
\begin{minipage}{0.49\linewidth}
  \centering
  \includegraphics[width=\linewidth]{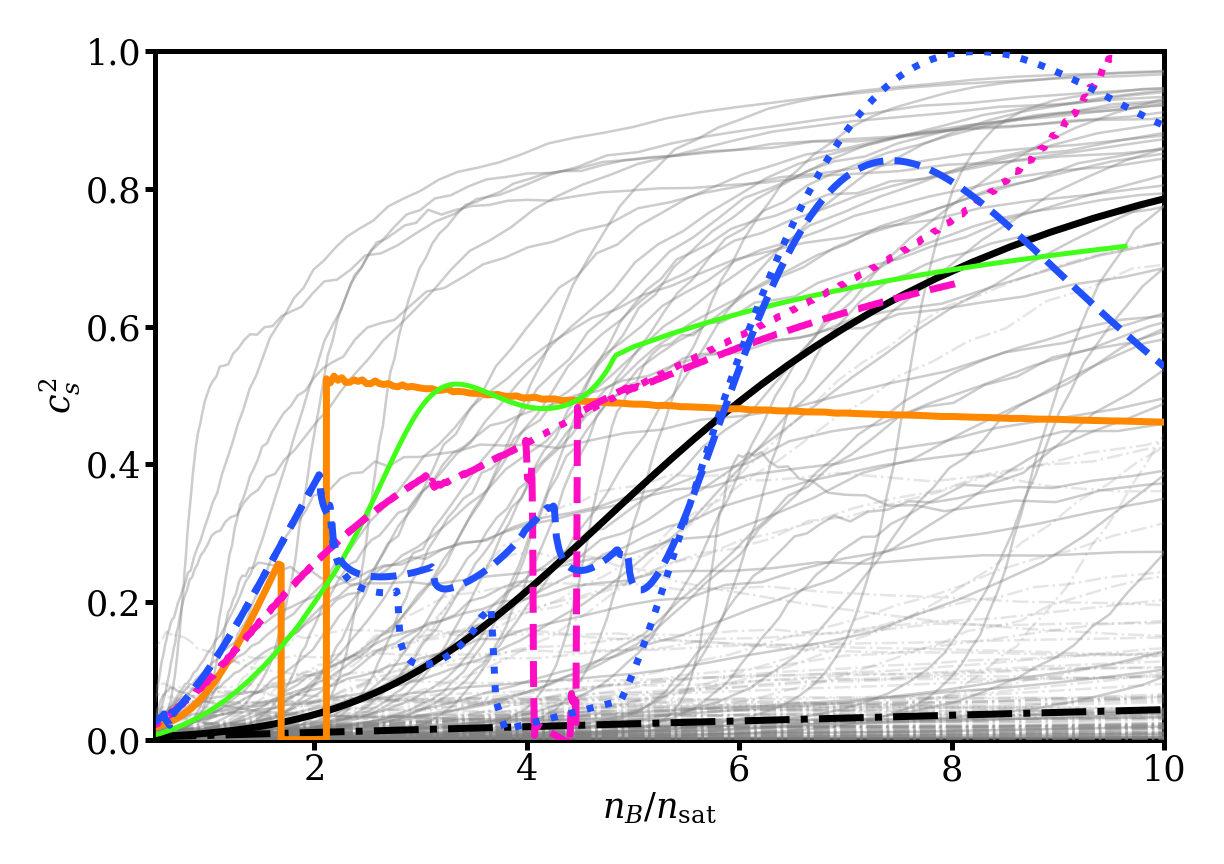} 
\end{minipage}

\begin{minipage}{\linewidth}
  \centering
  \includegraphics[width=\linewidth]{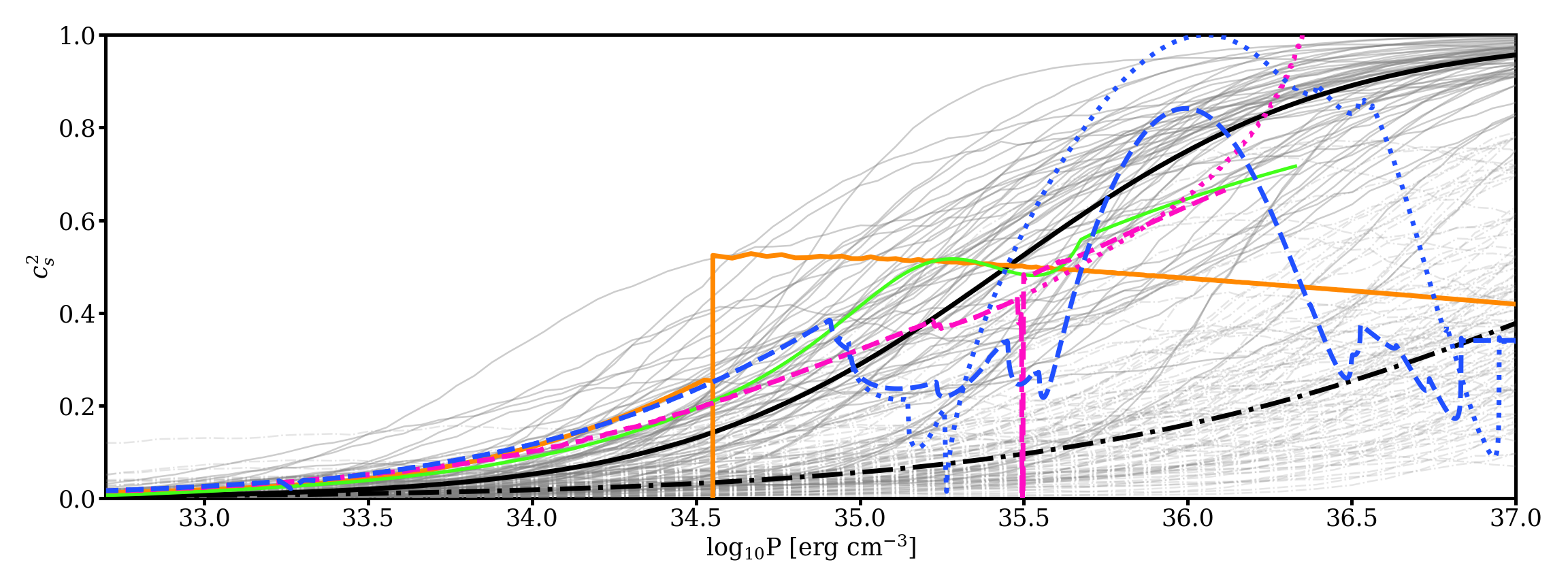}
\end{minipage}
\caption{The auxiliary variable  $\phi \equiv \ln((d\varepsilon/dP) - 1)$ as a function of ${\rm log}_{10} P$ in cgs units (left) and the speed of sound as a function of baryon number density in units of $n_{\rm sat}$ (right) for different parameterizations of the EoS from chiral mean field theory (CMF), which include nucleons (np), delta resonances (D), hyperons (H), quarks (q), and leptons \cite{Dexheimer:2008ax,Dexheimer:2017nse,Dexheimer:2009hi,Dexheimer:2020rlp,Clevinger:2022xzl,Dexheimer:2018dhb,Dexheimer:2014pea}, the QHC19 EoS (np + q + leptons)  \cite{Baym:2019iky}, and the relativistic density-functional (RDF) approach introduced in Ref.~\cite{Ivanytskyi:2022bjc} (np + q + leptons). In light gray we show a sample of 160 total functional forms for $c_s^2$ from the benchmark GPs using the ansatz in Eq. (\ref{eq:eosansatz}) and a squared-exponential kernel with $l = \sigma =1$. The functional forms from the hard and soft GP are shown in dot-dashed and solid lines, respectively. The benchmark GPs capture a wide range of behavior, but the \textit{a priori} requirement that functional forms display only long-range correlations across densities exponentially suppresses sharp and non-trivial features in $c_s^2$, which are observed in state-of-the-art nuclear physics simulations.}
\label{fig:nucEoS}
\end{figure*}

Given the joint probability density for $\phi$, we can construct a GP realization, or ``sample'', by selecting a range of pressures and then drawing the associated $\phi$ values using Eq.~\eqref{eq:Cholesky}~\cite{Reed:2021nqk}. We then invert Eq.~\eqref{eq:phitocs2} to find $c_s^2$ as a function of $p$. The definition of the speed of sound can then be used to specify a differential equation for the EoS, $c_s^2(p) = dp/d\varepsilon$, which can be solved in first quadrature as 
\begin{align}
    \varepsilon = \int \frac{dp}{c_s^2(p)}\,,
\end{align}
and then inverted to find $p(\varepsilon)$. We then obtain the baryon density using the first law of thermodynamics, which, at zero temperature and assuming charge neutrality, can be written as
\begin{align}
\frac{dn_B}{d\varepsilon}  &= \frac{n_B}{\varepsilon + p(\varepsilon)}  \,.  
\end{align}
Once these equations are solved, we have the set $\left\{\phi(p), c_s^2(n_B),p(\varepsilon)\right\}$, which defines an EoS sample from a GP. In practice, we build our EoS numerically by sampling on a finite set of pressures and baryon densities with a sufficiently fine discretization. As pointed out in Ref. \cite{Tan:2021ahl}, a simple check that the EoS is being recovered correctly is to calculate $c_s^2$ from $p$ and the reconstructed $\varepsilon$, and check that it matches the $c_s^2$ from the GP. The EoS samples generated through the GPs will only be used above 0.5 $n_{\rm sat}$~\cite{Hebeler:2013nza}, which we denote as the core-crust transition. Below this density, we model the crust through the QHC19 EoS~\cite{Baym:2019iky,Togashi:2017mjp}.

\subsection{Benchmark Gaussian Processes}  

Now that we have explained the idea behind constructing an EoS sample from a GP, we need to specify the input for the joint probability density function in Eq.~\eqref{eq: GPform}. The two main ingredients that define a GP are the {\bf means} $\{\mu_i\}$, which will determine the average trend for the function that is being sampled, and the {\bf covariance matrix} $\Sigma_{ij}$, which specifies the joint variability between two points $x_i$ and $x_j$.

Let us first discuss how we model the {\bf means}. Our goal at this stage is to create a benchmark model $c_s^2(p)$ \textit{without} any sharp, nontrivial features. To do so, we adopt the approach taken in Miller \textit{et al.} \cite{Miller:2021qha}, which looked at a collection of twelve cold neutron star EoS from the CompOSE data base \cite{Typel:2013rza,Oertel:2016bki} on the $\log_{10}p - \phi$ plane, and found that these EoS follow a linear trend over the domain  $32.7 \leq \log_{10}p_i (\textrm{erg cm$^{-3}$}) \leq 37$. This trend was empirically approximated as
\begin{align}\label{eq:eosansatz}
    \mu_i(\log_{10}p_i) = 5.5 -m(\log_{10}p_i - 32.7)\,,
\end{align}
where $m$ is the slope of the linear regression. Reference \cite{Miller:2021qha} fixes $m=2$ based on the spread of EoS from CompOSE. Other choices for the means are also possible~\cite{Essick:2019ldf}. 
Out of the total twelve EoS that this model is based on, seven were purely proton, neutron, and electron matter (npe) models \cite{Baym:1971pw,Douchin:2001sv,Baldo:1997ag,Danielewicz:2008cm,Gulminelli:2015csa,Agrawal:2005ix,Reinhard:1995zz,Agrawal:2003xb,Gaitanos:2003zg,Grill:2014aea,Douchin:2001sv,Glendenning:1991es}, one model included npe matter, heavy baryonic resonances, and a crossover transition to quarks (QHC18 \cite{Akmal:1998cf,Togashi:2017mjp,Baym:2017whm}), and four models included npe matter plus strange baryons \cite{Glendenning:1991es,Oertel:2014qza,Douchin:2001sv,Dexheimer:2008ax,Schurhoff:2010ph,Dexheimer:2015qha,Dexheimer:2017nse,Gaitanos:2003zg,Grill:2014aea}. These models largely approach the causal limit at high densities, which biases the behavior of the EoS in that regime. Notably, models that predict a softer EoS at large densities, such as quarkyonic models \cite{McLerran:2018hbz,Duarte:2023cki,Duarte:2021tsx,Duarte:2020kvi,Kojo:2015fua}, are missing from the collection of EoS that was used to determine Eq. \eqref{eq:eosansatz}. 

To test the assumption of Miller \textit{et al.} \cite{Miller:2021qha}, we use the relation in Eq.\ (\ref{eq:eosansatz}) with $m=2$ to create a set of EoS samples. As shown in the top left-panel of Fig.~\ref{fig:nucEoS} (solid, thin, gray lines), these EoS samples cluster around a mean (solid, thick, black line). The top right panel of this figure shows that speed of sound functional forms constructed using $m=2$ largely approach unity with increasing density. This behavior is highlighted in the bottom panel, which shows $c_s^2$ as a function of pressure. Note that different EoS have different ranges in baryon density for the same range of pressure, so the range of pressures in the top left panel does not correspond to the range of baryon densities in the top right panel. 
We now contrast this set of EoS samples with a new set, constructed from GPs with a softer mean. More specifically, we set $m=1.6$ in Eq.~\eqref{eq:eosansatz}, resulting in the functional forms shown in the top left panel of Fig.~\ref{fig:nucEoS} (dot-dashed, thin, gray lines). As expected by construction, the mean of these samples has a softer slope (dot-dashed, thick, black line). The effect of this softer mean is to reduce the speed of sound to values largely below $\approx 0.4$ in the neutron star range of baryon densities, as shown in the top right panel of Fig.~\ref{fig:nucEoS}. For baryon densities larger than what we expect in neutron stars, the distribution of speeds of sound has a mean of 1/3 (i.e.~the conformal limit), and a scatter that leads to $c_s^2$'s as large as 0.8 and as small as 0.1, as shown in the bottom panel.  From here on, we refer to the set of EoS samples resulting from GPs with $m=2$ and $m=1.6$ as ``hard GP'' and ``soft GP,'' respectively. Figure~\ref{fig:nucEoS} also presents specific realizations of nuclear physics simulations of the EoS, but we defer a discussion of those to Sec.~\ref{subsec:mGP}. 

Why consider a soft GP when astronomical observations seem to indicate that the conformal limit is broken at $n_B \approx 2 \ n_{\rm sat}$~\cite{Fujimoto:2019hxv,Legred:2021hdx,Miller:2021qha,Raaijmakers:2021uju}? Our motivation is to show the effect of softer means in the speed of sound, while at the same time generating a new benchmark model that can be modified through sharp features in a narrow baryon density range to make them consistent with astronomical observations. We will discuss such modifications in the next subsection. 

Before proceeding, let us discuss two other important modifications from the approach in Miller \textit{et al.}~\cite{Miller:2021qha}. The highest value sampled in $\log_{10} p$ is 37, instead of 36 in~\cite{Miller:2021qha}, and we use a significantly finer grid -- Miller \textit{et al.} samples 50 points, whereas we sample 100. These modifications are necessary because our procedure allows for softer EoS, which result in higher neutron star central pressures. Expanding the sampled domain ensures that the entire stable branch is captured, rather than cutting it off at an arbitrary, smaller value. Also, since in the next subsection we will introduce sharp features in $c_s^2$ that lead to rapid changes in the EoS, a finer grid is needed to keep numerical errors under control when recovering the EoS samples from a GP. 

Let us now discuss the second ingredient that defines a GP: the {\bf covariance matrix}. We assume that $\Sigma_{ij}$ is a matrix whose elements are determined through a kernel function of the pair $\{x_i,x_j\}$, where $x_i$ is the point at which the normal distribution is being sampled (i.e., in our case $x$ is the $\log_{10}$ of the pressure in cgs units) and $x_j$ is any other point, i.e.~$\Sigma_{ij} = K(x_i,x_j)$. We further assume a squared-exponential kernel,
\begin{align}\label{eq:kernel}
    K_{\textrm{se}}(x_i,x_j) = \sigma^2\exp{\left[-\dfrac{(x_i - x_j)^2}{2 \ell^2}\right]},
\end{align}
which depends only on the distance between $x_i$ and $x_j$, and on two hyperparameters, $\ell$ and $\sigma$. Specifically, $\ell$ determines the correlation length scale (e.g.,~when $\ell \rightarrow 0$ all points are independent of each other) and $\sigma$ represents the strength of the overall correlation (e.g.~when $\sigma \rightarrow 0$ all points go to the mean). The benchmark models should be smooth, meaning that $\ell$ should be compatible with longer correlations across domain points. In addition, because of the exponential map between $\phi$ and $c_s^2$, a $\sigma$ that is too large would lead to $c_s^2 \approx 0$ or $c_s^2 \approx 1$ more often. Nonetheless, $\sigma$ should not be too small, so that there is enough variability in the EoS samples from any given GP. In accordance with Miller \textit{et al.} \cite{Miller:2021qha}, we set $l=\sigma=1$ for both the hard and soft GP benchmark models. 

Let us now consider the effect of our choice of $\ell$ and $\sigma$ on our EoS samples. Figure \ref{fig:nucEoS} shows that, in the hard GP case, $\sigma = 1$ still allows for enough deviation from the mean to create variability in the EoS samples, without oversampling $c_s^2 \approx 1$ or $c_s^2 \approx 0$. 
In the case of the soft GP, a $\sigma = 1$ leads to oversampling $c_s^2 \approx 0$, since the mean is already at very low values of $c_s^2$. However, for both GPs, $\ell = 1$ heavily suppresses large deviations in $c_s^2$ from one value of pressure to pressures in a close neighborhood, resulting in EoS without sharp, non-trivial features. Our assumptions in the benchmark models do not force $c_s^2$ to increase monotonically; nevertheless, because $\ell=1$ imposes large-scale correlations, non-monotonic behavior is smeared out across a wide range of densities, a feature that is consistent only with a smooth (i.e.~wide) crossover.

\subsection{Modified Gaussian Process}\label{subsec:mGP}

Are the benchmark models discussed in the previous subsection enough to accurately represent nuclear-physics-derived EoS?  Figure~\ref{fig:nucEoS} shows a set of EoS derived from state-of-the-art simulations of chiral mean field (CMF) models \cite{Dexheimer:2008ax,Dexheimer:2017nse,Dexheimer:2009hi,Dexheimer:2020rlp,Clevinger:2022xzl,Dexheimer:2018dhb}, a simulation of the commonly-used quark-hadron crossover EoS framework (QHC19) \cite{Baym:2019iky}, and one example from the relativistic density-functional (RDF) model with density-dependent vector and diquark couplings ~\cite{Ivanytskyi:2022oxv,Ivanytskyi:2022bjc}. In particular, we include in this figure CMF models with delta resonances (D), hyperons (H), and quark (q) degrees of freedom, where the transition to quark degrees of freedom is a first-order phase transition (denoted CMF) or a crossover due to an excluded volume term (denoted CMF$_{\rm ex}$)  for two different parametrizations of the strange vector quark couplings \cite{Dexheimer:2014pea}. The RDF example included here corresponds to the onset of a two-flavor color-superconducting quark phase at $n_B = 0.287$ 1/fm$^3$, the central density of a $\sim1 \ M_\odot$ star in this model. As shown in Fig.~\ref{fig:nucEoS}, exotic degrees of freedom lead to kinks, spikes, and plateaus in $c_s^2$ that occur across short correlation lengths in baryon density \footnote{See also Fig. 2 in Ref.~\cite{Tan:2020ics} for more examples of nuclear physics simulations of EoS with exotic degrees of freedom and how non-trivial features appear in the speed of sound.}. None of the EoS samples drawn from either of the two benchmark GPs is able to reproduce these features. 

This discrepancy between the benchmark GPs and nuclear physics simulations motivates the creation of modified Gaussian processes (mGP). More specifically, we wish to create a modification to the benchmark GPs that lead to EoS samples that contain the short-length correlation structures in the speed of sound that are present in realistic nuclear-physics simulations, while maintaining long-length correlation scales driven by an overarching mean behavior. 
An mGP sample is built from a benchmark GP that serves as a \emph{baseline}, but that is \emph{modified} through the addition of a specific feature in a range of pressures. We do not introduce modifications below \nsat because a variety of experimental constraints (see \cite{Li:2019xxz,MUSES:2023hyz} for a recent review) and $\chi$EFT calculations~\cite{Drischler:2021kxf} require no such structure at these low densities.

Two main reasons drive our choice to introduce modifications to a baseline GP. The first is  direct control over the functional form of $c_s^2$ at a low computational cost. Each modification that is introduced can be directly related to a thermodynamic process and we have the ability to track where and how modifications appear without any post-processing. The second is \emph{a priori} multi-scale correlations in density. We note that GPs with a fixed, but sufficiently small correlation-length can converge to an EoS that displays long, medium, and short correlations in density \emph{a posteriori}, in which case convergence (i.e.~the posterior credible regions are small and centered around the true EoS at all density/mass scales) may require a large number of samples.\footnote{It is also important to note that data may not constrain large changes in the speed of sound over a short range in density very strongly. That is, the likelihood for any individual event may not be very informative, which would required many events to get an informative joint-likelihood. In that case, any tighter credible regions derived with priors constrained to display short-range correlations would be due to the prior rather than the likelihood \cite{reedprivate}.} However, if the GPs are constrained to larger correlation lengths \emph{a priori}, medium and short-range correlations will be exponentially suppressed and an even larger number of samples would be required to converge to a posterior that displays multi-scale correlations \cite{reedprivate}.

In the following paragraphs, we connect the types of modifications we introduce in the mGPs to the phase transition phenomenology from numerical simulations of nuclear physics models and general thermodynamic arguments.

\begin{figure}[]
\centering
\hspace{-0.5cm}\includegraphics[width=\linewidth]{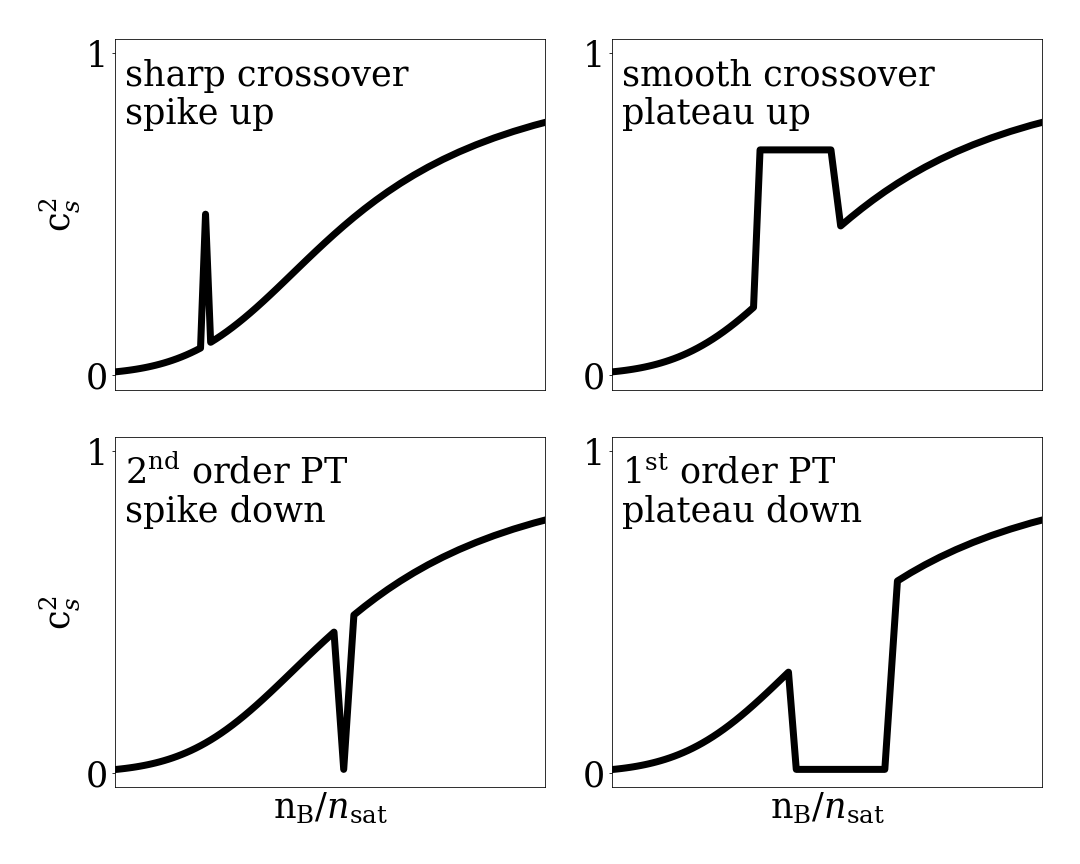} 
    \caption{Cartoon of $c_s^2$ as a function of $n_B$ to illustrate different physical features to incorporate in mGPs. A sharp crossover corresponds to a rapid change in degrees of freedom, where $c_s^2$ will first become stiffer -- due to the onset of, e.g., repulsive or excluded volume interactions -- and then quickly soften with the emergence of new degrees of freedom; this leads to a spike/sharp bump upwards in $c_s^2$ with respect to the baseline (top left). A smooth crossover corresponds to a slower change in degrees of freedom, which we model as a plateau upwards with respect to the baseline (top right). A second-order phase transition corresponds to the critical point at the end of a first-order phase transition line, leading to a very small region (approximately a \textit{point}) where $c_s^2 = 0$, which can be modeled as a spike downwards with respect to the baseline (bottom left). Lastly, a first-order phase transition separates two phases with distinct densities, leading to a gap in $c_s^2$, which can be represented as a plateau, or region, where $c_s^2 = 0 $ (bottom right).} 
    \label{fig:cartoonmods}
\end{figure}

There are a few different thermodynamic phenomena relevant to study of phase transitions. These phenomena are illustrated in Fig. \ref{fig:cartoonmods} as functional forms of the speed of sound as a function of baryon number density and Table \ref{tab:dof}, which connects the different features illustrated in Fig. \ref{fig:cartoonmods} to relevant physical processes and nuclear physics models. We will now discuss each of the categories of physical phenomena and modifications individually, and draw connections to Fig. \ref{fig:cartoonmods} and Table \ref{tab:dof}.

\begin{table*}[]
\normalsize
    \caption{Connection between phase transitions of different orders/crossover to corresponding physical processes in terms of the effect on the speed of sound in equilibrium and modifications in the mGP framework. Note that a first-order phase transition has a jump in baryon density across $\Delta n_B$. 
    }
    \centering
  \begin{tabular}{c | c | c | c}
    \hline
         Transition type & Physical Process & Representation in $c_s^2$ & Modification  \\
         \hline
         \hline 
\sr sharp crossover & \multirow{4}{*}{ \parbox{4.5cm}{quarkyonic matter \cite{McLerran:2018hbz,Fujimoto:2023ioi,Duarte:2023cki,Duarte:2021tsx,Duarte:2020kvi}, 
        percolation to quark matter \cite{Baym:2019iky,Kojo:2015fua},  quark-meson coupling \cite{Stone:2019blq,Guichon:1995ue}  heavy resonances \cite{Dexheimer:2008ax,Dexheimer:2017nse,Dexheimer:2009hi,Dexheimer:2020rlp,Clevinger:2022xzl,Dexheimer:2018dhb,Dexheimer:2014pea},  hyperons \cite{Dexheimer:2008ax,CMF_MUSES,Shahrbaf:2022upc}, chiral-superfluid transition \cite{Hippert:2021gfs}} } & \sr \parbox{4.1cm}{for $\delta \ll 1$,
        if $n_B=\tilde{n}_B \pm \delta$, then  $\left(c_s^2\right)^\prime=\pm \delta^{-1}$
        
        if $n_B=\tilde{n}_B$, then  $\left(c_s^2\right)^\prime=0$} &  \sr spike up,  $c_s^2 \neq 0$ \\ \cline{1-1} \cline{3-4}
\sr smooth crossover &     & \sr     \parbox{4.1cm}{for $\delta >0$,

if $n_B=\tilde{n}_B \pm \delta$, then  $\left(c_s^2\right)^\prime=\pm \delta^{-1}$

        if $n_B\sim \tilde{n}_B$, then  $\left(c_s^2\right)^\prime\sim 0$}        & \sr  plateau up,  $c_s^2 = 0$ \\\cline{1-1} \cline{3-4}
\sr $n^{\rm th}$-order PT, $n>2$ &    & \sr  \parbox{4.1cm}{if $n_B=n_{\rm crit}$, then $d^n p/d\mu_B^n \rightarrow \infty$ } & \sr \parbox{4.1cm}{spike or plateau down, $c_s^2 \neq 0$ }\\
        \hline
         2$^\textrm{nd}$ order PT & \parbox{4.1cm}{critical point due to exotic quark phases} & $c_s^2(n_{\rm crit.}) = 0$ & \parbox{4.1cm}{spike down toward $c_s^2 \approx 0$} \\
         \hline
        1$^\textrm{st}$ order PT & \parbox{4.3cm}{quark deconfinement \cite{Dexheimer:2008ax,Dexheimer:2009hi}, \\ color-superconductivity \cite{Ivanytskyi:2022oxv,Ivanytskyi:2022bjc},
          colorflavor-locking \cite{Alford:2017qgh}} & $c_s^2(n_B)=0$ with $n_B \in [n^*_B, n^*_B + \Delta n_B]$ & plateau down at $c_s^2 \approx 0$ \\
          \hline
    \end{tabular}
    \label{tab:dof}
\end{table*}

In general, a phase transition can be continuous (also known as a \emph{crossover}) or discontinuous. If continuous, all derivatives of the pressure are finite, i.e. 
\begin{equation}
  \left(  \frac{\partial^n p}{\partial \mu_B^n}\right)_{\rm crossover} \neq  \infty,
\end{equation}
and the pressure is an analytic function of the energy density. If a discontinuity exists, i.e.
 \begin{equation}
  \left(  \frac{\partial^n p}{\partial \mu_B^n}\right)_{\rm nth-order}\rightarrow  \infty,
\end{equation}
the phase transition is classified through its order $n$. An ($n^{\rm th}$)-order phase transition is one in which the ($n^{\rm th}$)-order derivative of the pressure with respect to the chemical potential  at the critical baryon density $n_B=n_{\rm crit}$. 

At vanishing temperatures, {\bf crossovers} lead to a non-monotonic peak-like behavior -- i.e.~a bump -- in the speed of sound (see Ref.~\cite{McLerran:2018hbz} for an example and explanation of this behavior).  Mathematically, we can define the center of this bump at $\tilde{n}_B$ and half the range in $n_B$ where this bump occurs will be defined as $\delta$. Then, if the peak of the bump is centered at $\tilde{n}_B$ (we make the assumption here that $\tilde{n}_B$ is centered at the middle of the bump but it does not necessarily have to be the case), we find that at $n_B=\tilde{n}_B$, $\left(c_s^2\right)^\prime\sim 0$.

Crossovers can be further classified into  {\bf smooth} (see, e.g., Ref.~\cite{McLerran:2018hbz}) or \emph{\bf sharp} (see, e.g., Ref.~\cite{Fujimoto:2023ioi}), depending on the abruptness of the non-monotonic behavior. A {\bf sharp} crossover is when the change in degrees of freedom happens rapidly (as expected in quarkyonic matter models~\cite{McLerran:2018hbz,Fujimoto:2023ioi,Duarte:2023cki,Duarte:2021tsx,Duarte:2020kvi}, for instance), such that $\delta\ll 1$ $n_{\rm sat}$ and the peak behavior becomes more of a ``spike'' (although all derivatives of the pressure remain finite). This kind of  crossover is summarized in the top row of Table~\ref{tab:dof}, is illustrated through the cartoon in the top left panel of Fig.~\ref{fig:cartoonmods}, and will be represented in the mGP framework by a spike that rises relative to the baseline. A {\bf smooth} crossover is when the change in degrees of freedom happens slowly so that $\delta >0$  is large and the peak behavior becomes more of a ``plateau'' such that there is no longer a single sharp point in baryon density where the derivative of $c_s^2$ is zero but rather a range of $n_B\sim \tilde{n}_B$ such that $\left(c_s^2\right)^\prime\sim 0$. This kind of crossover is summarized in the second row of Table~\ref{tab:dof},  is illustrated through the cartoon in the top right panel of Fig.~\ref{fig:cartoonmods}, and will be represented in the mGP framework by constant $c_s^2 \neq 0$ in a $\tilde{n}_B\pm \delta$ region.  For both sharp and smooth bumps, we define the bump as an increase of at least $10\%$ compared to the original benchmark functional sampled in the regime of $\tilde{n}_B\pm \delta$. 

Let us consider an example of a sharp crossover in more detail, taking the quarkyonic model as a reference~ \cite{McLerran:2018hbz,Fujimoto:2023ioi}. In this framework, the speed of sound squared is always below the conformal value of 1/3, except in a narrow range of densities where the crossover transition happens. The rapid stiffening is associated with repulsive, excluded volume interactions, followed by a softening of the speed of sound, once quark and gluon degrees of freedom appear in the system. A scenario like this one is equivalent to a soft GP baseline with a spike that rises up in a small baryon density region. Our mGP also includes more general cases where a spike up is added to a hard GP baseline, meaning that $c_s^2$ will not be required to stay below the conformal value neither before nor after the crossover, since those cases cannot yet be ruled out by the data.

Let us now consider a few examples of smooth crossovers in more detail. A minimal set of requirements to create a plateau or ``bump'' structure is discussed in Ref.~\cite{Hippert:2021gfs}, which found that in QCD this feature can be the result of a ``chiral-superfluid" transition, such as the condensation of diquarks or dibaryons. In quarkyonic frameworks, a plateau structure can appear when repulsive excluded-volume terms are partially balanced by the onset of quark degrees of freedom, which stiffen and soften the EoS, respectively \cite{Duarte:2023cki,Duarte:2021tsx,Duarte:2020kvi}. 
In the CMF model, implementing an excluded-volume term for the hadrons leads to a crossover transition to the quark phase \cite{Dexheimer:2014pea}. In the quark-hadron crossover EoS (QHC) \cite{Baym:2019iky}, or three-window modeling of the EoS \cite{Kojo:2015fua}, the crossover regime is constructed via a smooth interpolation of the hadronic and quark regimes, which must also respect thermodynamic constraints such as causality. Lastly, the Quark-Meson Coupling (QMC) class of models, which is based on baryons that interact via the exchange of virtual mesons between confined valence quarks, also gives rise to smooth crossover structure in $c_s^2$ \cite{Stone:2019blq,Guichon:1995ue}. Furthermore, QMC EoS soften rapidly with the onset of hyperons, leading to $c_s^2 < 1/3$ within neutron star densities even when no quarks are produced \cite{Stone:2019blq}, a feature that is relevant for recent discussions on the onset of a ``conformal" regime in the core of massive neutron stars \cite{Annala:2023cwx,Gorda:2022jvk,Annala:2019puf}. Scenarios like the ones described above lead to a rounded peak structure in the speed of sound (see Fig. \ref{fig:nucEoS}), which can be approximated as a plateau at some finite $c_s^2$ that rises above the baseline EoS. We make this approximation for simplicity, given that there are an infinite number of continuous functions that can be constructed in the crossover regime. Although we do not expect this approximation to affect macroscopic observables significantly, it would be valuable to quantify its impact in a future study, accounting for variables such as the width and height of the peak and density dependence. A smooth crossover can also be constructed phenomenologically (see, e.g., Refs.~\cite{Tews:2018kmu,Ivanytskyi:2022wln, Ayriyan:2021prr}).

Discontinuous {\bf phase transitions of order higher than two} lead to speeds of sound that resemble that of crossovers, although technically the higher derivatives of the pressure are not defined and the pressure is thus a non-analytic function of energy density. In fact, for some models, it is still an open question whether certain phase transitions are crossovers or discontinuous phase transitions of finite order \cite{CMF_MUSES}. For this reason, we will model discontinuous phase transitions of order higher than two through spikes and plateaus in the speed of sound that dip below the baseline but do not lead to vanishing $c_{s}^{2}$. The inclusion of these features leads to a variety of non-monotonic structure in $c_s^2$ across some finite range in $n_{\rm B}$.

A {\bf discontinuous phase transition of order two} is sometimes referred to as a ``critical point'' (illustrated in Fig. \ref{fig:cartoonmods}, bottom left) because it is the endpoint of a first-order phase transition line. As presented in Table \ref{tab:dof}, we are not aware of any models that predict a critical point at zero temperature for $\beta$-equilibrated nuclear matter (note that a zero temperature critical point is known as a quantum critical point). However, this possibility cannot currently be ruled out by the data, and thus, we choose to model it. At a critical point, $c_s^2 \approx 0$ only at a critical baryon density $n_{\rm crit.}$. In the mGP framework, we model a critical point as a spike that dips to exactly zero at a single value of baryon density, i.e. $c_s^2 \left(n_{\rm crit.}\right) = 0$. 

A discontinuous phase transition of order one, also known as a {\bf first-order phase transition}, occurs when the transition between two different phases of matter requires a non-zero latent heat. As a result, the two phases have different baryon densities. If the transition density is $n_B^*$,  then the speed of sound displays a gap, i.e.~a region where $c_s^2 \approx 0$, between $n_B^*$ and $n_B^* + \Delta n^*_B$ . The larger the gap, the stronger the phase transition.  This description assumes that the system is in equilibrium and a Maxwell construction was performed to remove any metastable region. In a dynamical system, the first-order phase transition would present as a metastable, or spinodal, region wherein one would see non-monotonic behavior in $c_s^2(n_B), \ n_B \in [n_B^*,n_B^* + \Delta n^*_B]$.  
Since neutron stars are in equilibrium,  a Maxwell construction across the phase transition is a good assumption, leading to a plateau in $p(\varepsilon)$ that results in a region of $c_s^2(n_B) = 0, \ n_B \in [n_B^*,n_B^* + \Delta n^*_B]$. 

First-order phase transitions arise in a variety of nuclear-physics models, as presented in Table \ref{tab:dof}. In the CMF framework \cite{Dexheimer:2008ax, Dexheimer:2009hi},  a first-order phase transition results from a Polyakov loop being used to describe the separation between the hadronic phase with deltas and hyperons from the quark phase. The Triplets model \cite{Alford:2017qgh} contains sequential first-order phase transitions that separate a density-dependent relativistic mean-field model with nucleons and hyperons phase from a 2-flavor quark color-superconducting phase (2SC) and a quark color-flavor-locked phase (CFL). The relativistic density-functional (RDF) model introduced in Ref.~\cite{Ivanytskyi:2022oxv}, generalized to include density-dependent vector and diquark couplings in Ref.~\cite{Ivanytskyi:2022bjc}, contains a first-order transition that separates a hadronic regime \cite{Shahrbaf:2022upc} from a 2SC phase. First-order phase transitions can also be constructed phenomenologically to separate phases from different descriptions using  $n_B^*$ and $\Delta n_B$ as a variable parameter to tune the transition density and the gap in baryon number density between the two phases (see, e.g., \cite{Mondal:2023gbf,Zhang:2023wqj,Han:2018mtj,Benic:2014jia,Ivanytskyi:2022oxv,Ivanytskyi:2022bjc}). Such transitions are straightforward to model with an mGP assuming a Maxwell construction by replacing a portion of the baseline $c_s^2$ with a segment for which $c_s^2 = 0$.  

\section{Choice of Priors} \label{sec:priors}

Now that we have described how EoS are created from the benchmark GP and the mGP models for $c_s^2(\log_{10} p)$, we will specify the relevant prior distributions.  Generally, what constitutes an appropriate prior will depend on the parameters being estimated in a Bayesian analysis. The implicit assumptions we make by modeling the EoS from a non-parametric framework are that (i) the speed of sound at each sampled value of pressure, $c_s^2(p_i)$, is an effective parameter, and (ii) that the method and the hyperparameters we choose for generating $c_s^2(p_i)$ dictate both the
prior distribution and the correlations across the effective parameter space. 

The prior distribution that we choose is, therefore, a statement on our prior beliefs of the allowed values of $c_s^2(p_i)$. When dealing with effective parameters of this type, there are two important aspects to consider. On the one hand, we must model a diverse set of functional forms of $c_s^2(p_i)$ to span a sufficiently large sample of its function space. On the other hand, we must also ensure that this diverse set of functional forms leads to $c_s^2(p_i)$ that are statistically consistent with astronomical observations of neutron stars, i.e.~that, to the best of our knowledge, our prior offers a reasonable description of neutron stars. 

In the next two subsections, we will discuss in detail the priors on the GP and mGP hyperparameters and how we ensure the sample size is large enough and consistent with reliable astronomical observations.

\subsection{Priors on GP and mGP hyperparameters}
Let us first discuss the priors that we choose on the benchmark GP hyperparameters. 
These hyperparameters correspond to the correlation length ($\ell$) and correlation strength ($\sigma$) at each $c_s^2(p_i)$, and the slope ($m$) of the mean function $\mu_i(\log_{10} p_i)$. For $\ell$ and $\sigma$, we choose delta-function priors that  fix these parameters to unity. For $m$, we choose two equal-probability delta-function priors, one peaked at $m=1.6$ and one at $m=2$, such that $50\%$ of the time the benchmark GP corresponds to a hard GP and $50\%$ of the time to a soft GP. 

We will now discuss the priors on the mGP hyperparameters. As noted in Sec.~\ref{subsec:mGP}, the mGP model introduces modifications on top of a baseline, which is modeled through the benchmark GP. The introduction of one spike is controlled by four hyperparameters: a true or false switch $q_{\rm sp}$, a spike magnitude $sp$, a spike direction $\widehat{sp}$, and a spike location $p_{\rm sp}$, such that $c_s^2(p_{\rm sp}) = sp$ if a spike is present. Similarly, the introduction of one plateau is controlled by five hyperparameters: a true or false switch $q_{\rm pl}$, a plateau magnitude $pl$, a plateau direction $\widehat{pl}$, a plateau width in pressure $\Delta p_{\rm pl}$, and a plateau starting location $p_{\rm pl}$, such that $c_s^2(p) = pl$ for $p \in [p_{\rm pl}, p_{\rm pl} + \Delta p_{\rm pl}]$ if a plateau is present.

In this work, we will consider the introduction of one spike, one spike and one plateau, and two plateaus. This implies that every modification is controlled by a choice of the hyperparameter vector $\vec{h} = \vec{h}_1 \cup \vec{h}_2$, where $\vec{h}_1 = \{q_{\rm sp}, q_{\rm pl 1}, q_{\rm pl 2}, \widehat{sp}, \widehat{pl}_1,\widehat{pl}_2\}$ determines whether spikes and plateaus are present, and $\vec{h}_2 = \{sp, p_{\rm sp}, pl_1, \Delta p_{\rm pl1}, p_{\rm pl1}, pl_2, \Delta p_{\rm pl2}, p_{\rm pl2} \}$ determines the properties of these modifications.

The hyperparameters in $\vec{h}_2$ are dependent on the choices made for $\vec{h}_1$ so we first focus on $\vec{h}_1$.
The switch hyperparameters $\{q_{\rm sp}, q_{\rm pl1}, q_{\rm pl2}\}$  determine whether a feature is present, and thus, they can only be $0$ or $1$. Similarly, the unit vector hyperparameters $\{\widehat{sp}, \widehat{pl}_1,\widehat{pl}_2\}$ indicate the direction of a spike or a plateau (i.e.~whether the modification increases or decreases the speed of sound with respect to the baseline) and can only take values of $\pm 1$. We consider the following seven configurations:
\begin{itemize}
\item $\vec{h}_1 = \{0,0,0,\widehat{sp},\widehat{pl}_1,\widehat{pl}_2\}$. No modification is introduced and the mGP reduces to the benchmark model.
\item $\vec{h}_1 = \{1,0,0,\pm 1,\widehat{pl}_1,\widehat{pl}_2\}$. A spike is introduced that goes either above or below the baseline. 
\item $\vec{h}_1 = \{1,1,0,\pm 1,\mp 1,\widehat{pl}_2\}$. A spike is introduced that goes either above or below the baseline, and a plateau is introduced, which goes in the direction opposite to the spike. 
\item $\vec{h}_1 = \{0,1,1,\widehat{sp},\pm 1, \pm 1\}$. Two plateaus with different magnitude are introduced, with both plateaus being allowed to go above or below the baseline. 
\end{itemize}
We assign equal prior probability to each of these four options, implying that 25\% of our samples are from the benchmark GPs and the remaining 75\% come from the mGPs (i.e., out of the total number of samples, 25\% contain a single spike, 25\% contain a spike and a plateau, and 25\% contain a double plateau.) 

The remaining hyperparameters $\vec{h}_2$ have specific allowed ranges, which depend on which of the above four options is drawn. In the single spike case ($\vec{h}_1 = \{1,0,0,\pm 1,\widehat{pl}_1,\widehat{pl}_2\}$), we must choose the height of the spike $sp$ and its location $p_{\rm sp}$, such that at the spike $c_s^2(p_{\rm sp}) = sp$. For the location of the spike, $p_{\rm sp}$, we choose a flat prior with edges at $p(n_B = 1.1 n_{\rm sat})$ and $p = 10^{37}$ erg cm$^{-3}$. For the height parameter, $sp$, we choose different priors depending on whether the spike goes above or below the baseline. If the spike is up, then we choose a flat prior with edges at $sp= 1.1 \, c_{s,\rm benchmark}^2(p_{\rm sp})$ and $sp=1$. If the spike is down, then we choose a flat prior with edges $sp=0$ and $sp= 0.9 \, c_{s,\rm benchmark}^2(p_{\rm sp})$. This choice of prior guarantees that the modified speed of sound squared is never negative, never exceeds unity, and always introduces at least a 10\% modification. 

In the spike plus a plateau case ($\vec{h}_1 = \{1,1,0,\pm 1,\mp 1,\widehat{pl}_2\}$), we must first choose the properties of the plateau to guarantee that there are no spikes within the plateaus. That is, because the plateau has a width $\Delta p_{\rm pl1}$, when sampling the location of the spike, $p_{\rm sp}$, we must not sample within $[p_{\rm pl1}, p_{\rm pl1} + \Delta p_{\rm pl1}]$. Given the above, for the plateau width, $\Delta p_{\rm pl1}$ we sample on $\log_{10} \Delta p_{\rm pl1}$ from a flat prior on the interval $[0.12,1.2]$, with pressure in units of ${\rm erg \; cm}^{-3}$. For the plateau location, $p_{\rm pl1}$, we use a flat prior with edges $p(n_B = 1.1 n_{\rm sat})$ and $p = (10^{37} {\rm erg \; cm}^{-3} - \Delta p_{\rm pl1})$ to ensure the entire plateau falls within the allowed pressure range. Thus, the range of the prior for the location of the spike, $p_{\rm sp}$, must be modified (from the case when there is no plateau) to [$p(n_B = 1.1 n_{\rm sat}), p_{\rm pl1}$) $\cup \ (p_{\rm pl1} + \Delta p_{\rm pl1}, 10^{37}$ erg cm$^{-3}$]. The prior on the plateau magnitude, $pl_1$, is chosen in the same way as the prior on the spike magnitude $sp$ (see paragraph above), but with one modification: instead of setting the edge at 0.9 or 1.1 of $c_{s,\rm benchmark}^2(p_{\rm sp})$, we use 0.9 or 1.1 of $c_{s,\rm benchmark}^2(p_{\rm pl1})$. With this adjustment, the plateau always introduces at least a 10\% modification to the benchmark from the starting point of the plateau, $p_{\rm pl1}$.

In the case with two plateaus ($\vec{h}_1 = \{0,1,1,\widehat{sp},\pm 1, \pm 1\}$), we follow the same procedure as above, but with the following modifications. After drawing a location and width for the first plateau, we must ensure the second one is distinct (i.e.~non-overlapping), and thus, it must be placed to the left or to the right of the first plateau. We enforce this constraint by first drawing $\Delta p_{\rm pl2}$ from the same flat prior as that used for $\Delta p_{\rm pl1}$. We then remove $\Delta p_{\rm pl2}$ from the right side of the intervals [$p(n_B = 1.1 n_{\rm sat}), p_{\rm pl1}$) and $(p_{\rm pl1} + \Delta p_{\rm pl1}, 10^{37}$ erg cm$^{-3}$] and draw $p_{\rm pl2}$ from a flat prior in the interval [$p(n_B = 1.1 n_{\rm sat}), p_{\rm pl1}- \Delta p_{\rm pl2}$) $\cup \ (p_{\rm pl1} + \Delta p_{\rm pl1}, 10^{37}$ erg cm$^{-3} - \Delta p_{\rm pl2}$]. This procedure is computationally efficient and it guarantees the two plateaus do not overlap. 

\begin{figure}
   \centering
\begin{tabular}{c}
\includegraphics[width=0.9\linewidth]{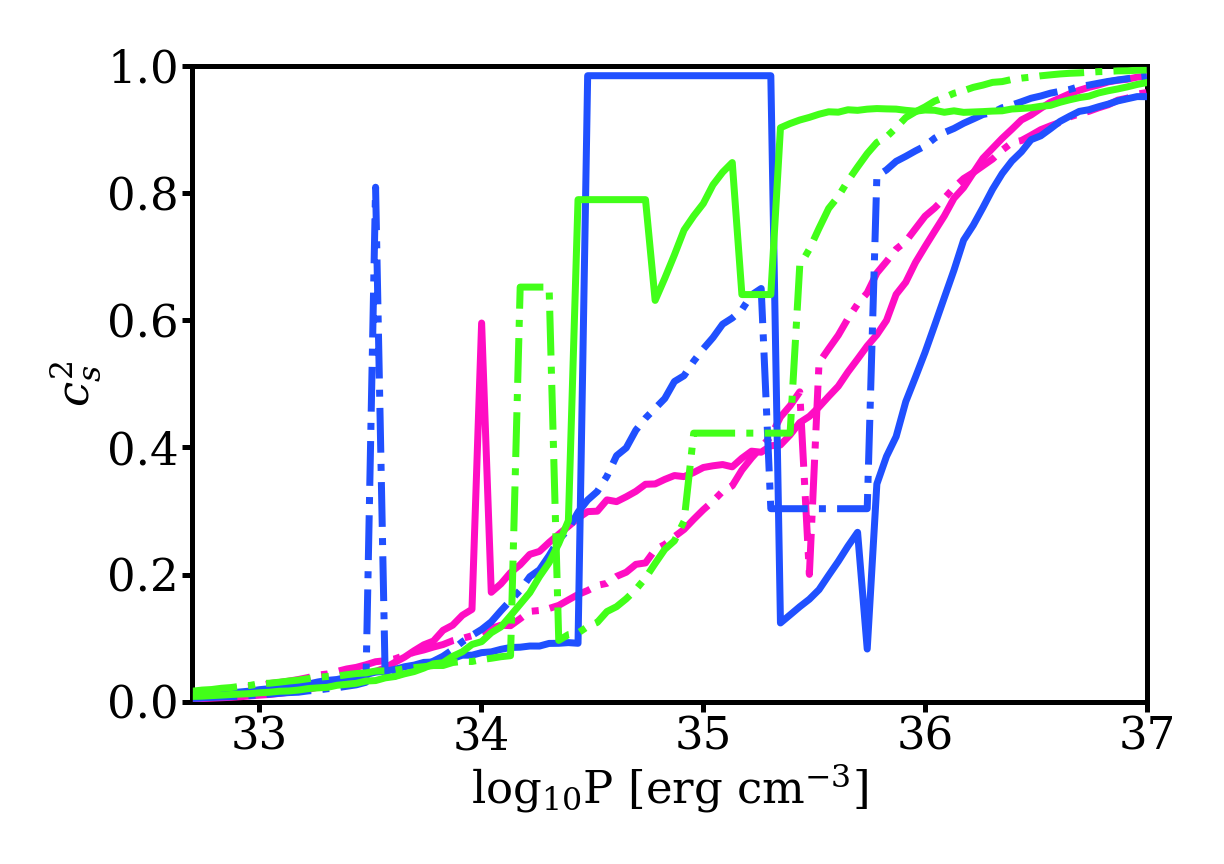} \\
\includegraphics[width=0.9\linewidth]{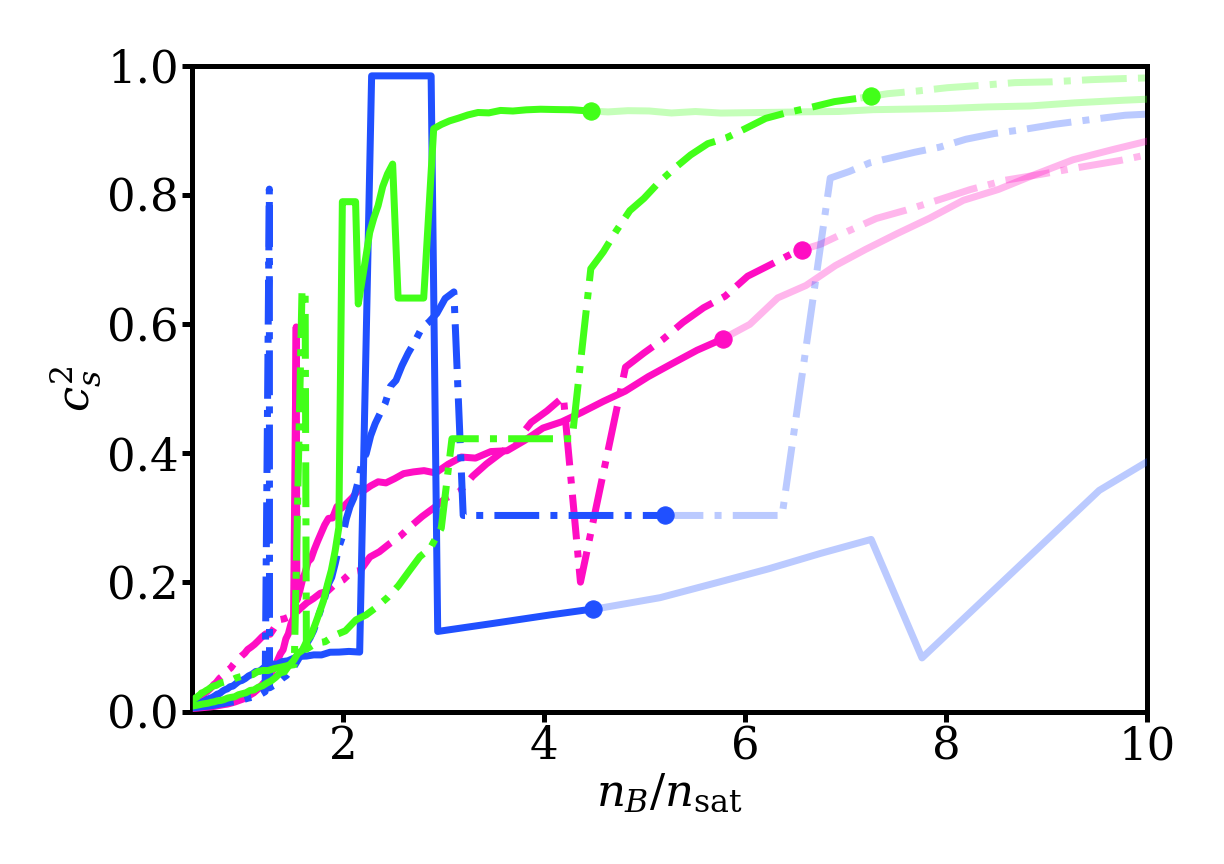} \\
\includegraphics[width=0.9\linewidth]{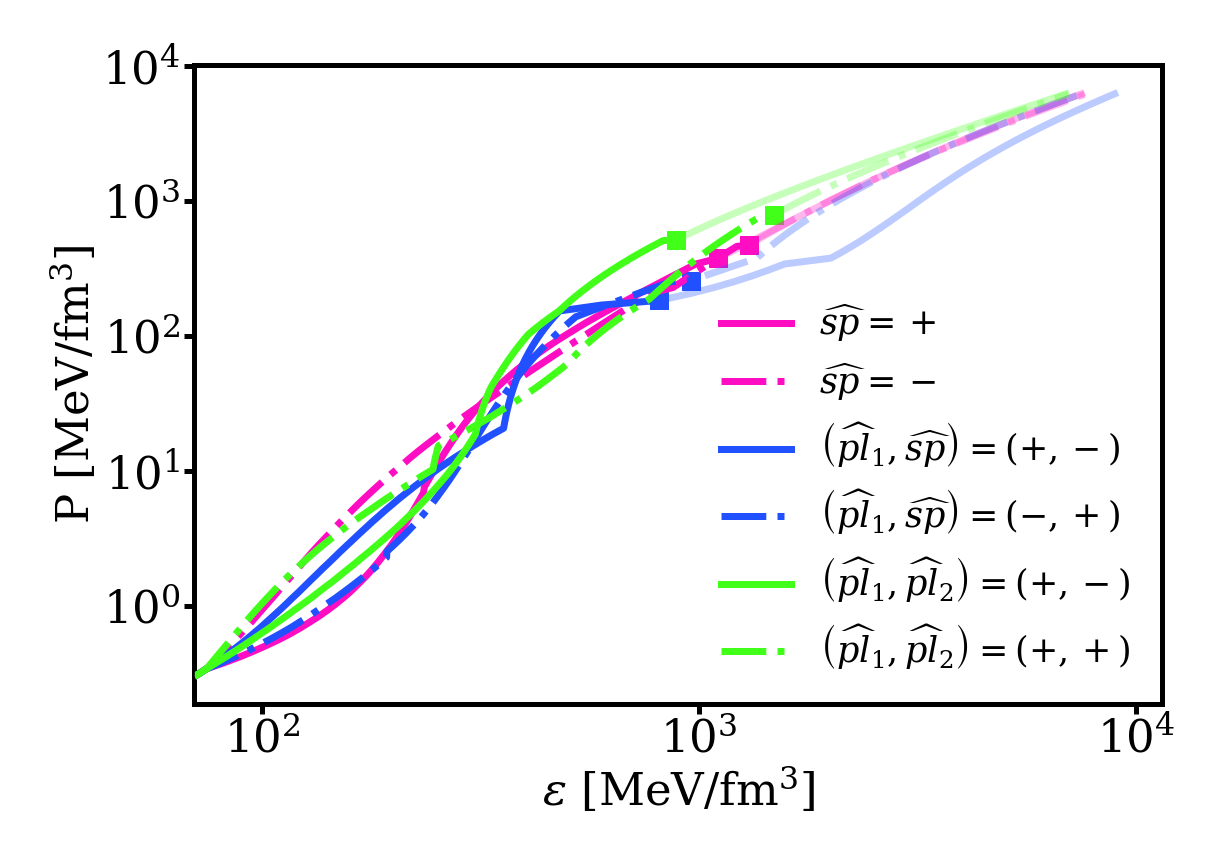}
\end{tabular}
    \caption{Top to bottom: The speed of sound squared in units of $c^2$ as a function of the pressure in units of erg cm$^{-3}$, the speed of sound squared in units of $c^2$ as a function of the baryon number density in units of $n_{\rm sat}$, the pressure as a function of energy density, both in units of MeV fm$^{-3}$, for a representative set of samples generated using the mGP framework. The circles in the middle panel represent the maximal central density predicted for a stable, nonrotating star. The squares in the bottom panel are the corresponding maximal values for the pressure and energy density for a stable, nonrotating star. Samples that contain a spike are shown in pink, samples that contain a spike and a plateau are shown in blue, and samples that contain two plateaus are shown in green. Solid and dot-dashed lines differentiate between different features for samples in the same category, as indicated by the switch parameters $\widehat{sp},\, \widehat{pl}_1$ and $\widehat{pl}_2$, which are defined in Sec.~\ref{subsec:hyperprior}. The mGP framework produces a diverse set of EoS which contain multi-scale correlations across densities at a low computational cost. }
    \label{fig:samples}
\end{figure}

In Fig. \ref{fig:samples}, we show examples from each of the three groups within the mGP framework: a spike modification ($\widehat{sp}=\pm $), a spike and a plateau modification ($(\widehat{pl}_1,\widehat{sp})=(\pm,\mp)$), and a two-plateau modification ($(\widehat{pl}_1,\widehat{pl}_2)=(\pm,\pm)$). In the top panel, the samples are represented by $c_s^2$ as a function of ${\rm log}_{10} P$, where $P$ is in units of erg cm$^{-3}$, exactly as they were generated by the mGP. We can see that the mGP framework succeeds in introducing multi-scale correlations to the speed of sound functional form. Once the samples are generated, the EoS $p(\epsilon)$ is extracted by solving the differential equation $dp/d\epsilon = c_s^2(\log_{10}p)$. In the middle panel, the same samples of $c_s^2$ are shown but now as a function of $n_B/n_{\rm sat}$. In this panel, we also calculate the maximal central density for a stable, nonrotating star and denote it with a circle.

From these two panels, we can make several observations. First, note that in the middle panel the structure in $c_s^2$ is more condensed at low densities and more spread out at large densities, relative to the structure in the top panel. This is because the pressure increases more rapidly as a function of density in the outer layers of the star, where the densities are low. As a consequence, even the smallest structure in $c_s^2$ at low densities introduces structure over a large range of pressures. In contrast, at higher densities, the pressure increases slowly with respect to density. Therefore, dramatic features in $c_s^2$ at high densities translate into structure that arises over a small range in pressures. 

From this comparison, we also arrive at an important conclusion regarding the optimal variable to sample over when introducing modifications. As shown in the middle panel, most stable, nonrotating stars (those with darker lines) will reach at most $n_{\rm B} \sim 7 \ n_{\rm sat}$. The density regime between $1-7 \ n_{\rm sat}$ is precisely where structure in $c_s^2$ is spread out over a larger interval in $P$. This implies that sampling over $P$ will allow us to resolve this structure better than if we sampled over $n_B$, for a finite resolution. This is the choice, i.e.~to sample in pressure, that we will make henceforth in this paper. 

The bottom panel of Fig.~\ref{fig:samples} displays the EoS samples on the $\varepsilon-P$ plane, with squares denoting the maximal central pressure and energy density for each EoS. From this panel, it is clear that the modifications introduced in $c_s^2$ are not causing the EoS to significantly deviate from each other. In fact, we see that large changes in the speed of sound translate into rather small changes in the EoS, leading to clustering around a region in the $\varepsilon-P$ plane. We stress here that Fig.~\ref{fig:samples} only shows six representative samples of $c_s^2$ and EoS out of the nearly one million samples that we study in this paper.

\subsection{Astronomical observables}\label{subsec:astro_obs}

We have specified a prior for the EoS, a quantity that we cannot directly measure. Thus, we need to translate the information carried by the EoS into astronomical observables if we wish to infer properties of the former. For each EoS, we can calculate the mass-radius ($M-R$), moment of inertia-mass ($I -M$), quadrupole moment-mass ($Q-M$), and tidal deformability-mass curves ($\Lambda-M$), which encode the properties of neutron stars of different central densities. These properties can then be compared to astronomical observations to determine the validity of the EoS. In this subsection, we will discuss briefly how these quantities are obtained, following mostly Ref.~\cite{Yagi:2016bkt}. In the next subsection, we will explain how we use astronomical observations of these quantities to inform our prior sample size. 

Millisecond pulsars rotate slowly compared to their mass-shedding limit, and any radius corrections resulting from rotation are significantly below measurement precision for the NICER mission \cite{Miller:2021qha}. Therefore, for calculating the mass-radius curve, we can approximate milisecond pulsars as nonrotating, isolated objects. The condition that a stable nonrotating star must be in hydrostatic equilibrium yields the well-known Tolman-Oppenheimer-Volkoff (TOV) equation. For each EoS, given a central pressure $p_c$, the TOV equation will describe how the mass and pressure vary with the radial coordinate up to a limiting value, $R_*$, where $p(R_*) \approx 0$. The mass of the star is given by $M=M(R_*)$ and the stellar radius is $R = R_*$. By repeating this process for multiple values of $p_c$, we get the $M-R$ curve for a given EoS. 

At some value of central pressure, the $M-R$ sequence will become unstable, which occurs when $dM/dp_c<0$. This value of the central pressure marks the end of what we refer to as the (stable) neutron star EoS, and establishes a maximal allowed value for the central pressure, $p^{\rm max}_c$, for astrophysically realistic stars. Note that the zero-temperature QCD EoS continues beyond $p^{\rm max}_c$, but we do not expect to observe stable, isolated, non- or slowly-rotating pulsars with central pressures exceeding $p^{\rm max}_c$, assuming the EoS used in the calculation is correct. We also note that the $M-R$ sequence can have multiple stable branches separated by unstable regimes where $dM/dp_c<0$. When that is the case, $p^{\rm max}_c$ is still the largest density corresponding to the end-point of a stable branch, but it might not correspond to the central density of a maximally massive star for a given EoS.

Now, let us consider an isolated star that rotates with dimensionless angular velocity $\Omega$, where we have normalized the stellar angular velocity by the mass-shedding limit. We assume $\Omega$ is small enough that the solutions to the Einstein equations can be expanded in powers of this dimensionless angular frequency. 
Note that to $\mathcal{O}(\Omega^0)$, the $(t,t)$ and $(r,r)$ components of the Einstein equations, in conjunction with the stress-energy conservation equation, yield the TOV and continuity equations. At $\mathcal{O}(\Omega)$, the Einstein equations only modify the gravito-magnetic sector of the metric, whose exterior behavior is characterized by the moment of inertia $I$. At $\mathcal{O}(\Omega^2)$, we obtain a correction to the total mass and to the mass distribution within the star, which now acquires an oblate spheroidal shape due to rotational motion. The latter is controlled by the quadrupole moment $Q$. Both $I$ and $Q$ vary with the central density of the star, such that we can obtain solutions for a range of central pressures and relate them to the $M-R$ curve to obtain the $I-M$ and $Q-M$ curves.

Finally, we will consider a non-rotating star in the presence of a companion compact object, which causes the star to tidally deform. We can study the redistribution of mass due to the external perturbation through a multipolar expansion. 
The deformation at leading order in perturbation theory is dominated by the quadrupole moment. The quadrupolar deformation is controlled by the Love number, or its dimensionless version, the tidal deformability $\Lambda$. The tidal deformability can be calculated by solving the linearized Einstein equations combined with continuity and differentiability arguments. Once again, the exact solution for $\Lambda$ requires an EoS and is dependent on the central pressure of the star \cite{Hinderer:2007mb}. The calculation can be repeated for a range of central pressures to obtain the $\Lambda-M$ curve.

\subsection{Prior sample size}\label{subsec:hyperprior}

The priors on neutron star observables, such as their mass, radius and tidal deformability are determined by the prior on the EoS. But how do we know that we have chosen a good EoS prior? 
For each measurement available, we want enough samples in the prior that offer a reasonable match to observations as predictions. Due to the functional complexity allowed by both the GP and mGP frameworks, we expect that most EoS generated will not meet basic requirements based on neutron star observations. With that in mind, we use three metrics to gauge the how well samples in the prior describe astronomical measurements. First, we check that $M_{\rm max} \geq 1.8$ $M_\odot$ based on the observation of three high-mass pulsars \cite{arzoumanian2018nanograv, antoniadis2013massive,cromartie2020relativistic}. Second, based on the inference of the tidal deformability of a 1.4 $M_\odot$ star, $\Lambda_{1.4}$, from GW170817 \cite{de2018tidal}, we check that $ 10 \leq \Lambda_{1.4} \leq 2000$. Third, based on NICER's inference of the mass-radius posterior for PSR J0030+0451 \cite{Miller:2019cac,Riley:2019yda}, we check that $ 8.0 \leq R_{1.4} \leq 16.0$ km. We note that these bounds are far outside the 90\% credible region for the most constraining estimates of the maximum mass \cite{NANOGrav:2023hde}, radius of a 1.4 $M_\odot$ star \cite{Miller:2021qha}, and tidal deformability of a 1.4 $M_\odot$ star \cite{de2018tidal}. 

For every sample EoS that we draw from our GP or mGP framework, we keep tally of whether the three checks specified above are passed or not. We emphasize that these metrics are \textit{not} used to cut the sample size or to modify our priors in any way. We simply track how many samples pass these checks to ensure that we have enough strong candidate EoS in our prior sample. In particular, we continue drawing samples until we have obtained a subset of at least 100,000 candidate EoS that pass all three checks. This requires that we sample about 1,000,000 times from the benchmark GP and mGP frameworks.  

Another benefit of checking our priors in terms of astronomical observables is that we can assign zero likelihood to EoS that fall outside the intervals we defined above. This is because those sample EoS are already in significant conflict with the observations discussed above, and thus, their likelihood will be very close to zero. We can justify this approach as follows. Consider an observable $Y$ at the value $y_k$ predicted by EoS $k$, which we will parametrize in terms of a vector $\vec{\phi}_k$ (see Eq.~\eqref{eq:phitocs2}, where here the vector symbol denotes the 100 values of $\phi$ that we sample at each point in pressure). Let us also consider a set of $N$ total number of EoS, such that $k$ is between 1 and $N$. Then, the conditional probability of $y_k$ given EoS $k$ is  
\begin{equation}\label{eq:posterior_obs}
    P(y_k) = \dfrac{q(\vec{\phi}_k)\mathcal{L}(\vec{\phi}_k)}{\sum_i^N q(\vec{\phi}_i)\mathcal{L}(\vec{\phi}_i)},
\end{equation}
where $q(\vec{\phi}_k)$ is the prior probability assigned to EoS $k$ and $\mathcal{L}(\vec{\phi}_k)$ is the likelihood of the data given EoS $k$. Let us now order the values of $y_k$ from smallest to largest, such that 
\begin{eqnarray}
          P(y_1 \leq y \leq y_2) = \sum_k P(y_k), \, y_k\in[y_1,y_2],
\end{eqnarray}
defines the credible region delimited by $y_1$ and $y_2$. For most observables, only a limited domain in $y_k$ will have nonzero likelihood. That is,
\begin{eqnarray}
     P(y_{\rm low} \leq y \leq y_{\rm high}) = \sum_k P(y_k)\approx 1 , \, y_k\in[y_{\rm low},y_{\rm high}],
\end{eqnarray}
where $P(y_k) \approx 0 $ for $y_k$ outside the interval $[y_{\rm low},y_{\rm high}]$. 
For $N_{\rm obs}$ different observables (e.g., maximum mass, radius at $1.4~M_\odot$, or dimensionless tidal deformability at $1.4~M_\odot$), we then simply have a $N_{\rm obs}$-dimensional region $[y_{\rm low}^1,y_{\rm high}^1] \cup [y_{\rm low}^2,y_{\rm high}^2] \cup \ldots \cup [y_{\rm low}^{N_{\rm obs}},y_{\rm high}^{N_{\rm obs}}]$ outside of which we expect $P(y_k) \approx 0 $.  
In our case, we consider three observables, so the region outside of which the posterior is approximately zero is [$M_{\rm max, low} = 1.8 \, M_\odot, M_{\rm max, high} = \infty) \cup  [R_{1.4,{\rm low}} = 8.0 \textrm{ km}, R_{1.4,{\rm high}} = 16.0 \textrm{ km}] \cup [\Lambda_{1.4,{\rm low}} = 10, \Lambda_{1.4,{\rm high}} = 2000]$.

We can implement the condition that the posterior is zero outside of the above region as follows. First, we divide our prior sample into two subsets, one for which each EoS meets all the requirements specified above ($\Phi_\checkmark$), where
\begin{eqnarray}
    \Phi_\checkmark = & \, \{\vec{\phi}_k : M_{\rm max, k} \geq 1.8 M_\odot \, \wedge R_{1.4,k} \in [8.0\textrm{ km},16.0\textrm{ km}] \nonumber \\
    & \wedge \Lambda_{1.4,k} \in [10,2000]\},
\end{eqnarray}
and one for which all EoS fail at least one of the checks ($\Phi_\times$), such that
\begin{equation}
    \Phi = \Phi_\checkmark \cup \Phi_\times.
\end{equation} 
With this in hand, we now define the likelihood of the data given EoS $k$ to be $\mathcal{L}(\vec{\phi}_k)_\checkmark = \mathcal{L}(\vec{\phi}_k)$ if $\vec{\phi}_k\in \Phi_\checkmark$, and we define $\mathcal{L}_\times(\vec{\phi}_k) = 0$ if $\vec{\phi}_k\in \Phi_\times$. We emphasize again that the prior distribution remains unchanged, so this procedure is in no way equivalent to performing cuts on the prior. 

Our goal is to generate enough samples to ensure that $\Phi_\checkmark$ contains at least $\sim$100,000 EoS. Using the $M_{\rm max}, R_{1.4}$, and $\Lambda_{1.4}$ checks as a guide, we generate 900,000 EoS. Out of this total sample, 104,594 EoS passed the checks, and therefore, contribute non-negligibly to the posterior distribution of the observables discussed later in Sec.~\ref{sec:results}. Note that the number of samples in $\Phi_\checkmark$ is roughly 10\% of the total number of samples generated. Based on this result, we argue that studies using non-parametric methods, or any method that allows for a vast functional space, should implement similar checks, or at least verify the robustness of results for different prior sample sizes. 

\section{Statistical Methods}\label{sec:statmethods}

There should be a unique EoS that correctly describes all neutron stars in the universe. However, honing in on this exact EoS would only be possible with infinitely precise observations. A more common and realistic approach is to obtain posteriors that describe the probability of a given EoS by comparing its predictions against data. Using an ensemble of theoretical models for the EoS, each with a corresponding posterior probability, we can extract credible regions for the EoS that occurs in nature. This method requires us to first state our prior beliefs about the EoS, which then get updated as we gain knowledge of the EoS through data. 

We have introduced in Sec.~\ref{sec:EOSgen} two frameworks for generating theoretical models for the EoS: benchmark GPs and mGPs. Those frameworks reflect two different beliefs about the EoS. The GP assumes the EoS displays long-range correlations in pressure, resulting in smooth functional forms for $c_s^2(p)$. This belief is compatible with nuclear physics simulations for hadronic models, or models that display a smooth crossover into an exotic phase, where the change in the degrees of freedom happens over a wide range in density. On the other hand, the mGP framework assumes the EoS contains nontrivial degrees of freedom or interactions that lead to sudden changes in $c_s^2(p)$ in the form of kinks, spikes, and plateaus. These features are predicted by many state-of-the-art nuclear physics simulations with exotic degrees of freedom. With these two distinct prior beliefs in mind, our goal is to assess if one framework is better at accounting for observations than the other. 

We attempt to answer this question using a fully Bayesian approach. In Sec.~\ref{sec:priors}, we detailed how we generate a prior distribution using the benchmark GP and the mGP as theoretical frameworks. Now, we need to discuss how we calculate posterior distributions by incorporating constraints on the EoS of neutron stars from astronomical observations, controlled terrestrial experiments, and perturbative QCD calculations \footnote{For a detailed discussion on current constraints on the QCD EoS across different regimes, see Ref.~\cite{MUSES:2023hyz}.}, and how we quantify each framework's ability to describe observations.  

We begin this section with a Bayesian ``primer," where we review a generic approach for implementing our knowledge about the EoS into a posterior distribution and how we can determine whether observations favor one of the EoS frameworks over the other using the Bayes factor. Obtaining posterior distributions requires specific choices and assumptions for which observations and associated likelihood factors are used. Those are discussed in a dedicated likelihood subsection. Similarly, we devote a separate subsection to explaining how the model evidence for the benchmark GP and the mGP are determined -- a requirement for calculating the Bayes factor.

\subsection{Bayesian primer}\label{subsec:primer}

Consider an EoS $k$ which is represented by a set of values sampled from either the benchmark GP or the mGP, $\vec{\phi}_k$. Bayes' theorem states that the posterior probability of EoS $k$ is proportional to the product of a prior term and the likelihood of the data given $\vec{\phi}_k$, 
\begin{equation}\label{eq:bayes}
  P_{\rm EoS}(\vec{\phi}_k) \propto q(\vec{\phi}_k) \mathcal{L}(\vec{\phi}_k),
\end{equation} 
where we recall that $q(\vec{\phi}_k)$ is the prior probability distribution encoding our prior beliefs about how likely EoS $k$ is to occur, and we recall that the likelihood term $\mathcal{L}(\vec{\phi}_k)$ reflects how well predictions from EoS $k$ match observed properties. 

In this paper, we assume that all observations are independent of each other, which means the likelihood of a set of observations $(i,j)$ for EoS model $k$ can be written as

\begin{align}\label{eq:likelihood}
    \mathcal{L}(\vec{\phi}_k)= \prod_i \left [\prod_{j=1}^{j(i)} \mathcal{L}_k(i,j) \right ],
\end{align}
where $i$ is a type of measurement (e.g.~mass, radius) and $j$ is an independent measurement of type $i$ (e.g.~two independent measurements of the mass of one object). 
For each of the measurements, we must make a choice for how it will be incorporated into the analysis via a likelihood function, $\mathcal{L}_k(i,j)$. We will discuss our choice of likelihood functions and specify which measurements we include in our analysis in the next subsection.

Let us now instead return to Eq.\ (\ref{eq:bayes}). The normalizing factor that would make Eq.\ (\ref{eq:bayes}) an equality is called the model evidence. The model evidence assesses the ability of a set of prior beliefs to account for observations. In the case of nonparametric EoS, the evidence can be defined as
\begin{equation}\label{eq:evidence}
    \mathcal{E}_m = \int_{\Phi_\textrm{m}} \mathcal{L}(\vec{\phi}_k)q(\vec{\phi}_k) dk,
\end{equation}
where $\Phi_m$ is the set of samples in the prior that were generated using a specific theoretical framework $m$. The goal is often to have competing frameworks such that we can compute the model evidence for each one and then take the ratio between them. This ratio between model evidences is known as the Bayes factor, 
\begin{equation}
    K = \dfrac{\int_{\Phi_{\textrm{m}1}} \mathcal{L}(\vec{\phi}_{k_1})q(\vec{\phi}_{k_1}) dk_1}{\int_{\Phi_{\textrm{m}2}} \mathcal{L}(\vec{\phi}_{k_2})q(\vec{\phi}_{k_2}) dk_2},
\end{equation}
where m1 and m2 indicate distinct theoretical frameworks with different equations of state in their samples (indexed here by $k_1$ and $k_2$). When the Bayes factor deviates significantly from unity, it indicates that the data prefer one model and prior over the other. 

\subsection{Likelihood}
As stated in Eq.~(\ref{eq:likelihood}), we assume that all measurements we take into account are independent, and that systematic errors can be neglected such that the total likelihood is a product of individual likelihood factors for each measurement. In particular, we will consider estimates of the nuclear symmetry energy, the three highest reliably measured pulsar masses, two NICER simultaneous mass and radius measurements, and tidal deformability estimates from two gravitational-wave events. Additionally, we will incorporate a perturbative QCD weight~\cite{Gorda:2022jvk}, which accounts for the behavior of the EoS at very large ($\sim40$ $n_{\rm sat}$) densities from pQCD calculations. In summary, Eq.~(\ref{eq:likelihood}) can then be written as

\begin{equation}
    \mathcal{L}(\vec{\phi}_k)= \mathcal{L}_{S}(\vec{\phi}_k) \mathcal{L}_{\rm Mmax}(\vec{\phi}_k) \mathcal{L}_{\rm M-R}(\vec{\phi}_k)  \mathcal{L}_{\Lambda}(\vec{\phi}_k)w_{\rm pQCD}(\vec{\phi}_k),
\end{equation}
where $S$ denotes the likelihood factor associated with symmetry energy measurements, $M_{\rm max}$ that associated with high-mass pulsar measurements, $M-R$ that associated with simultaneous mass-radius measurements, and $\Lambda$ that associated with tidal deformability measurements. We represent input from pQCD not as an additional likelihood factor but as a weight, $w_{\rm pQCD}$. We make this choice because the uncertainty in the pQCD input stems from its poorly constrained regime of applicability and uncertainty around the missing higher-order term error in truncated results, in contrast to traditional measurements with quantifiable statistical uncertainties that can be consistently included in a Bayesian framework. Our approach for incorporating observational and experimental constraints on the EoS is based on Refs.~\cite{Miller:2019nzo,Miller:2019cac,Miller:2021qha}, while the use of pQCD input is based on Ref.~\cite{Gorda:2022jvk}. We review and discuss the most important aspects of these approaches as they pertain to our analysis below, and refer the reader to the corresponding original works for further detail. 

\subsubsection{Symmetry energy}

In terrestrial experiments, it is possible to probe the $T \rightarrow 0$ limit of dense nuclear matter with low-energy collisions of heavy-ions \cite{MUSES:2023hyz}. However, the nuclei used in these experiments have a charge fraction (the ratio of proton number to baryon number or, in other words, electric charge density $n_{\rm Q} $ over baryon density $n_{\rm B} $) of $Y_{\rm Q}=n_{\rm Q}/n_{\rm B}  \sim 0.4-0.5$. A value of $Y_{\rm Q}=0.5$ is known as symmetric nuclear matter, SNM, because there are an equal number of protons and neutrons in the system. On the contrary, neutron stars are primarily neutron-rich, with $Y_{\rm Q}\sim0.001-0.2$, thus probing the asymmetric nuclear matter regime. Pure neutron matter (PNM) is the limit where $Y_{\rm Q}=0$. 

The densities probed in these low-energy heavy-ion experiments are at or near $n_{\rm sat}$, and in that regime, experiments and \CEFT  calculations can extract properties that are relevant to the EoS.  A nucleus is composed of $Z$ protons and $A-Z$ neutrons, where $A$ is the total number of protons and neutrons in the nucleus. The mass of the nucleus $m_A$ that contains $A$ nucleons is always less than the masses of the individual protons, $m_p$, and neutrons, $m_n$, summed together, i.e.,
\begin{equation}
    m_A<Zm_p+(A-Z)m_n,
\end{equation}
because a finite amount of energy is released in the formation of a nucleus.  That difference in the rest mass energy per nucleon is known as the binding energy and is defined as (we remind the reader that the speed of light is $c=1$ in this work)
\begin{equation}
    B=\frac{1}{A}\left[m_A -\left(Zm_p+(A-Z)m_n\right)\right].
\end{equation}
It is often assumed that the mass of the proton and neutron are identical since their masses differ by just over $1$ MeV.  Setting the neutron and proton mass to be the same and calling this the nucleon mass for simplicity, $m_p = m_n = m_N$, the above simplifies to
\begin{equation}
    B=\frac{m_A}{A}-m_{N}\equiv \frac{E}{A}\, ,
\end{equation}
where $E/A$ is the nucleonic energy per particle.
As defined here, the binding energy does not depend explicitly on the mass of the nucleon because the $m_N$ dependence in the first term of the above equation cancels the second term exactly. 

The binding energy for SNM at $n_{\rm sat}$ is estimated to be $B\sim-16$ MeV from previous global analyses\footnote{Uncertainty quantification was not performed in these studies.} that extracted the volume term of the liquid drop model from a large sample of nuclei, which reported values of $B= - 15.77$ \cite{Myers:1966zz} and $B= - 16.24$ MeV \cite{Myers:1995wx}. One can also use \CEFT tuned to a large number of experimentally measured nuclei wherein one obtains $B=-15.86\pm 0.57$ MeV \cite{Drischler:2020yad}. However, in this work we assume $B=-16$ MeV is exact. 

The next quantity that can be measured from nuclear experiments is known as the symmetry energy, which we denote as $S$. At $n_{\rm sat}$, the symmetry energy is the difference in total energy between the SNM and PNM limits, i.e.
\begin{equation}
    S(n_{\rm sat})=\frac{1}{A}\left(E_{\rm PNM}-E_{\rm SNM}\right),
\end{equation}
or, in terms of energy densities, we can write
\begin{equation}\label{eqn:PNM_Esym}
    S(n_{\rm sat})=\frac{1}{n_{\rm sat}}\left(\varepsilon_{\rm PNM}-\varepsilon_{\rm SNM}\right),
\end{equation}
where we can relate the energy densities to the total energy via
\begin{equation}\label{eqn:convEA_den}
    \frac{\varepsilon}{n_B}=\frac{E}{A}+m_{N}.
\end{equation}

Neutron stars are not exactly in the limit of PNM since a small fraction of protons exists is also present.  Thus, for asymmetric nuclear matter (ANM), where the the system is at finite value of $Y_Q$, the symmetry energy can be expanded about $Y_Q = 1/2$ to obtain
\begin{equation}\label{eqn:symE_expan}
    S_2(n_{B})\left[1-2Y_Q\right]^2+\mathcal{O}\left[1-2Y_Q\right]^4=\frac{1}{n_{B}}\left(\varepsilon_{\rm ANM}-\varepsilon_{\rm SNM}\right),
\end{equation}
where the factor of $1/2$ in the Taylor expansion is reabsorbed into the quadratic term $S_2(n_B)$. 
The quadratic coefficient of the Taylor expansion, $S_2(n_B)$, can then be further Taylor expanded about $n_B = n_{\rm sat}$, but in this paper we will retain only the leading-order term in this expansion and set $S(n_{\rm sat}) = S_2(n_{\rm sat})$ in Eq.~\eqref{eqn:convEA_den} (see \cite{Li:2019xxz} for a derivation and further details). 

We can relate the symmetry energy to the binding energy by substituting in Eq.\ (\ref{eqn:convEA_den}) into Eq.\ (\ref{eqn:symE_expan}) at $n_B= n_{\rm sat}$ for the $\varepsilon_{\rm SNM}/n_{\rm sat}$ term to find
\begin{eqnarray}
    S(n_{\rm sat})\left[1-2Y_Q\right]^2&\approx&\frac{\varepsilon_{\rm ANM}}{n_{\rm sat}}-\left[\frac{E}{A}+m_{N}\right]\\
    &=&\frac{\varepsilon_{\rm ANM}}{n_{\rm sat}}-\left[B+m_{N}\right].
\end{eqnarray}
In an extreme extension of Taylor expansions, however, we will evaluate the above expression at very small $Y_Q$, because state-of-the-art \CEFT models predict a value for $S(n_{\rm sat})$ and indicate that $Y_Q\sim 0.05$ at $n_{\rm sat}$ for $\beta$-equilibrated, cold nuclear matter \cite{Drischler:2020fvz}. Therefore, setting $Y_Q = 0$, $B=-16$ MeV and $m_N = m_n= 939.6$ MeV in the above equation, we obtain
\begin{eqnarray}\label{eq:Esym}
    S(n_{\rm sat})&\sim&\frac{\varepsilon_{\rm PNM}}{n_{\rm sat}}-923.6 \mathrm{ [MeV]}.
\end{eqnarray}

Given all of the above, we assume a likelihood factor associated with the symmetry energy of the form 
\begin{equation}
    \mathcal{L}_S(\vec{\phi}_k) = \mathcal{L}(S_0 |, S_k (n_{\rm sat})) = \dfrac{1}{\sqrt{2\pi(\sigma_S)^2}}\exp\left[-{\dfrac{( S_k(n_{\rm sat}) - S_0)^2}{2(\sigma_S)^2}}\right],
\end{equation}
where $S_k(n_{\rm sat})$ now denotes the EoS $\vec{\phi}_k$ from which the $S(n_{\rm sat})$ is obtained using Eq.~(\ref{eq:Esym}), assuming that $\varepsilon_{\rm PNM} = \varepsilon_k(n_{\rm sat})$, and we take the observed value of the symmetry energy at nuclear saturation density to be $S_0 = 32$ MeV with a standard deviation of $\sigma_S = 2 \MeV$ \cite{Tsang:2012se,Li:2019xxz}. 

Future work could consider other available constraints~\cite{Li:2019xxz,Drischler:2020hwi} on the symmetry energy. Additionally, one could fold into the analysis the uncertainty on the value of the binding energy, the uncertainty on the value of $n_{\rm sat}$ itself, the systematic error introduced by neglecting deviations from exact charge fraction asymmetry (i.e.~the small $Y_Q$ that we set to zero in this derivation).\, and the slope of the symmetry energy. The impact of uncertainties around the binding energy, saturation density, and small deviations from PNM at $n_{\rm sat}$ mostly affect stars with masses below 1.0 $M_\odot$ \cite{Miller:2019nzo}. A more promising direction is to include the slope of the symmetry energy, $L$, which would provide constraints when $n_B\neq n_{\rm sat}$. However, the exact value of the slope of the symmetry energy is poorly constrained with current theory and experiments. We also refer the reader to Refs.~\cite{Essick:2021ezp,Essick:2021kjb}, which extract from a GP prior and neutron star measurements values for $S_0$, $L$, and the neutron skin thickness of \textsuperscript{208}Pb.

\subsubsection{High-mass pulsars}
In principle, we could gain information about the EoS from any neutron star mass measurement. Given a fully parameterized model for neutron star birth and accretion and a prior for the neutron star EoS, we could make population predictions that can be compared to measurements. However, due to both computational and theoretical challenges, such a complete analysis is not currently feasible. Without any information about how stars form and accrete, we can focus instead on the maximum mass, since any realistic EoS must be able to support the highest reliably measured masses. As discussed in Sec.~\ref{subsec:astro_obs}, in the slow rotation regime, the maximum mass is a function only of the EoS. We implement a likelihood factor for EoS $k$ given a likelihood function for the mass of star $j$ obtained from a radio observation of a binary pulsar, $\mathcal{L}(M_j|M(\rho_c))$, that can be written as
\begin{equation}
    \mathcal{L}(M_j | k) = \int_{\rho_{\rm min}}^{\rho_{\rm max}} q(\rho_c)\mathcal{L}(M_j|M(\rho_c))d\rho_c,
\end{equation}
where $\rho_c$ is the central density. The prior on the central density, $q(\rho_c)$, is calculated for each EoS from a distribution that is quadratic between the central density of a 1 $M_\odot$ star, which we denote $\rho_{\rm min}$, and that of a maximally massive star, $\rho_{\rm max}$, assuming EoS $k$. That is, we sample uniformly between $0 \leq x \leq 1$ for $\rho_c = \rho_{\rm min} + x^2 (\rho_{\rm max} - \rho_{\rm min})$. We use a quadratic prior to avoid giving greater prior weight to more massive stars, since the central density changes more rapidly near the maximum mass \cite{Miller:2021qha}. Outside the interval $[\rho_{\rm min},\rho_{\rm max}]$ the prior probability is zero. We also assign zero prior probability for any $\rho_c$ resulting in an unstable star, such that if two or more stable branches are connected by an unstable branch, the unstable branch does not contribute to the likelihood.

We model the posterior probability distribution for the mass of an observed pulsar with a Gaussian function, namely
\begin{equation}
    P(M_j|M) = \frac{1}{\sqrt{2 \pi (\sigma_{M_j})^2}} \exp\left[-\frac{(M - M_j)^2}{2(\sigma_{M_j})^2} \right],
\end{equation}
where $M_j$ is the maximum likelihood estimate for the mass of a given pulsar, and $\sigma_{M_j}$ is the standard deviation for that observation. 
The total likelihood associated with heavy pulsar mass measurements is then
\begin{equation}
    \mathcal{L}_{M{\rm max}}(\vec{\phi}_k) = \prod_j \mathcal{L}({M_j\, | \,M_{{\rm max}, k}}).
\end{equation}
This form for the likelihood only disfavors EoS with $M_{{\rm max}, k}$ less than existing observations. We do not want to disfavor EoS with $M_{{\rm max}, k}$ higher than existing observations because observations may be biased towards lower masses for a variety of reasons unrelated to the EoS.

We incorporate in our analysis the two highest mass measurements of neutron stars in relativistic binaries, which allow for measurements of post-Keplerian parameters, such as the Shapiro delay, pericenter precession, and orbital decay due to the emission of gravitational radiation (currently the gold standard for mass measurements of neutron stars \cite{Miller:2019nzo}). In particular, we use measurements of $M_{J1614-2230} = 1.908 M_\odot$ with $\sigma_{M_{J1614-2230}} =  0.016 M_\odot$ for PSR J1614-2230 \cite{demorest2010two,arzoumanian2018nanograv} and $M_{J0348+0432} = 2.01 M_\odot$ with $\sigma_{M_{J0348+0432}} = 0.04 M_\odot$ for PSR J0348+0432 \cite{antoniadis2013massive}. There is a third pulsar, PSR J0740+6620, with a measured mass of $M_{J0740+6620} = 2.07~M_\odot$ and $\sigma_{M_{J0740+6620}} = 0.08~M_\odot$ \cite{cromartie2020relativistic,Fonseca:2021wxt} that we will also include in the next sub-subsection as a joint mass-radius measurement. We also note that the mass measurement for PSR J1614$-$2230 was recently updated and reported to be $M_{J1614-2230} = 1.937 M_\odot$ with $\sigma_{M_{J1614-2230}} =  0.014 M_\odot$ \cite{NANOGrav:2023hde}, but we do not expect this update to affect our results significantly. In Section~\ref{subsec:not_inlcuded}, we explain why we leave out of our analysis the even-higher neutron star masses that have been suggested in the literature. 

Lastly, we note that the likelihood-based approach used here (emphasized in Ref.~\cite{Miller:2019nzo}) is preferable in a Bayesian analysis compared to imposing a lower bound on the maximum mass for two reasons. First, a hard bound does not account for uncertainties in the mass measurements. As Ref.~\cite{Miller:2019nzo} illustrates, if we consider the mass estimate for PSR J0740+6620, $M_{J0740+6620} = 2.07\pm 0.08$ $M_\odot$, a 1-$\sigma$ maximum mass cut at $M_{\rm max} \geq 2.04$ $M_\odot$ predicts that EoS with maximum masses of 2.04 $M_\odot$ and 2.14 $M_\odot$ are equally viable, when in reality, assuming Gaussian statistical uncertainties, the latter is substantially more consistent with observations. Second, a hard bound does not allow for the incorporation of information from multiple measurements. Thus, although widely practiced in the literature, imposing maximum mass cuts is statistically inconsistent and discards important information. 

\subsubsection{NICER}

Still assuming slow rotation, the $M-R$ curve for an EoS specifies a prediction for the radius given a stellar mass which is only a function of the EoS itself. Thus, given a joint $M-R$ posterior, we can integrate over central densities and the full $M-R$ sequence predicted by EoS $k$ to obtain the likelihood factor associated with an independent radius measurement $l$,

\begin{equation}
     \mathcal{L}(R_l \, | \, R_k) = \int_{\rho_{\rm min}}^{\rho_{\rm max}} q(\rho_c)\mathcal{L}_l(M(\rho_c), R_k(M(\rho_c)))d\rho_c,
\end{equation}
where $\mathcal{L}_l(M(\rho_c), R_k(M(\rho_c)))$ is the likelihood of a mass $M(\rho_c)$ and a radius $R(M(\rho_c))$ given measurement $l$, $R_k(M(\rho_c))$ is the circumferential radius for a star with gravitational mass $M$ given EoS $k$ and central density $\rho_c$, and $q(\rho_c)$ is the prior on the central density, which is determined for each EoS in the same way as described in the previous section. 

The total likelihood associated with simultaneous mass and radius measurements is then
\begin{equation}\label{eq:NICERlikelihood}
    \mathcal{L}_{M-R}(\vec{\phi}_k) = \prod_l \mathcal{L}(R_l \, | \, R_k).
\end{equation}
This particular form for the likelihood is equivalent to integrating the full mass and radius likelihood over the full $M-R$ sequence predicted by an EoS. It accounts for measurement uncertainties and possible correlations between radius and mass,\footnote{Correlations between mass and radius are system dependent and may or may not be present, see Refs.~\cite{lo2013determining,miller2015determining} for more details. } as well as the entire $M-R$ sequence, not just an individual $R_k(M)$. 

We adopt as constraints on the radius the posteriors obtained from NICER measurements for  PSR J0030-0451 \cite{Miller:2019cac} and PSR J0740+6620 \cite{Miller:2021qha} (again, see, respectively, \cite{Riley:2019yda} and \cite{Riley:2021pdl} for independent analyses of these two pulsars from a separate group within the NICER collaboration). 

Though other neutron star radii estimates are available, there are potentially significant systematic errors that have not been resolved \cite{Miller:2013tca,Miller:2016pom}. In contrast, NICER posteriors rely on fits of rotating hot spot patterns for which studies using synthetic waveforms found no significant mass or radius bias in statistically good fits \cite{miller2015determining,lo2013determining}. Lastly, we highlight that NICER posteriors for both pulsars are non-trivial shapes on the $M-R$ plane and display significant correlations between mass and radius, further emphasizing the importance of this particular approach to calculating the likelihood.

\subsubsection{Gravitational waves}

As discussed in Sec.~\ref{subsec:astro_obs}, for each EoS $k$, we can calculate the $\Lambda-M$ curve, such that for a star of gravitational mass $M$ and equatorial radius $R$, the tidal deformability is
\begin{equation}
    \Lambda_k = \dfrac{2}{3}k_2\left(\dfrac{R c}{G M}\right)^5,
\end{equation}
where $k_2$ is the tidal love number, which depends intrinsically on the EoS \cite{hinderer2008tidal,hinderer2009erratum}. 

In practice, it is more constraining to incorporate input from gravitational-wave observations using information from the binary tidal deformability, which can be measured to higher accuracy. In the Taylor family of post-Newtonian waveforms, given a binary neutron star system of stars with masses $M_1$ and $M_2 \leq M_1$ with tidal deformabilities $\Lambda_1$ and $\Lambda_2$, the most easily measurable quantity is \cite{wade2014systematic}
\begin{equation}
    \tilde{\Lambda} = \dfrac{16}{13}\dfrac{(M_1 + 12M_2)M_1^4\Lambda_1 + (M_2 + 12M_1)M_2^4\Lambda_2}{(M_1 + M_2)^5},
\end{equation}
which is sometimes called the binary or effective tidal deformability. Similarly, it can be difficult to extract individual masses from gravitational-wave events, but the chirp mass, $M_{\rm ch}  = (M_1 M_2)^{3/5}/(M_1 + M_2)^{1/5}$, can be measured with high precision since it relates directly to the gravitational-wave frequency during the inspiral phase. 

Assuming a gravitational-wave event $n$ results in a full posterior in $(M_1,M_2,\tilde{\Lambda})$ space, our procedure for incorporating it is as follows. The total likelihood factor has the form 
\begin{widetext}
\begin{equation}\label{eq:tidallikelihood}
   \mathcal{L}_\Lambda = \prod_n \mathcal{L}(\tilde{\Lambda}_n \, | \,\tilde{\Lambda}_k) = \prod_n \int dM_1 q(M_1)\int q(M_2\, | \,M_{{\rm ch},n},M_1)\mathcal{L}_n(M_1,M_2, \tilde{\Lambda}_k)dM_{{\rm ch},n}\, ,
\end{equation}
\end{widetext}
where $q(M_2\, | \,M_{\rm ch},M_1)$ is the prior probability density for $M_2$ at the value of $M_2$ implied by $M_{\rm ch}$ and $M_1$, and the integral is over the probability distribution for $M_{\rm ch}$ obtained from the gravitational-wave analysis. Since there is a limited width for $M_{\rm ch}$ which is dependent both on the EoS and the prior for the masses, our implementation is as follows. For a binary event involving two masses $M_1 \geq M_2$, we select the central density of a 1 $M_\odot$ star for the lower-mass star, $\rho_{c,2} = \rho_{\rm min}$. Then, we calculate the value of $M_1$ implied by $M_{\rm ch}$, which we know to the precision that we know the chirp mass. There is a range in $\rho_{c,1}$ that, given the value of $M_2$ implied by $\rho_{c,2}$, leads to $M_{\rm ch}$ within the 68\% credible region inferred for event $n$. That is the range we integrate over for $\rho_{c,1}$, using the same prior as before (quadratic between $\rho_{\rm min}$ and $\rho_{\rm max}$). We then select a new $\rho_{c,2}$, also following the quadratic prior, and repeat the same process for $\rho_{c,1}$ that we just outlined. We continue to increase $\rho_{c,2}$ up to the density at which $M_{\rm ch}$ implies $M_1 = M_2$. That means we can rewrite Eq.~(\ref{eq:tidallikelihood}) for a single event $n$ in terms of the central densities of the two objects,
\begin{widetext}
\begin{equation}
   \mathcal{L}(\tilde{\Lambda}_n \, | \,\tilde{\Lambda}_k) = \int d\rho_{c,1} q(\rho_{c,1})\int q(\rho_{c,2}\, | \,(M_{{\rm ch},n},M_1(\rho_{c,1})))\mathcal{L}_n(M_1(\rho_{c,1}),M_2(\rho_{c,2}), \tilde{\Lambda}_k \, | \, M_1(\rho_{c,1}), M_2(\rho_{c,2}))d\rho_{c,2}\,
\end{equation}
\end{widetext}
where $q(\rho_{c,2}\, | \,(M_{{\rm ch},n},M_1(\rho_{c,1}))) = 0$ outside of the 68\% credible region for $M_{{\rm ch},n}$. We highlight that even though $M_{\rm ch}$ is typically measured to high precision, it is not statistically consistent to write the integral over $\rho_{c,2}$ as a delta function. That is because the range of $M_1$ allowed for a given $M_2$ and $M_{\rm ch}$ depends on both the EoS and the prior for the central densities. Consequently, it will vary between individual EoS and must be calculated separately for each EoS \cite{Miller:2019nzo}.

We include in our analysis binary tidal deformability estimates from GW170817 \cite{LIGOScientific:2017vwq,De:2018uhw,LIGOScientific:2018cki} and GW190425 \cite{LIGOScientific:2020aai}. We use the publicly available posteriors over the full model parameter space \cite{170817data,190425data} as input for a kernel density estimate of the marginalized posterior for $M_1$ and $\tilde{\Lambda}$. Since the combined mass in GW190425 is high enough that one of the objects might have been a black hole, we check whether for the EoS and central density under consideration the higher-mass object is a neutron star. If so, we compute the tidal deformabilities of both stars, using the same EoS, following the procedure outlined in Ref.~\cite{hinderer2008tidal, hinderer2009erratum}. However, if the EoS predicts a black hole at the central density under consideration for the higher-mass object, we set its tidal deformability to zero. 
Lastly, we note that some EoS predict more than one stable branch in the $M-R$ sequence, and that we assign a prior probability of zero to all central densities corresponding to an unstable branch for a given EoS.

\subsubsection{Perturbative QCD}\label{subsec:pqcd_input}

Because of asymptotic freedom, QCD can be treated perturbatively at high densities ($\sim$40 $n_{\rm sat}$) \cite{Andersen:2002jz}. It has recently been argued that perturbative QCD offers nontrivial constraints to the neutron star EoS when state-of-the-art N3LO perturbative results \cite{Gorda:2021znl,Gorda:2021kme} are extended to neutron star densities using stability, causality, and consistency arguments \cite{Gorda:2022jvk,Komoltsev:2021jzg}. The formalism was initially introduced in Ref.~\cite{Komoltsev:2021jzg}. We briefly review its key components here, but refer the reader to the original work for additional details. 

Suppose an EoS can be characterized by a correlated set of values $\vec{\beta} \equiv \{p(\mu),n(\mu),\mu\}$, where $p$ is the pressure, $n$ is the number density, and $\mu$ is the chemical potential. Also suppose that we have knowledge of the EoS at some low-density limiting value, $\mu_{\rm low}$, and a high-density limiting value, $\mu_{\rm high}$, meaning that we know
\begin{eqnarray}
   \vec{\beta}_{\rm low} = & \{p_{\rm low},n_{\rm low},\mu_{\rm low}\}  \equiv &\{p(\mu_{\rm low}),n(\mu_{\rm low}),\mu_{\rm low}\}, \label{eq:betalow}\\
    \vec{\beta}_{\rm high} = & \{p_{\rm high},n_{\rm high},\mu_{\rm high}\} \equiv &\{p(\mu_{\rm high}),n(\mu_{\rm high}),\mu_{\rm high}\}. \label{eq:betahigh}
\end{eqnarray}

There are an infinite number of EoS that can connect $\vec{\beta}_{\rm low}$ and $\vec{\beta}_{\rm high}$, but any such EoS must respect thermodynamic stability, causality, and consistency. Thermodynamic stability implies that the the grand-canonical potential from which the EoS is derived ($\Omega$) is concave with respect to $\mu$, meaning that $\partial_\mu^2 \Omega \leq 0$. At $T=0$, we also have 
\begin{eqnarray}
    \Omega(\mu) = & -p(\mu),\\
    n = & \dfrac{\partial p}{\partial \mu}.
\end{eqnarray}
Therefore, stability results in a constraint on the slope of $n(\mu)$,
\begin{equation}
    \dfrac{\partial n}{\partial \mu} \geq 0.
\end{equation}
Moreover, the causality requirement constrains $c_s^2 \leq 1$, which, at $T=0$, relates to $n(\mu)$ and $\partial_\mu n$ such that
\begin{equation}
    c_s^{-2} = \frac{\mu}{n}\dfrac{\partial n }{\partial \mu } \geq 1.
\end{equation}
Combining stability and causality, we have that, at each point in $\mu - n$ space, the slope of the curve passing through that point corresponding to a maximally stiff ($c_s^2 = 1$) EoS is $\partial n / \partial \mu = n/\mu$.

Finally, because we must also ensure that at $(\mu_{\rm low},n_{\rm low})$ the pressure is $p_{\rm low}$ and, similarly, that at $(\mu_{\rm high},n_{\rm high})$ the pressure is $p_{\rm high}$, it must also be true that
\begin{equation}\label{eq:integral_constraints}
    \int_{\mu_{\rm low}}^{\mu_{\rm high}} n(\mu)d\mu = p_{\rm high} - p_{\rm low} = \Delta p. 
\end{equation}
We can derive constraints on $\Delta p$ based on stability and causality constraints on $n(\mu)$. We can place a lower bound on $\Delta p$ by asking which curve connecting $(\mu_{\rm low},n_{\rm low})$ to $(\mu_{\rm high},n_{\rm high})$ minimizes the integral in Eq.~\eqref{eq:integral_constraints} while still respecting stability and causality. We will call that quantity $\Delta p_{\rm min}$. Equivalently, we can construct the curve which maximizes the integral in Eq.~\ref{eq:integral_constraints} and still respects stability and causality, and denote that $\Delta p_{\rm max}$. Assuming $c_s^2$ is bounded from above only by the causal limit, we have \cite{Komoltsev:2021jzg}
\begin{equation}\label{eq:plow}
    \Delta p_{\rm min}  = \dfrac{1}{2}\left(\dfrac{\mu_{\rm high}^2}{\mu_{\rm low}} - \mu_{\rm low}\right)n_{\rm low}
\end{equation}
\begin{equation}\label{eq:phigh}
    \Delta p_{\rm max}  = \dfrac{1}{2}\left( \mu_{\rm high} - \dfrac{\mu_{\rm low}^2}{\mu_{\rm high}}\right)n_{\rm high}.
\end{equation}
All these constraints combined imply that for any $\vec{\beta}_{\rm high}$ and for a fixed $(\mu_{\rm low}, n_{\rm low})$, $p_{\rm low}$ must be between $[p_{\rm high} - \Delta p_{\rm min}, p_{\rm high} - \Delta p_{\rm max}]$.

These guidelines for connecting two arbitrary regimes via an EoS which respects stability, causality, and consistency can be used to extrapolate pQCD results to densities relevant to neutron stars. That is because, if we know $\vec{\beta}_{\rm high} = \vec{\beta}_{\rm pQCD}$, we can check if an EoS for which we only have knowledge up to a lower matching density $n_{\rm low} = n_{\rm match}$ can be connected to $\vec{\beta}_{\rm pQCD}$ through a causal and stable EoS.   

Our knowledge from pQCD is derived from current state-of-the-art calculations in Refs.~\cite{Gorda:2018gpy,Gorda:2021gha}, which report a partial next-to-next-to-next-to leading order (N3LO) calculation of the zero-temperature, high-density QCD grand-canonical potential. Because these results arise from a series expansion in the QCD coupling constant and are then truncated at a finite order, we have to estimate the error introduced by the missing higher order (MHO) terms. In the case of QCD, the MHO error depends on a residual, unphysical renormalization scale, $\Bar{\Lambda} \propto \mu$, which is underdetermined. Instead, the standard approach is to vary $\Bar{\Lambda}$ around a fiducial scale by some fixed factor. We follow Ref.~\cite{Gorda:2022jvk}, which adopted a scale-averaging approach. That means that pQCD results are given as a family of independent predictions $\vec{\beta}_{\rm pQCD}(X)$, where $X \equiv 3\Bar{\Lambda}/2\mu_{\rm high}$. We set $\mu_{\rm high} = 2.6$ GeV based on Ref.~\cite{Fraga:2013qra}, which points out that the uncertainty estimation for pQCD calculations at this value is similar to that of $\chi$EFT at 1.1 $n_{\rm sat}$ (about $\pm$24\% variation around the mean value \cite{Gorda:2022jvk}). We consider $X \in [1/2,2]$, the same range that was implemented in Ref.~\cite{Gorda:2022jvk} and that has been suggested by phenomenological models \cite{Schneider:2003uz,Rebhan:2003wn,Cassing:2007nb,Gardim:2009mt} as well as the large-flavor limit of QCD \cite{Ipp:2003jy}.

Now that we have defined the theoretical input from high densities, we need to discuss how we define the low-density input from GP and mGP EoS. For any neutron star EoS that we generate with the GP or the mGP, $\vec{\phi}_k$, we have to check that it can be connected to $\vec{\beta}_{\rm pQCD}(X)$, for a given X, from $\vec{\beta}_{\rm low} = \{p_k(n_{\rm match}), n_{\rm match}, \mu_k(n_{\rm match}) \}$. In practice, we check that $p_k(n_{\rm match})$ leads to $\Delta p \ \in [\Delta p_{\rm min}, \Delta p_{\rm max}]$, given $p_{\rm high}$ from $\vec{\beta}_{\rm pQCD}(X)$. Since the relevant scale for the neutron star EoS is the central density of a maximally massive star, $n_B^{\rm max}$, we set $n_{\rm match} = n_{B,k}^{\rm max}$, which varies for each EoS. For the renormalization scale parameter, we follow Ref.~\cite{Gorda:2022jvk}, and sample 1000 values of $X \in [1/2,2]$, evenly spaced in $\log(X)$. Hence, the pQCD weight associated with EoS $k$ is
\begin{equation}\label{eq:w_pQCD_def}
    w_{\rm pQCD}(\vec{\phi}_k) = \dfrac{1}{1000}\sum^{1000}_{i= 1} \mathbf{1}_{X}(\vec{\phi}_k),
\end{equation}
where $\mathbf{1}_{X}(\vec{\phi}_k)$ is the indicator function
\begin{equation}
    \mathbf{1}_{X}(\vec{\phi}_k) = \begin{cases}
    1,& \text{if } \Delta p_k \in [\Delta p_{\rm min}, \Delta p_{\rm max}]\\
    0,              & \text{otherwise}
\end{cases},
\end{equation}
and $\Delta p_k = p_{\rm pQCD}(X) - p_k(n_{B,k}^{\rm max})$. Recall that $\Delta p_{\rm min}$ and $\Delta p_{\rm max}$ can be obtained from Eqs.~(\ref{eq:plow}, \ref{eq:phigh}), using 
\begin{eqnarray*}
    \mu_{\rm high}  & = & 2.6 \textrm{ GeV}, \\
    \mu_{\rm low}  &=  & \mu_k(n_{B,k}^{\rm max}),\\
    n_{\rm high}  &=  & n_{\rm pQCD}(\mu = 2.6 \textrm{ GeV},X), \\
    n_{\rm low}  &= & n_{B,k}^{\rm max}.
\end{eqnarray*}
Effectively, $w_{pQCD}(\vec{\phi}_k)$ captures how often, out of the 1000 values for $X$, $\vec{\phi}_k$ can be connected to $\vec{\beta}_{\rm pQCD}(X)$ with an EoS that respects thermodynamic stability, causality, and consistency. This procedure defines a weighting scheme associated with input from pQCD, which suppresses EoS that are in tension with pQCD results by a factor proportional to the strength of the disagreement under the scale-averaging assumption.

\subsubsection{Observational measurements not included in our analysis}\label{subsec:not_inlcuded}

We make the choice here to not include in our analysis other recent claims of very heavy or very light neutron stars. For example, there have been recent claims of pulsars heavier than the ones considered here, namely PSR J1810+1744 at $2.13\pm0.04$ \Msol \cite{Romani:2021xmb} and PSR J0952-0607 at $2.35\pm 0.17$ \Msol \cite{Romani:2022jhd}, but possible systematic errors for these measurements are not as well-understood as those involved in Shapiro time-delay-based measurements, such as those for PSR J0740+6620 and PSR J1614$-$2230. Specifically, the fit residuals in Ref.~\cite{Romani:2021xmb} for the properties of the companion to spider-pulsar PSR J1810+1744 are clearly not a random scatter (see Fig.~1 in Ref.~\cite{Romani:2021xmb}), which suggests that the fit values and inferred mass are subject to systematic errors we do not currently understand. The picture is more promising for the inferred mass of PSR J0952-0607, where at least the residuals do not seem to indicate problems with the fit (see Fig.~1 in Ref.~\cite{Romani:2022jhd}). But there is still the question about whether the good fit indicates that the system is well-understood from a theoretical perspective, and whether the inferred mass is not just precise, but also accurate. 

In a separate measurement, the central compact object of the supernova remnant HESS J1731-347 was recently estimated to have a mass of $0.77^{+0.20}_{-0.17}$ \Msol and radius of $10.4^{+0.86}_{-0.78}$ km \cite{doroshenko2022strangely}, possibly making it the lightest neutron star ever observed. Here, the low estimated mass stems from the use of a low distance to the source combined with the assumption that the surface radiates uniformly, which tends to favor a carbon atmosphere over a hydrogen or helium atmosphere. Moreover, in the fitting, it was assumed that surface magnetic fields can be ignored.  However, nonuniform emission is consistent with data on several similar sources \cite{Alford:2023waw}, making hydrogen and helium atmospheres possible and making it plausible that the neutron star in HESS~J1731-347 could have a standard mass, well above one solar mass \cite{MUSES:2023hyz}. Thus, to remain conservative on the data we use in this work, we do not consider PSR J1810+1744, PSR J0952-0607, and the center compact object in HESS J1731-347 in our analysis.

\subsection{Model evidence}\label{subsec:model_evidence}

We have two distinct set of prior beliefs, or models, for the EoS. We combine samples from these two models into one prior, which we introduced in Sec.~\ref{subsec:hyperprior} as $\Phi$, and which can be represented as the union of samples from the mGP and the benchmark GP, $\Phi = \Phi_{\rm mGP} \cup \Phi_\textrm{benchmark GP}$.

We defined the evidence ($\mathcal{E}$) in Sec.~\ref{subsec:primer}, where the integral in Eq.~(\ref{eq:evidence}) is over all possible samples that can be generated from a model. In practice, we only have access to a finite number of samples, and $\mathcal{E}$ is approximated as
\begin{equation}\label{eq:evidencefinite}
    \mathcal{E}_m \approx \sum_i^{N_{m}}\mathcal{L}(\vec{\phi}_k)q(\vec{\phi}_k), \quad \vec{\phi}_k\in \Phi_{m},
\end{equation}
where $N_{m}$ is the number of samples from model $m=$ \{benchmark GP, mGP\}, including the samples in $\Phi_m\cap\Phi_\times$ for which $\mathcal{L}(\vec{\phi}_k) = 0$. Therefore, a key assumption is that we sample \emph{enough} EoS to correctly approximate $\mathcal{E}$. This aspect further emphasizes the importance of the prior sample size checks we introduced in Sec.~\ref{subsec:hyperprior} -- if we do not have enough samples in the regions where $\mathcal{L}(\vec{\phi}_k)$ is non-negligible, we cannot correctly approximate $\mathcal{E}$. 

Another assumption we make is that each EoS has an equal prior probability. This implies that Eq.~(\ref{eq:evidencefinite}) is now
\begin{equation}
    \mathcal{E}_m \approx \dfrac{1}{N_{m}}\sum_i^{N_{m}}\mathcal{L}(\vec{\phi}_k), \quad \vec{\phi}_k\in \Phi_{m}\,.
\end{equation}
That is, the evidence becomes a simple average over the likelihoods of all the EoS from model $m =$ \{benchmark GP, mGP\}. 

A reasonable objection to this assumption is to question whether the different hyperparameters in each framework should have been included explicitly as hyperpriors. In general, we expect that with an increased number of parameters we also increase our chances of describing the data, but also that simplicity should be rewarded over complexity. Here, the mGP is more complex than the benchmark GP, so why is there not a penalty in the calculation of the model evidence for mGP EoS? Actually, the penalty exists, and it is included implicitly. To understand this argument, recall the discussion in Sec.~{\ref{sec:priors}} regarding assumptions that are implied in a non-parametric framework. We stated that our effective parameters are the value of the speed of sound at each sampled value of pressure, $c_s^2(p_i)$, and that the method and the hyperparameters we choose for generating $c_s^2(p_i)$ dictate both the prior distribution and the correlations across the \emph{effective} parameter space. That means that the mGP covers a bigger space in terms of the possible functional forms for $c_s^2(p_i)$, which results in more bad predictions ($\mathcal{L}(\vec{\phi}_k) \approx 0$) compared to the simpler benchmark GP. Therefore, in order to be competitive with the benchmark GP and make up for a larger number of bad predictions, the mGP must be more accurate in describing the data than the benchmark GP. We do not need to include the different hyperpriors in Eq.~(\ref{eq:evidencefinite}) because (i) the benchmark GP and mGP hyperparameters are not the parameters being estimated, and (ii) the  mGP model is implicitly penalized for its increased functional complexity in the effective parameter space because it covers a larger volume in that space where $\mathcal{L}(\vec{\phi}_k) \approx 0$ compared to the benchmark GP. 

\section{Bayesian analysis of nontrivial features in The speed of sound inside neutron stars}\label{sec:results}

Now that we have established methods for generating EoS that display long (benchmark GP) and multi-scale (mGP) correlations in $c_s^2(n_B)$, we can implement the constraints discussed in Sec.~\ref{sec:statmethods} and begin to answer specific physics questions from a Bayesian perspective.  

We will begin with an important sanity check -- does our new framework provide reasonable agreement with data, even when multi-scale correlations and nontrivial features appear in $c_s^2(n_B)$? We will answer that question by looking at the mass-radius posteriors. Following a discussion of the mass-radius posteriors, we can explore other questions such as: are the EoS posteriors sensitive to the structure in $c_s^2(n_B)$? What is the maximum central density of a neutron star? Do new pQCD constraints have a strong influence on our analysis? Is a peak in $c_s^2(n_B)$ supported by existing constraints? Finally, is the GP or the mGP framework favored by the data?  

To better understand our results, we use a plotting method for our priors and posteriors that, to our knowledge, has not been used to infer properties of the EoS in the literature. Let us first describe our approach for plotting the prior, because there are subtle differences between its plotting method and that of the posterior. We bin our 2-dimensional (variables $X$ and $Y$ e.g. mass and radius or $c_s^2$ and $n_B/n_{\rm sat}$) prior in bins of a certain width $\Delta X$, $\Delta Y$. We denote a particular bin as a pair $(X_i,Y_i)$, such that a point $(x,y)$ is in bin $i$ if $X_i \leq x < X_i + \Delta X$ and $Y_i \leq y < Y_i + \Delta Y$. A given EoS $k$ is characterized on the X-Y plane by a set of $l$ total pairs of points $\{(x_k^1,y_k^1),\ldots, (x_k^l,y_k^l)\}$, which produce a curve that passes through $N_k$ of these bins. Every time the EoS passes through a bin $i$, we count that as a hit: $h_k(X_i,Y_i)=1$. Otherwise, if the EoS does not pass through that location, we assign $h_k(X_i,Y_i)=0$. We sum over all hits within a bin (i.e.~count all the EoS that pass through it) to obtain the total hits for that specific bin, $\mathcal{H}_{prior}(X_i,Y_i)$, where
\begin{equation}\label{eqn:Hprior}
    \mathcal{H}_{\rm prior}(X_i,Y_i)=\sum_k^{N_{\rm EoS}} h_k(X_i,Y_i), 
\end{equation}
and
\begin{equation}
    h_k(X_i,Y_i)= \begin{cases}
    1,& \text{if } X_i \leq x^{l^*}_k < X_i + \Delta X \wedge Y_i \leq y^{l^*}_k < Y_i + \Delta Y,\\ 
    & l^*\in\{0,\ldots,l\}\\
    0,& \text{otherwise.}
\end{cases}
\end{equation}
To normalize the hits within a bin we determine the total number of hits across all of our EoS samples across all bins:
\begin{equation}\label{eqn:Htot}
    \mathcal{H}_{\rm tot}=\sum_{i} \mathcal{H}_{\rm prior}(X_i,Y_i)\,, 
\end{equation}
such that our normalized distribution for the prior within a given bin $(X_i, Y_i)$ is
\begin{equation}\label{eq:normalizedprior}
    \mathcal{N}_{\rm prior}(X_i,Y_i)=\frac{\mathcal{H}_{\rm prior}(X_i,Y_i) }{\mathcal{H}_{\rm tot}}.
\end{equation}

Our plotting method is similar for the posterior. However, because the posterior probability is proportional to the prior probability multiplied by the likelihood, we have to include the likelihood, $\mathcal{L}(\vec{\phi}_k)$, of each EoS when calculating the hits of the bin. More concretely,
\begin{equation}
    \mathcal{H}_{\rm posterior}(X_i,Y_i)=\sum_k^{N_{\rm EoS}} \mathcal{L}(\vec{\phi}_k) h_k(X_i,Y_i), 
\end{equation}
and 
$\mathcal{N}_{\rm posterior}$ has the same form as Eq.~(\ref{eq:normalizedprior}), 
\begin{equation}
    \mathcal{N}_{\rm posterior}(X_i,Y_i)=\frac{\mathcal{H}_{\rm posterior}(X_i,Y_i) }{\mathcal{H}_{\rm tot}}
\end{equation}
but we use the posterior values instead of prior values to calculate $\mathcal{H}_{\rm tot}$.

The procedure described above for representing prior and posterior distributions is equivalent to showing the prior as a normalized 2-D histogram, and the posterior as a weighted, normalized 2-D histogram. We will refer to these plotting methods as the binned joint prior/posterior probability density. Representing probability densities in this way is advantageous as long as the bin sizes are chosen appropriately because we do not have to rely on kernel density estimates that can smear out important information. Kernel density estimates also perform poorly near sharp boundaries, and may predict a finite probability density in regimes where data do not exist, or, in cases where such constraints exist, beyond physical boundaries (i.~e.~ $c_s^2 < 0$ or $c_s^2 > 1$). The only aspect that requires care when applying this method is the interplay between the size of the bins, $(\Delta X, \Delta Y)$, and the number of samples. Using a fine grid with too few samples can result in bin heights that fluctuate significantly between neighboring points. Similarly, using a grid that is too coarse when there are an adequate number of samples risks smearing out important features in the distribution, as can be the case with kernel density estimates.

\begin{figure*}[t]
   \centering
   \begin{tabular}{cc}
    \includegraphics[width=0.49\linewidth]{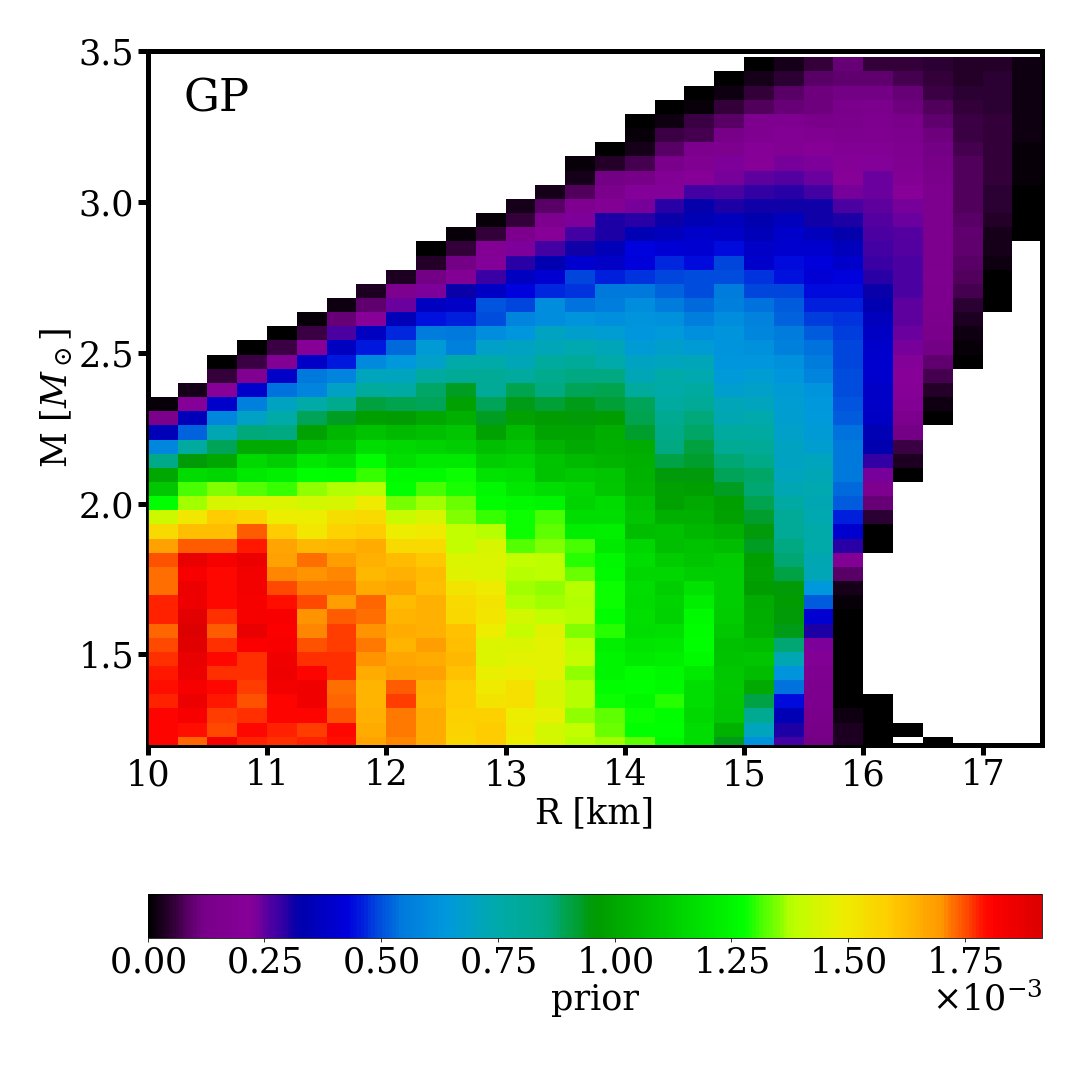} & 
    \includegraphics[width=0.49\linewidth]{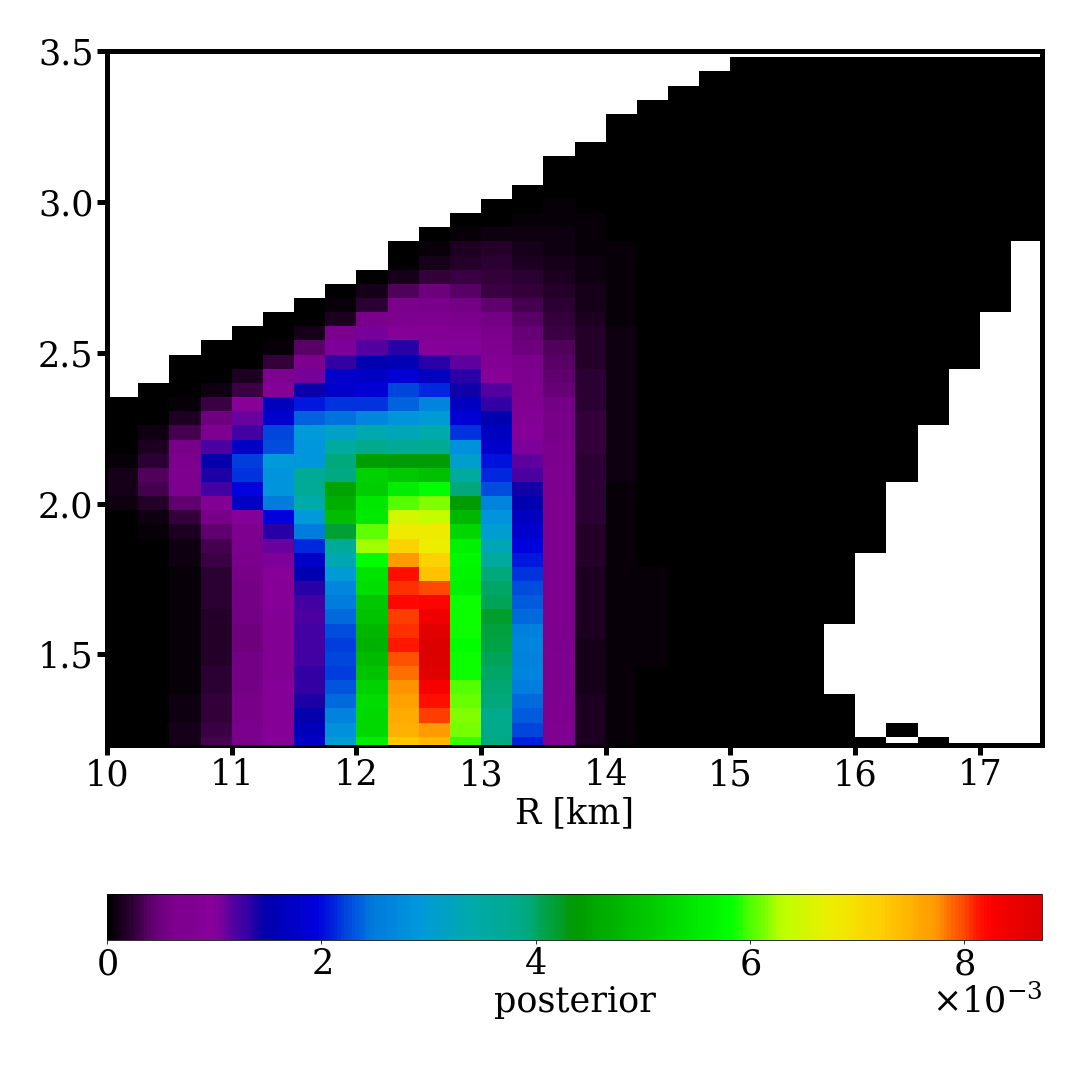}
    \\        
    \includegraphics[width=0.49\linewidth]{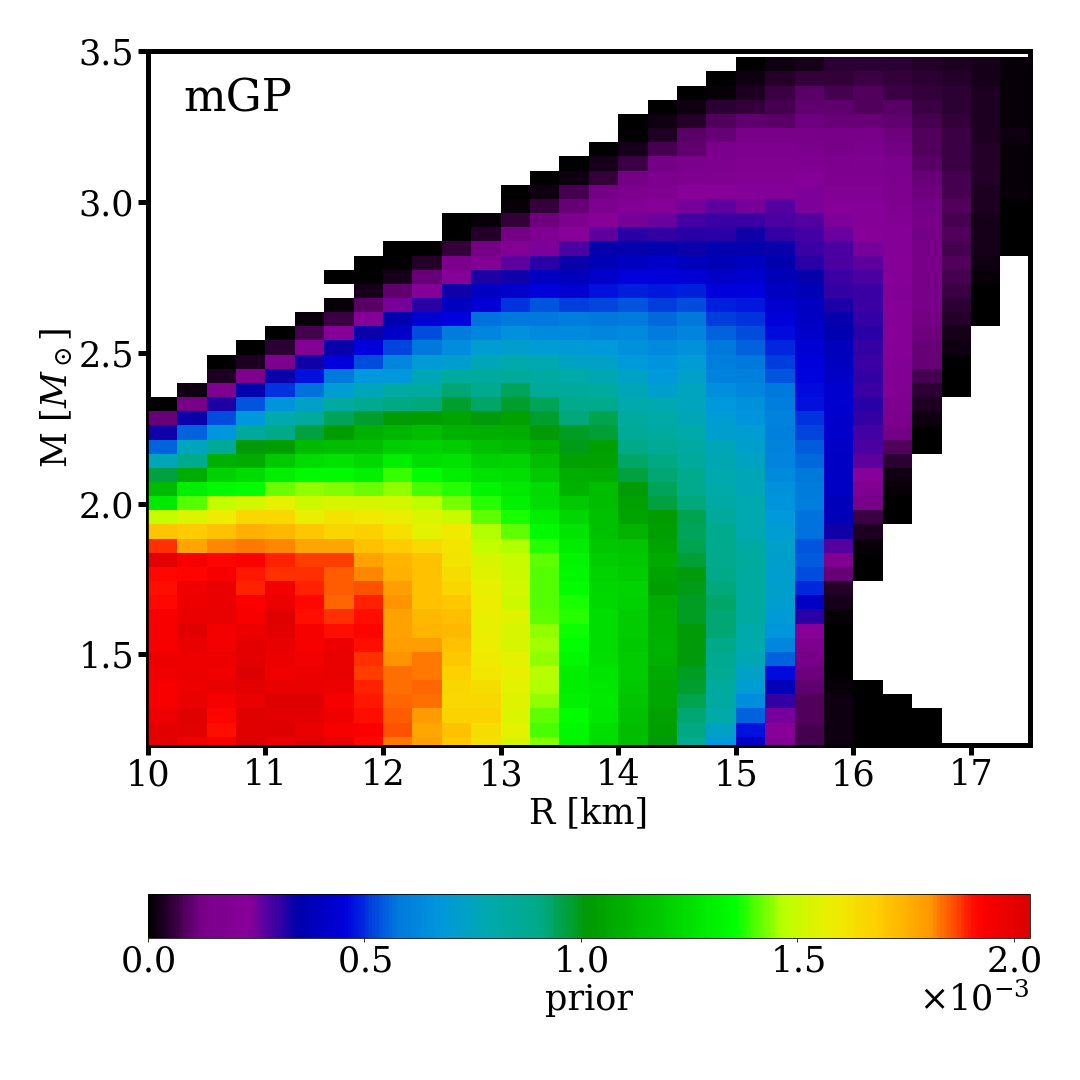} &
    \includegraphics[width=0.49\linewidth]{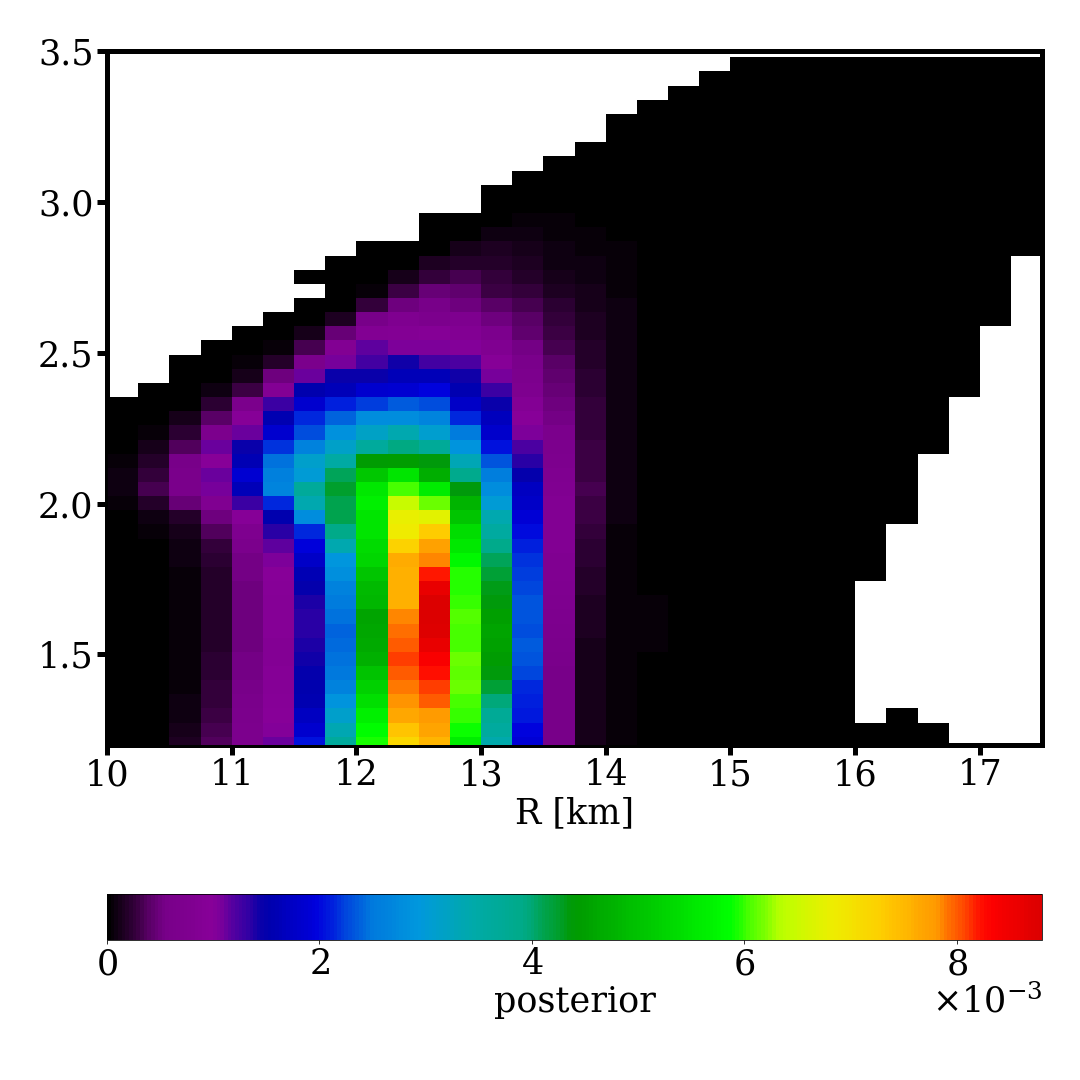}
   \end{tabular}
  \vspace{-0.5cm}
  \caption{Mass-radius prior (left) and posterior (right) probability distributions for GP (top) and mGP (bottom) EoS. Both the prior and posterior probability distributions are produced by binning the EoS by mass and radius and then normalizing the heights of the bins such that their sum is equal to one. For the posteriors, each EoS is weighted by the corresponding likelihood. Observe that the joint posteriors are similar when using the GP or mGP models, with masses larger than $2.7 M_\odot$ and radii larger than $14$ km disfavored.}
  \label{fig:massradius}
\end{figure*} 


\begin{table*}
    \centering
\caption{Median, 68\%, and 90\% credible regions from the benchmark GP and mGP posteriors for the equatorial radius of a $1.4 \ M_\odot$ star ($R_{1.4}$), the equatorial radius of a $2.1 \ M_\odot$ star ($R_{2.1}$), and maximum stellar mass ($M_{\rm max}$), assuming an isolated, slowly-rotating star.}
\label{tab:table2}
    \begin{tabular}{c|c|c|c} \hline 
         EoS &  $R_{1.4}$ [km]&  $R_{2.1}$ [km]&  $M_{\rm max}$ [$M_\odot$]\\ \hline 
         \hline
         Benchmark GP& 12.55, (12.07, 13.05), (11.56, 13.37) & 12.26, (11.52, 12.90), (11.02, 13.34) &2.26, (2.10, 2.51), (2.04, 2.68)\\ \hline 
         mGP& 12.55, (12.04, 13.04), (11.54, 13.33) & 12.34, (11.62, 12.98), (11.12, 13.39)  & 2.25, (2.09, 2.50), (2.03, 2.67) \\ \hline
         
    \end{tabular}

\end{table*}


\subsection{Are mass-radius posteriors sensitive to structure in $c_s^2(n_B)$?}

In order to fully explore the phase space of the EoS of neutron stars, it is important to have a broad prior in the mass-radius relation.  In the left panels of Fig.\ \ref{fig:massradius}, we show the stable branches of the mass-radius prior for the benchmark GP (top) and the mGP (bottom) using the plotting method described in Eqs.\ (\ref{eqn:Hprior}-\ref{eqn:Htot}). Here, we only plot the priors corresponding to samples in the set that meets the basic checks we outlined in Sec.~\ref{subsec:hyperprior} ($\Phi_{\checkmark}$). We find that both priors produce a similarly diverse set of mass-radius curves, with the highest prior regimes passing through $R=10-14$ km and up to masses $M\sim 2-2.2 \ M_\odot$. We also see that both priors allow for maximum masses up to $M\sim 3.5 \ M_\odot$, although the prior disfavors $M\gtrsim 2.2 \ M_\odot$. White regions in Fig.\ \ref{fig:massradius} indicate that no samples in $\Phi_{\checkmark}$ reach that region in the $M-R$ plane. The bottom right-hand side of the left panels in Fig.\ \ref{fig:massradius} -- the large-radius, low-mass regime -- is primarily ruled out by constraints on the symmetry energy, which predict a soft EoS in that regime. The top left-hand side of the left panels in Fig.\ \ref{fig:massradius} -- the small-radius, high-mass regime -- is ruled out mostly because it is beyond the point of stability for the $M-R$ sequences that reach such high masses. Comparing our priors for the benchmark GP and the mGP, we find they are nearly identical, despite the significant differences in how the EoS are constructed.

After applying the constraints outlined in Sec.~\ref{sec:statmethods}, we then produce our posteriors, which are shown on the right panels of Fig.\ \ref{fig:massradius}. We use the plotting technique described in Eqs.\ (\ref{eqn:Hprior}-\ref{eqn:Htot}) to present the posterior distribution for the benchmark GP (top panels) and the mGP (bottom panel) models. In addition, in Table \ref{tab:table2}, we present the median, 68\%, and 90\% credible regions from the benchmark GP and mGP posteriors for the equatorial radius of a $1.4 \ M_\odot$ star ($R_{1.4}$), the equatorial radius of a $2.1 \ M_\odot$ star ($R_{2.1}$), and maximum stellar mass ($M_{\rm max}$), assuming an isolated, slowly-rotating star. We find that in terms of the observables highlighted in Table \ref{tab:table2}, as well as the binned joint posteriors, the results for the benchmark GP and the mGP are nearly identical. 

Observe that both benchmark GP and mGP EoS support neutron star masses up to $M\sim 2.7 \ M_\odot$ (albeit with a smaller likelihood), but neutron stars with higher masses are extremely disfavored. Observe also that, from the credible bands alone, one would reasonably assume that all neutron stars heavier than $M\sim 2.7 \ M_\odot$ must have larger radii at high masses. However, from the joint posterior, it is clear that a number of mass-radius curves for heavy neutron stars may either be nearly straight or even  bend slightly to the left. In any case, we see that, although specific M--R curves may be affected by structure in the speed of sound, the latter does not affect the M--R posteriors, given currently available observations.  

\begin{figure*}
   \centering
   \begin{tabular}{cc}
        \includegraphics[width=0.49\linewidth]{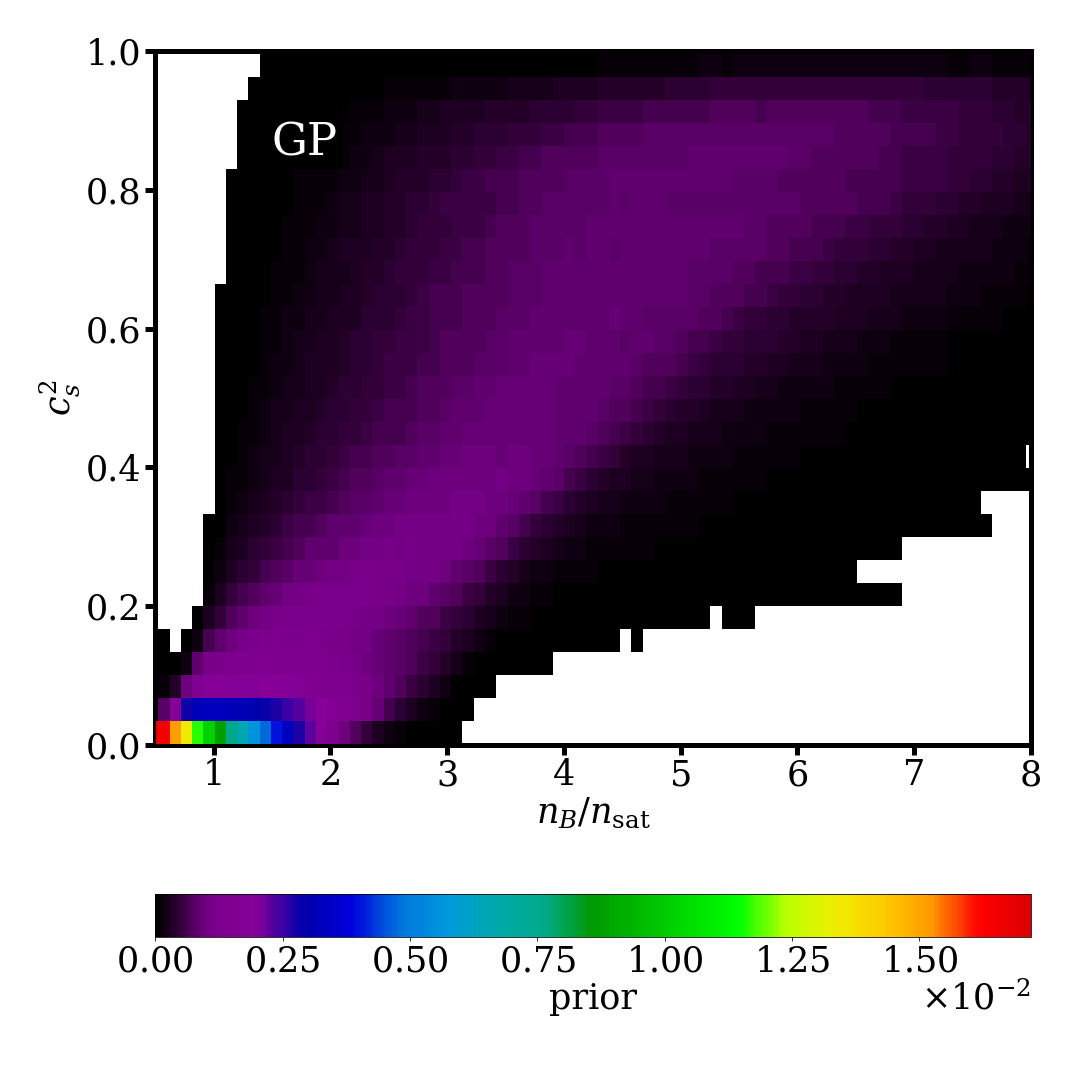} & \includegraphics[width=0.49\linewidth]{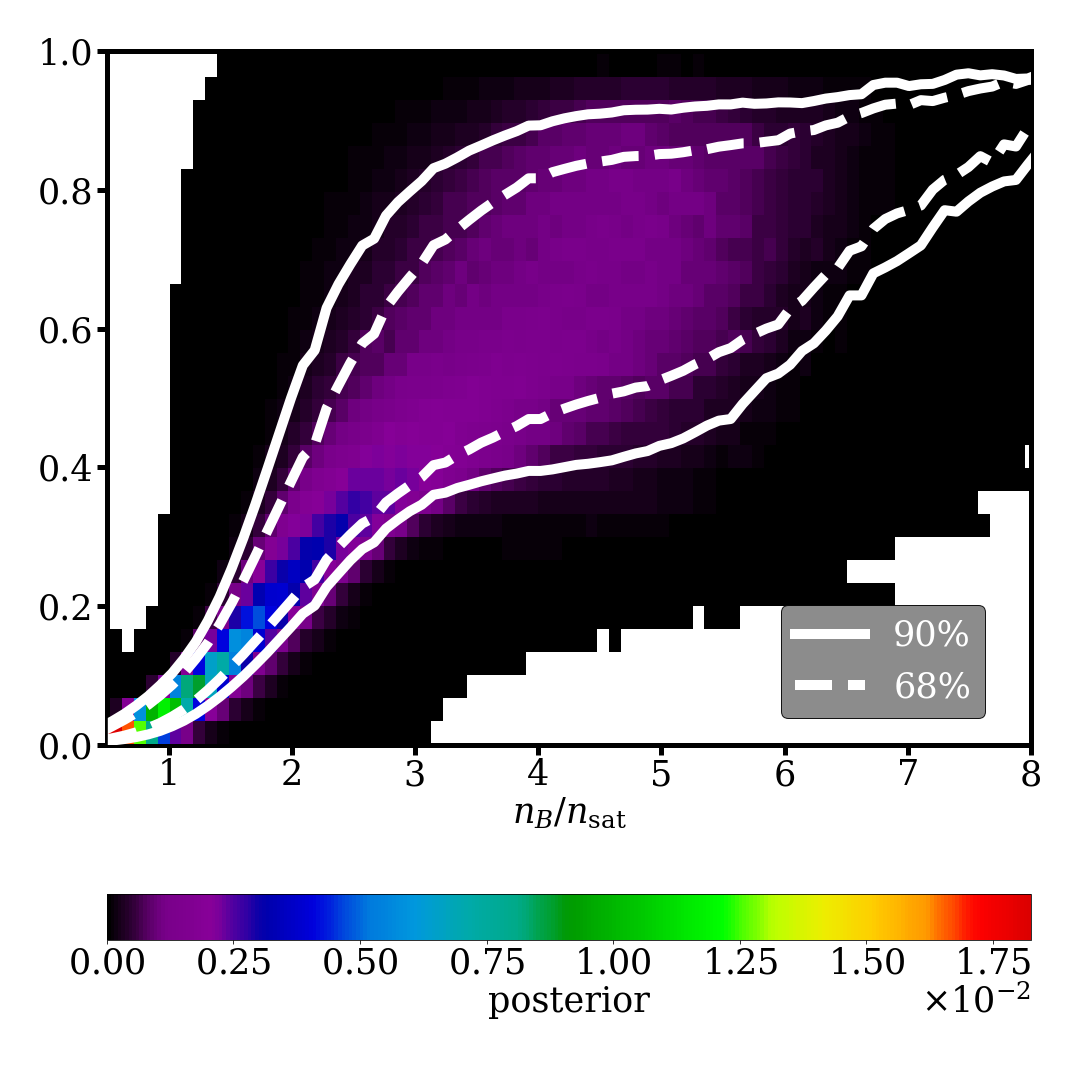}\\
        \includegraphics[width=0.49\linewidth]{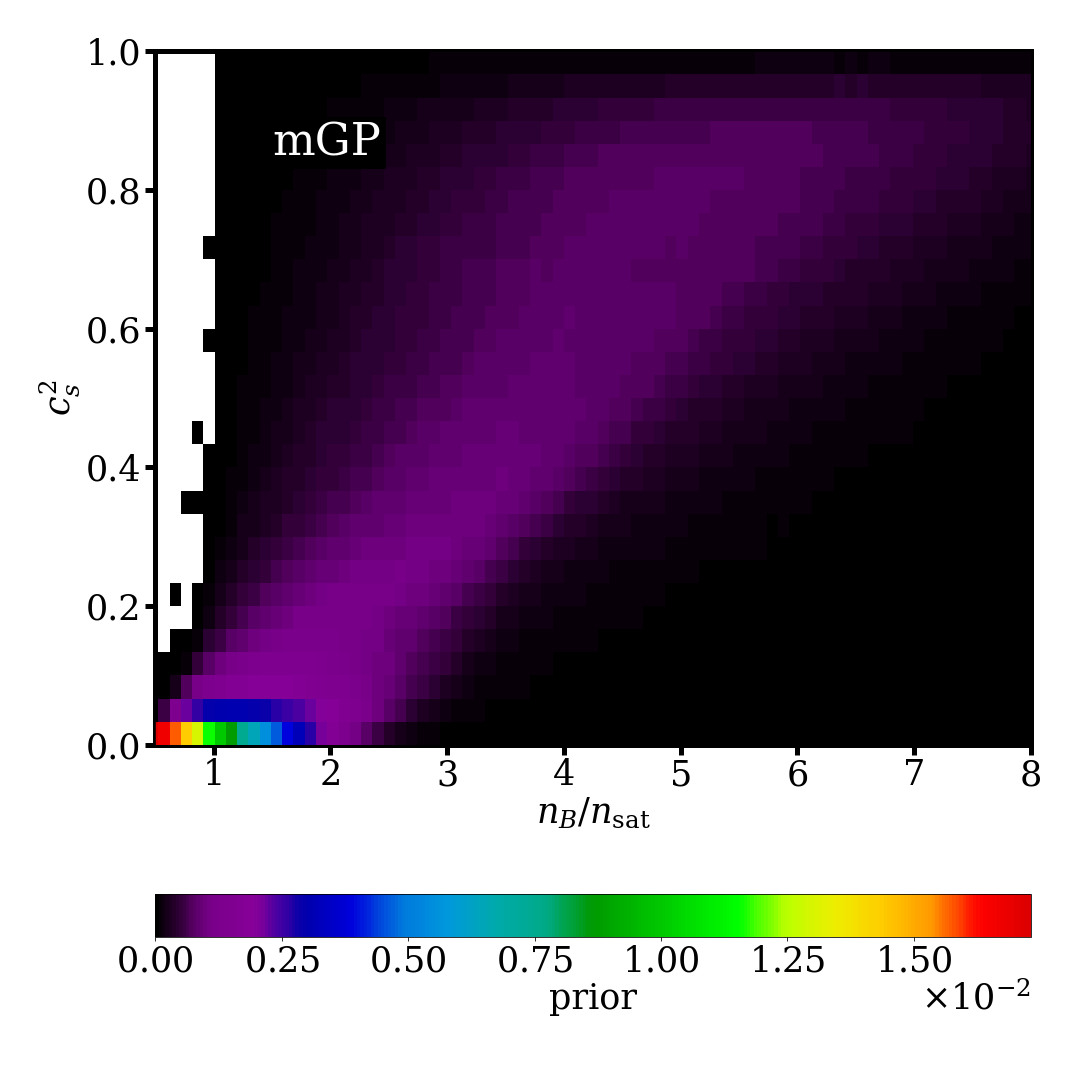} & \includegraphics[width=0.49\linewidth]{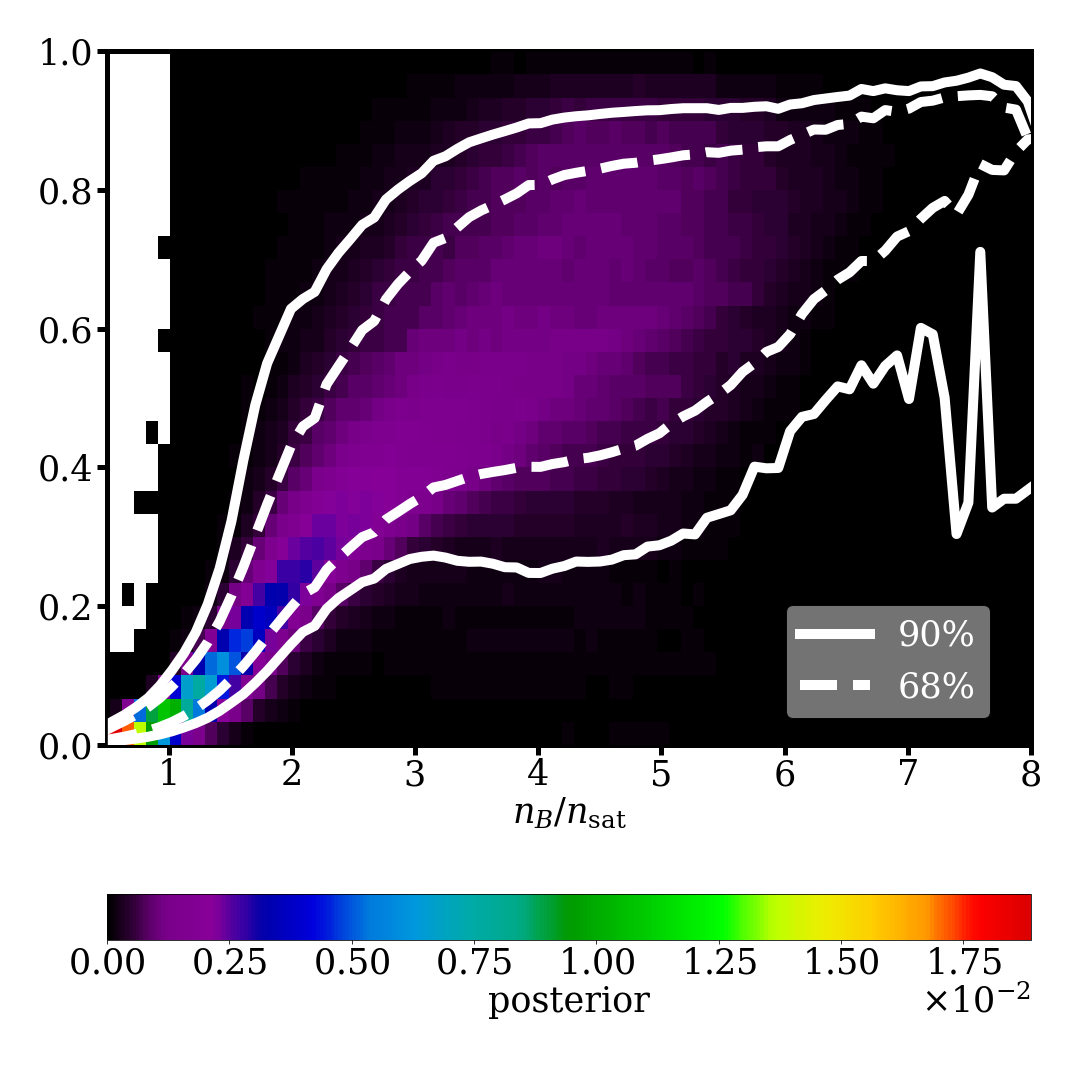}
   \end{tabular}
  \caption{EoS prior (left) and posterior (right) probability distributions for GP (top) and mGP (bottom) EoS. The EoS are represented by the speed of sound squared in units of $c^2$ as a function of baryon number density in units of $n_\textrm{sat}$. The prior and posterior probability distributions are produced by binning the EoS by the speed of sound and number density and then normalizing the heights of the bins such that their sum is equal to one. For the posteriors, each EoS is weighed by the corresponding likelihood. Also shown in the posterior plots are the 90\% and 68\% credible regions for the speed of sound squared at a given density for $0.5 \leq n_B \leq 8.0 \ n_{\rm sat}$. The posterior probability that the central density for a maximally massive star is greater than $\sim6.0 \ n_{\rm sat}$ is negligible in both cases. We note that probability densities are low in the regime between 2-6 $n_{\rm sat}$ because of the wide spread in the allowed behavior for $c_s^2$. Observe that at the 90\% level, the mGP posterior is wider than the GP one for all densities above $n_{\rm sat}$.
  }
  \label{fig:cs}
\end{figure*} 

\subsection{Are EoS posteriors sensitive to structure in $c_s^2(n_B)$?}

Next, we test if the different prior assumptions made about correlations across densities in the benchmark GP versus the mGP model lead to any significant differences in the posterior for $c_s^2(n_B)$. Recall that the benchmark GP produces smooth $c_s^2$ curves with uniform correlations across densities, whereas the mGP display sharp features in $c_s^2$ and multi-scale correlations across densities (refer also to Figs.\ \ref{fig:nucEoS} and \ref{fig:samples}).

\begin{figure}
   \centering        \hspace{-0.5cm}\includegraphics[width=\linewidth]{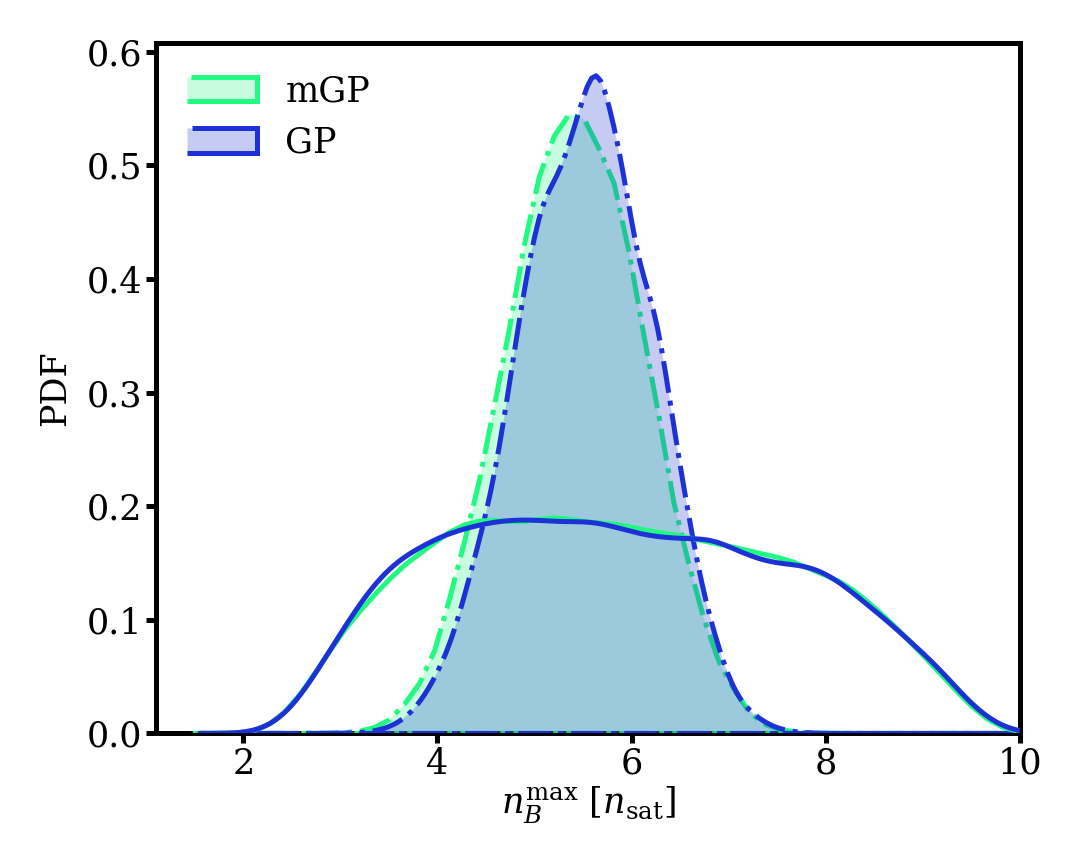}
  \caption{The estimated prior (solid lines) and estimated posterior (dot-dashed, filled lines) probability density distributions for the maximal central density of a stable, nonrotating star in units of $n_{\rm sat}$ for EoS from the benchmark GP (green) and mGP (blue). The priors for GP and mGP produce nearly identical PDFs for $n_B^{\rm max}$ such that the lines are indistinguishable from each other. The vast majority of EoS from the posterior predict a maximal central density between $\sim 4 - 8$ $n_{\rm sat}$.}
  \label{fig:nBmax}
\end{figure}

In Fig.~\ref{fig:cs}, we show the priors (left panels) and the posteriors (right panels) for $c_s^2$ as a function of $n_B$ in units of $n_{\rm sat}$, up to $n_B^\textrm{max}$ for each EoS. Again, benchmark GP and mGP results are shown on the top and bottom panels, respectively. The priors are shown using the binning method outlined in Eqs.\ (\ref{eqn:Hprior}-\ref{eqn:Htot}). The posteriors are shown both in terms of the binned joint posteriors, as well as constant density  $68\%$ and $90\%$ credible bands. In these plots, we only show $c_s^2$ up to the maximum central baryon density for a stable star in the slow-rotation regime, $n_B^{\rm max}$. Most EoS lead to $n_B^{\rm max}$ around 4-7 $n_{\rm sat}$ with a handful that extend up to 8 $n_{\rm sat}$.  Thus, we plot only up to $n_B = 8.0 \ n_{\rm sat}$. We will discuss the posterior for $n_B^{\rm max}$ separately in Fig.~\ref{fig:nBmax}. We note that Fig.~\ref{fig:cs} only includes EoS in $\Phi_\checkmark$ (the set of EoS that pass the constraints discussed in Sec.~\ref{sec:priors}), which clearly affects the priors. This selection leads to $c_s^2(n_B)$ functional forms that favor high values ($c_s^2 \geq 1/3$) at large densities ($n_B\geq 2\ n_{\rm sat}$). 

Let us first discuss the priors. Observe that the highest probability regions look nearly identical between the benchmark GP and mGP models when we only consider EoS in $\Phi_\checkmark$. In both cases, at very low densities ($n_B \leq n_{\rm sat}$), there is a strong preference for a nearly vanishing $c_s^2$. This result is unsurprising because we use the same crust at low densities for both models and also impose that no modifications are introduced in $c_s^2$ below $n_B< 1.1$ $n_{\rm sat}$ for mGP EoS. Progressing to intermediate densities ($n_{\rm sat} \leq n_B \leq$ 3 $n_{\rm sat}$), we find a general trend in both priors to larger $c_s^2$, but this trend has a rather wide spread, allowing for diverse behavior in $c_s^2$, hence the lower probability within the credible bands. At densities above 3 $n_{\rm sat}$, we find a general trend upward in $c_s^2$, but again with an even larger spread. 

One key difference, however, does exist between the benchmark GP and mGP priors. The benchmark GP EoS in $\Phi_\checkmark$ do not contain
\emph{any} samples that have a low $c_s^2$ at large $n_B$ (notice the large white region in the bottom, left corner of the top, left panel in Fig.~\ref{fig:cs}). On the other hand, the same region in the mGP prior has a nonzero prior probability density. The key difference is that the benchmark GPs are smooth and domain points are correlated over a long range in density. Therefore, the speed of sound functional forms from the benchmark GP cannot easily fluctuate downward to this region (especially since they need to support neutron stars with $M$ $\geq 1.8 M_\odot$). In contrast, functional forms from the mGP model can have fluctuations to larger $c_s^2$, followed by regions of lower $c_s^2$. In this way, it is clear that the mGP model allows us to explore a wider regime in EoS parameter space.  

Next, let us discuss the binned joint $n_B-c_s^2$ posteriors, and the constant density 90\% and 68\% credible regions, shown in the right panels of Fig.~\ref{fig:cs}. The posteriors and credible bands are similar between the benchmark GP and the mGP, but they are not identical. We find that, at the 90\% level, the mGP posterior is wider than the GP one for all densities above $n_{\rm sat}$. Notably, the mGP posterior allows for slightly stiffer EoS in the regime $1.5 \lesssim n_B \lesssim 3.0\ n_{\rm sat}$, and slightly softer EoS above $3.0\ n_{\rm sat}$. For instance, at twice nuclear saturation density and 68\% credibility, we extract $c^2_s =0.29^{+0.27}_{-0.11}$ using EoS from the GP posterior, and $c^2_s =0.29^{+0.34}_{-0.14}$ using the mGP posterior. At four times nuclear saturation density, the GP EoS range is $c^2_s =0.63^{+0.27}_{-0.23}$, while the mGP EoS allows for $c^2_s =0.59^{+0.31}_{-0.34}$. Thus, we find that the GP leads to slightly stiffer EoS and has slightly narrower posterior credible bands. As we will discuss later in Fig.~\ref{fig:nBmax}, very few EoS reach beyond $n_B > 6\ n_{\rm sat}$, so the statistics in that regime are not sufficient to draw conclusions about differences between the two frameworks. This is evident from the highly oscillatory behavior of the mGP credible bands in that region.

We can draw further conclusions from the binned joint probability density posteriors. In both posteriors, there is a strong preference (blue regions) for a sharp rise in the speed of sound between 1-2 $n_{\rm sat}$. Furthermore, at densities larger than 2 $n_{\rm sat}$, large $c_s^2$ is favored, well beyond the conformal limit of $c_s^2=1/3$.  However, $c_s^2(n_B)$ is significantly less constrained in that regime. Lastly, we also see from the binned joint posteriors that a large fraction of EoS must reach their maximum central densities around $\sim 5$ $n_{\rm sat}$, because the probability density decreases significantly beyond that point.  

While the relative differences in the posteriors are not huge between the GP and mGP models, the point still holds that when the $c_s^2$ is allowed to display sharp features, the posteriors are wider than when a smooth EoS is presumed. This is because implicitly imposing smoothness in the EoS through benchmark EoS translates to a prior that disallows low speeds of sound at high densities. We, therefore, argue that such sharp features should be adequately represented in priors for the extraction of the EoS in neutron star regimes. Furthermore, it is clear that if one is especially interested in studying whether low values of $c_s^2$ at high densities are allowed by nature, it is even more important to allow for sharp features in $c_s^2$. 

\subsection{Are maximum central density posteriors sensitive to structure in $c_s^2(n_B)$?}
\label{subsect:max-central-densities}

We will now discuss the priors and posteriors for the maximal central baryon density reached by a stable neutron star in the slow rotation regime, $n_B^{\rm max}$. From Fig.~\ref{fig:cs}, we already saw hints that $n_B^{\rm max}$ must peak around 5 $n_{\rm sat}$ because the posterior probability densities are low at baryon densities higher than that. To study this systematically, we have plotted the range of $n_B^{\rm max}$ for both our priors and posteriors in Fig.~\ref{fig:nBmax} using a kernel density estimate\footnote{For this observable, the probability densities vary smoothly with $n_B^{\rm max}$, and hence, the use of kernel density estimates instead of the binning method we discussed earlier is safe.}. 

Observe that the $n_B^{\rm max}$ priors for the benchmark GP (blue line) and mGP (green line) are essentially identical and mostly overlapping, ranging between $n_B^{\rm max}=2-10$ $n_{\rm sat}$ (again, recall that our priors shown here represent only the samples in $\Phi_\checkmark$).  However, there is a significant change in the posterior probability density for both the benchmark GP and the mGP compared to their priors. Both posteriors are sharply peaked at $n_B^{\rm max}=5-6$ $n_{\rm sat}$ with a range between $n_B^{\rm max}=4-8$ $n_{\rm sat}$ (with essentially no EoS that produce $n_B^{\rm max}>8$ $n_{\rm sat}$). The posterior for the mGP model peaks at a slightly smaller $n_B^{\rm max}$ than that of the GP model, but the difference between the two distributions is very small. These results are consistent with what is shown in Fig.~\ref{fig:cs}, where the probabilities are compatible with zero for $n_B\geq \ 8$ $n_{\rm sat}$.

\begin{figure}[ht]
   \centering
        \includegraphics[width=\linewidth]{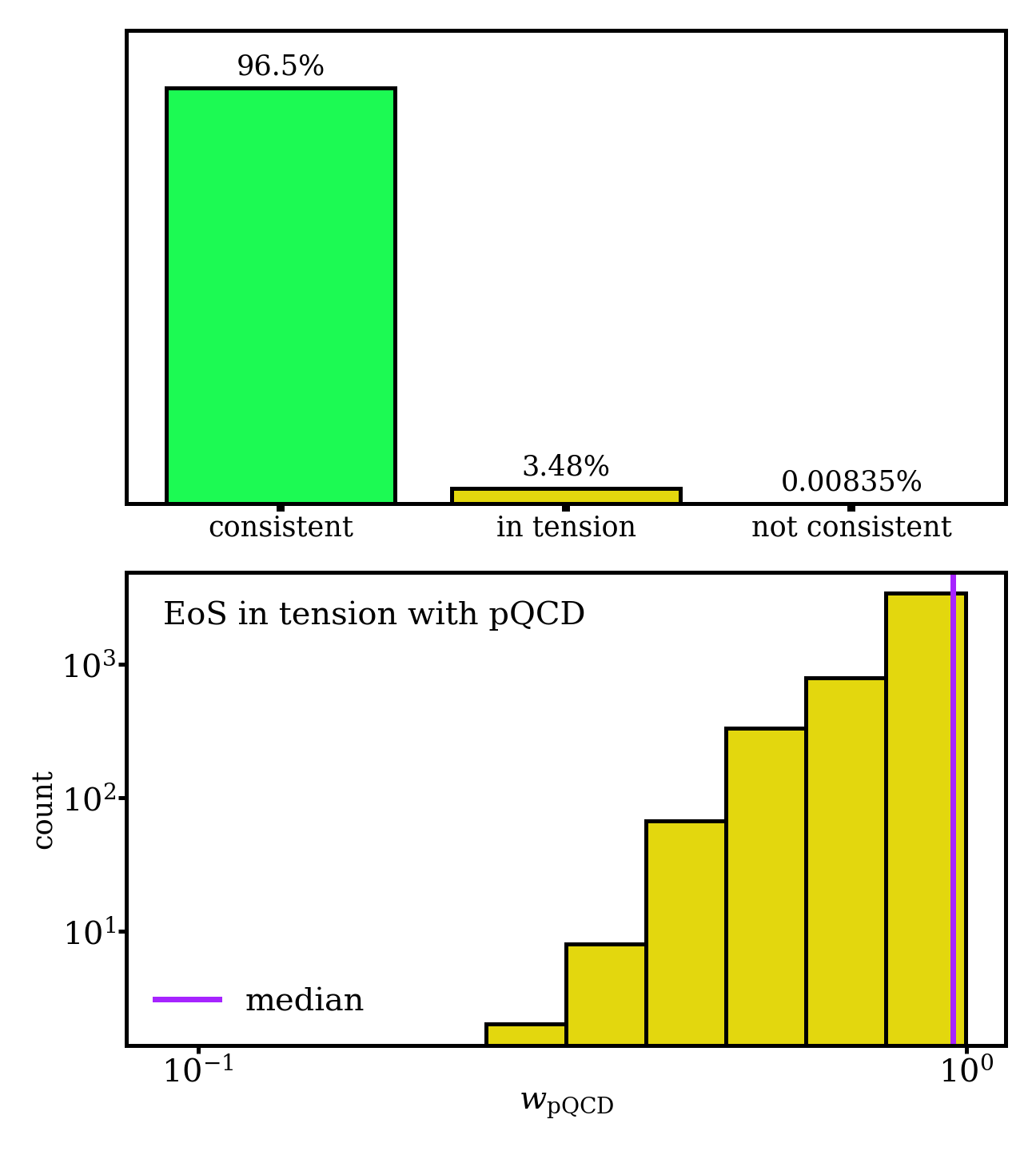}
  \caption{Top: bar chart of the percentage of EoS that are consistent ($w_{\rm pQCD}=1$), in tension ($0 < w_{\rm pQCD} < 1$), and not consistent ($w_{\rm pQCD}=0$) with pQCD input based on the formalism in Ref.~\cite{Gorda:2022jvk}. Bottom: histogram of the $w_{\rm pQCD}$ for the 4,592 EoS  ($\sim3.5\%$ of the total number of samples) in the combined benchmark GP and mGP prior that are in tension with pQCD input. The impact of pQCD input when $n_{B,k}^{\rm max}$ is used as the matching density is negligible (see Sec.~\ref{subsec:pqcd_input} for definitions). 
  }
  \label{fig:tension}
\end{figure} 

\begin{figure*}[ht]
      \centering
   \begin{tabular}{cc}
        \includegraphics[width=0.49\linewidth]{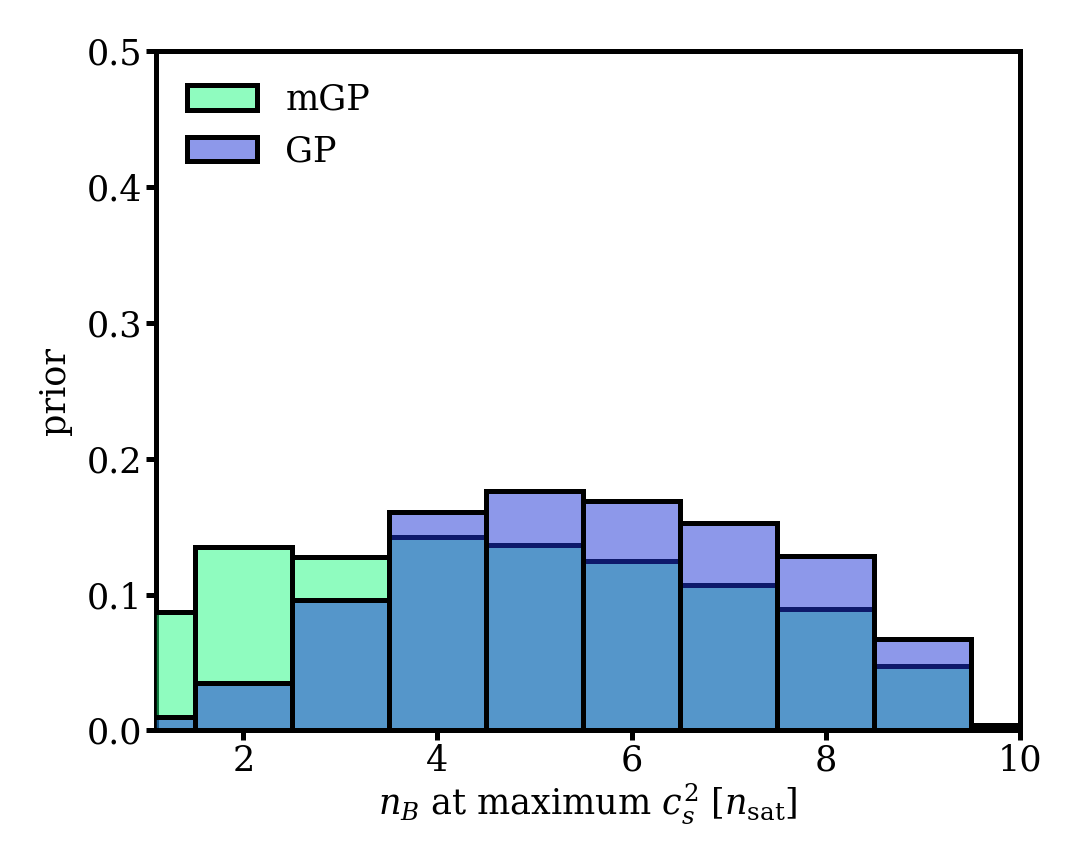} &
        \includegraphics[width=0.49\linewidth]{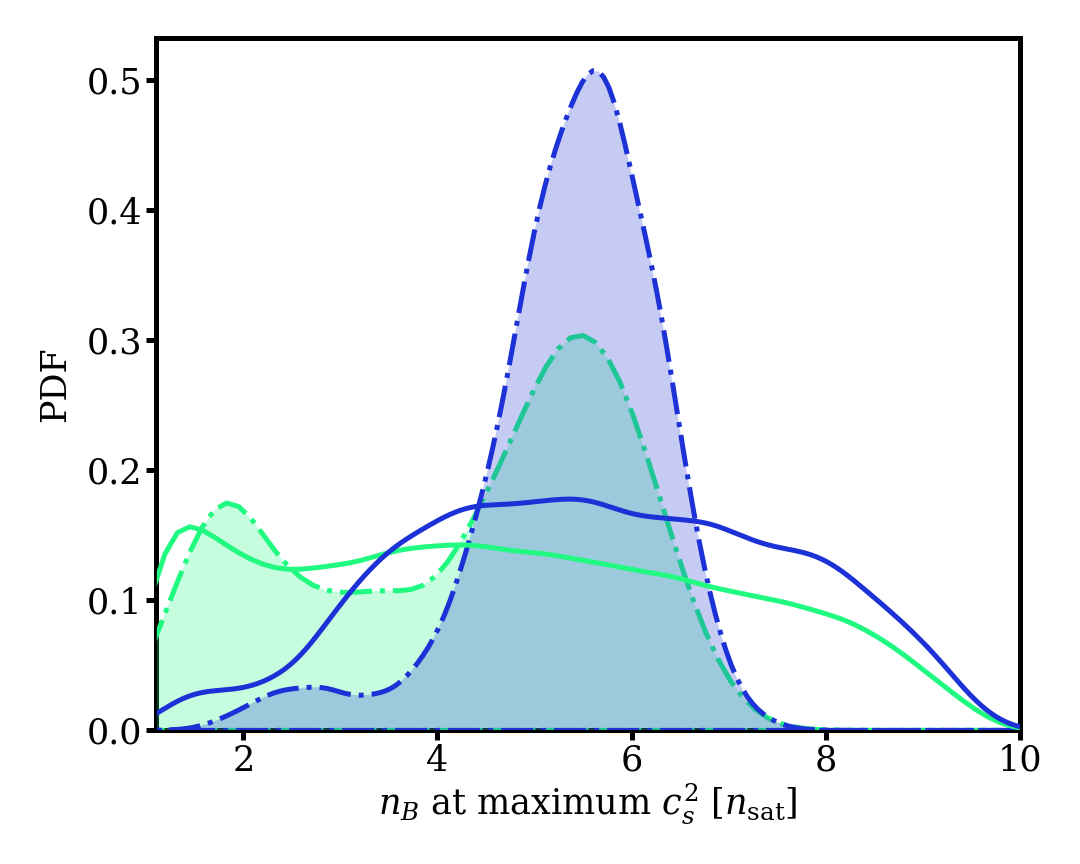}
   \end{tabular}
  \caption{Left: histograms of the prior distributions for the GP and mGP EoS of the value of the baryon density $n_B$ in units of $n_{\rm sat}$ at the global speed of sound maximum for a stable nonrotating neutron star. The increased flexibility of mGP EoS allows for a global $c_s^2$ maximum to occur at lower densities compared to the GP EoS. Right: the estimated prior probability density distributions corresponding to the histograms on the left (solid curves) and estimated posterior probability density distribution after constraints on the symmetry energy, mass, radius, binary tidal deformability are imposed along with pQCD input (dot-dashed curves). Data and theoretical constraints do not rule out a global $c_s^2$ maximum at densities below 3 $n_{\rm sat}$, but are also not yet informative enough to favor or disfavor it over a global $c_s^2$ maximum above 3 $n_{\rm sat}$.}
  \label{fig:posteriornBatglobalmax}
\end{figure*} 

\subsection{What is the impact of pQCD constraints?}\label{subsec:pqcdimpact}

As discussed in Sec.~\ref{subsec:pqcd_input}, in order to include pQCD constraints, one must choose a matching density, $n_{\rm matching}$. This matching density reflects how pQCD results are being propagated via stability, causality, and consistency constraints. In this work, we used $n_{\rm matching} = n_{B,k}^{\rm max}$, the maximal central density predicted by EoS $k$ for a stable, non-rotating neutron star. Although  $n_{B,k}^{\rm max}$ changes for each EoS, we saw in Sec.~\ref{subsect:max-central-densities} and Fig.~\ref{fig:nBmax} that the posterior probability density for $n_{B}^{\rm max}$ drops sharply above $\sim 6 \, n_{\rm sat}$ and is negligible beyond $8 \, n_{\rm sat}$.

In Fig.~\ref{fig:tension} (top), we quantify how pQCD constraints affected our inference of the EoS. In the bar chart, we show the proportion of EoS in our prior that are consistent ($w_{\rm pQCD} = 1$ exactly), in tension ($0<w_{\rm pQCD} < 1$), and inconsistent ($w_{\rm pQCD} = 0$ exactly) with pQCD input (see Eq.~(\ref{eq:w_pQCD_def}) for the definition of $w_{\rm pQCD}$). A vast majority of the EoS in the prior ($\sim96.5\%$) are consistent with pQCD results, $\sim3.5\%$ are in tension, and only $0.0083\%$ (11 total) EoS are inconsistent with pQCD results. Because pQCD only completely excludes a \textit{very} small fraction of EoS, these constraints cannot contribute strongly to the $M-R$ and $c_s^2(n_B)$ posteriors, shown in the previous section. 

We observe, however, that a non-negligible population ($\sim3.5\%$) that is in tension with pQCD results indeed exists. In Fig.~\ref{fig:tension} (bottom), we show a histogram for $w_{\rm pQCD}$ only for the samples that were found to be in tension with pQCD. Recall that $w_{\rm pQCD}$ is proportional to the strength of the disagreement between a given EoS and pQCD results over the range of $\Bar{\Lambda}$ we sampled. Thus, we can quantify the impact of these samples on the posterior probability by looking at their $w_{\rm pQCD}$ distribution. As shown in Fig.~\ref{fig:tension}, the median of the distribution is $\sim 0.95$, meaning that more than half of the samples in tension with pQCD are only marginally suppressed. 

The results shown here suggest that the impact of pQCD input on the inference of the neutron star EoS is minimal. Given that the opposite has been reported in the literature recently \cite{Gorda:2022jvk,Annala:2023cwx,Altiparmak:2022bke}, we would like to address why that is the case. 
Let us recall Sec.~\ref{subsec:pqcd_input}, where we discussed the assumptions associated with including pQCD input in the analysis. We assume we know the EoS at a low-density limit (see Eqs.~(\ref{eq:betalow},\ref{eq:betahigh})), each determined by a set of three values that fix the EoS in those limits. These values correspond to the number density ($n$), the chemical potential ($\mu$), and the pressure ($p$) at each limit. In the low density regime, we get $\mu_{\rm low}$ and $p_{\rm low}$ from the EoS, but we need to make a choice for $n_{\rm low}$. This choice is important because $n_{\rm low} = n_{\rm matching}$, meaning that pQCD results will be propagated down to $n_{\rm low}$. This choice is in principle arbitrary, but given that the largest density scale relevant to the EoS of isolated, slowly-rotating neutron stars is $n_B^{\rm max}$, it is reasonable to impose $n_{\rm low} = n_B^{\rm max}$ for each EoS. This is the choice that we make in this work and the choice that was made in Ref.~\cite{Somasundaram:2022ztm}, which also reported that pQCD only affected a very small number of EoS in the prior. Works that found a robust softening of the EoS at high densities, leading to a peak in the $c_s^2(n_B)$ posterior, used $n_{\rm matching} = 10$ $n_{\rm sat}$, where pQCD is more constraining because it is closer to $\mu_{\rm high} = 2.6$ GeV. However, as shown in Fig.~\ref{fig:nBmax}, the $n_B^{\rm max}$ posterior is essentially zero for $n_B^{\rm max}> 7$ $n_{\rm sat}$. That means that in Refs.~\cite{Gorda:2022jvk,Annala:2023cwx,Altiparmak:2022bke} pQCD constraints are being imposed far beyond the densities where most realistic EoS predict a stable neutron star. There are no constraints from astronomical observations in those densities, so the impact of pQCD on the posterior will depend on prior-imposed assumptions about the correlations in $c_s^2(n_B)$ in the regime above $n_B^{\rm max}$. We note here, as Refs.~\cite{Essick:2023fso,Brandes:2023hma} also noted, that such results are very sensitive to the prior. 

Imposing pQCD constraints at $n_B^{\rm max}$ with $X = [1/2,2]$ results in only 3.5\% of the EoS being affected. Only 11 are completely ruled out. Again, we highlight that in Fig.~\ref{fig:tension}, which shows a histogram of the pQCD weights assigned to EoS that were suppressed by pQCD, we see that the vast majority of the EoS affected were only marginally suppressed. With these results, we conclude that our posteriors are dominated by astrophysical observations, which is why we do not see a softening of the EoS at larger densities.
Nonetheless, pQCD offers nontrivial constraints even when incorporated exclusively at densities where an EoS predicts stable, slowly-rotating neutron stars exist. Lastly, we point out that our findings are in agreement with those of Ref.\cite{Somasundaram:2022ztm}, which found that pQCD affects the EoS mainly beyond the densities realized in neutrons stars.

\subsection{Does $c_s^2(n_B)$ display a peak within neutron star densities?}

Given that mGP EoS lead to reasonable mass-radius and $c_s^2(n_B)$ posteriors, we can now begin to explore the existence of structure in $c_s^2(n_B)$. One way to study the latter is to look for a bump (i.e.~$c_s^2(n_B)$ rises and reaches a global maximum at some $n_B<n_B^{\rm max}$ before decreasing again) that would signify a crossover phase transition. This type of structure is being actively discussed in the literature \cite{Marczenko:2022jhl,Annala:2019puf,Annala:2023cwx,Pang:2023dqj,Takatsy:2023xzf,Gorda:2022jvk,Altiparmak:2022bke} as a signature of quark matter in the core of massive neutron stars. Such structure has become especially relevant after studies that use pQCD constraints applied beyond densities realized in most realistic neutron star EoS found a posterior for $c_s^2(n_B)$ that displays a peak within neutron star densities \cite{Annala:2023cwx,Gorda:2022jvk}.

One caveat here is that, as seen in Fig.~\ref{fig:nucEoS}, $c_s^2(n_B)$ can oscillate or contain first-order phase transitions. Thus, the \emph{absence} of a global maximum in $c_s^2(n_B)$ before $n_B^{\rm max}$ does not imply that  a phase transition does not occur within neutron star densities. Similarly, the \emph{presence} of a global maximum in $c_s^2(n_B)$ before $n_B^{\rm max}$ does not confirm a transition to quark matter occurs in the core of massive neutron stars because, as shown in Fig.~\ref{fig:nucEoS}, the onset of degrees of freedom other than quarks, such as heavy resonances or hyperons, can also cause the EoS to soften in a similar way. With this caveat in mind, we define the density at which the maximum in $c_s^2$ is reached as $n_B(c^2_{s,{\rm max}})$, or $n_B$ at maximum $c_s^2$.

\begin{figure*}[ht]
   \begin{tabular}{cc}
        \includegraphics[width=0.49\linewidth]{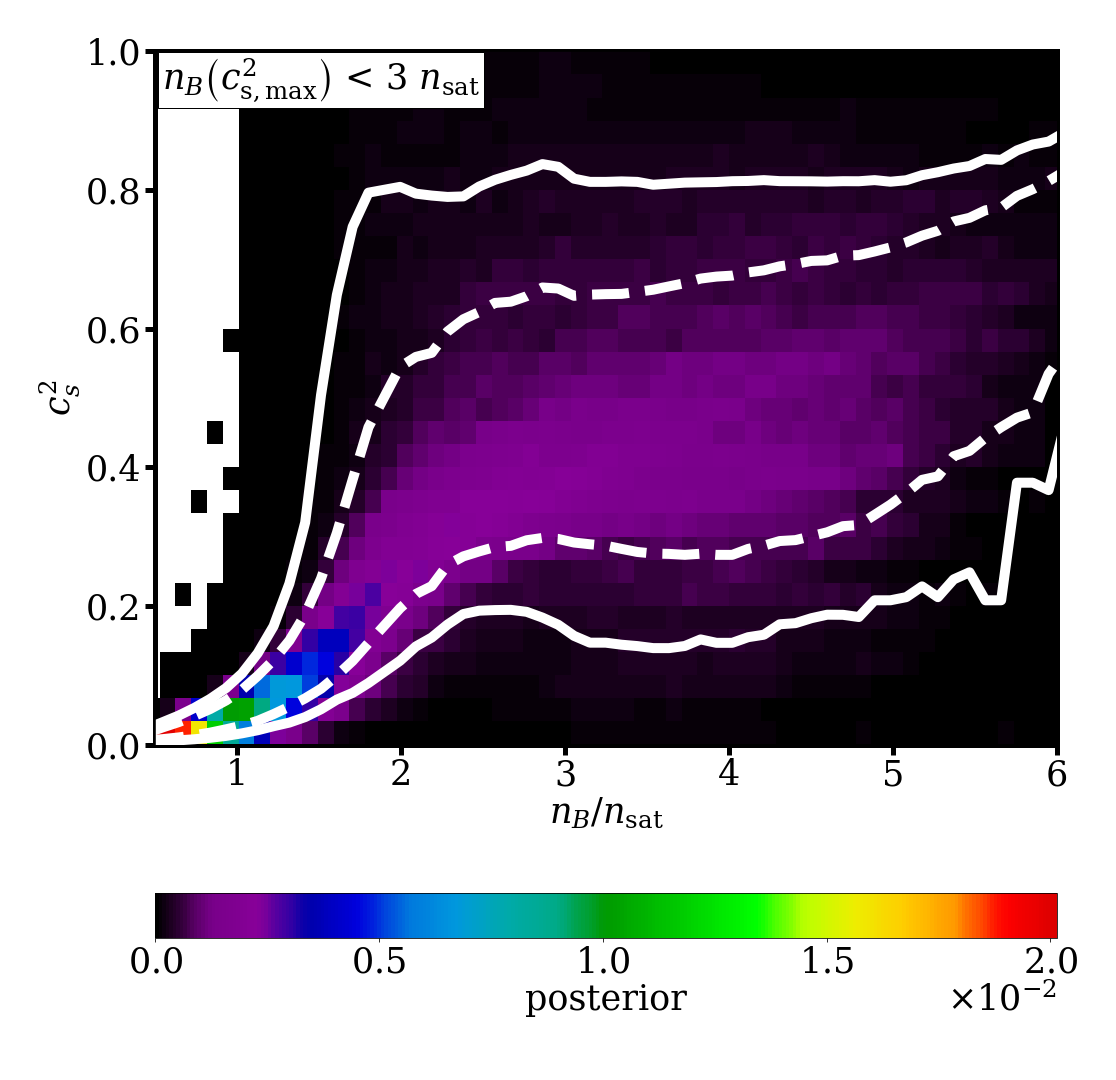} &
        \includegraphics[width=0.49\linewidth]{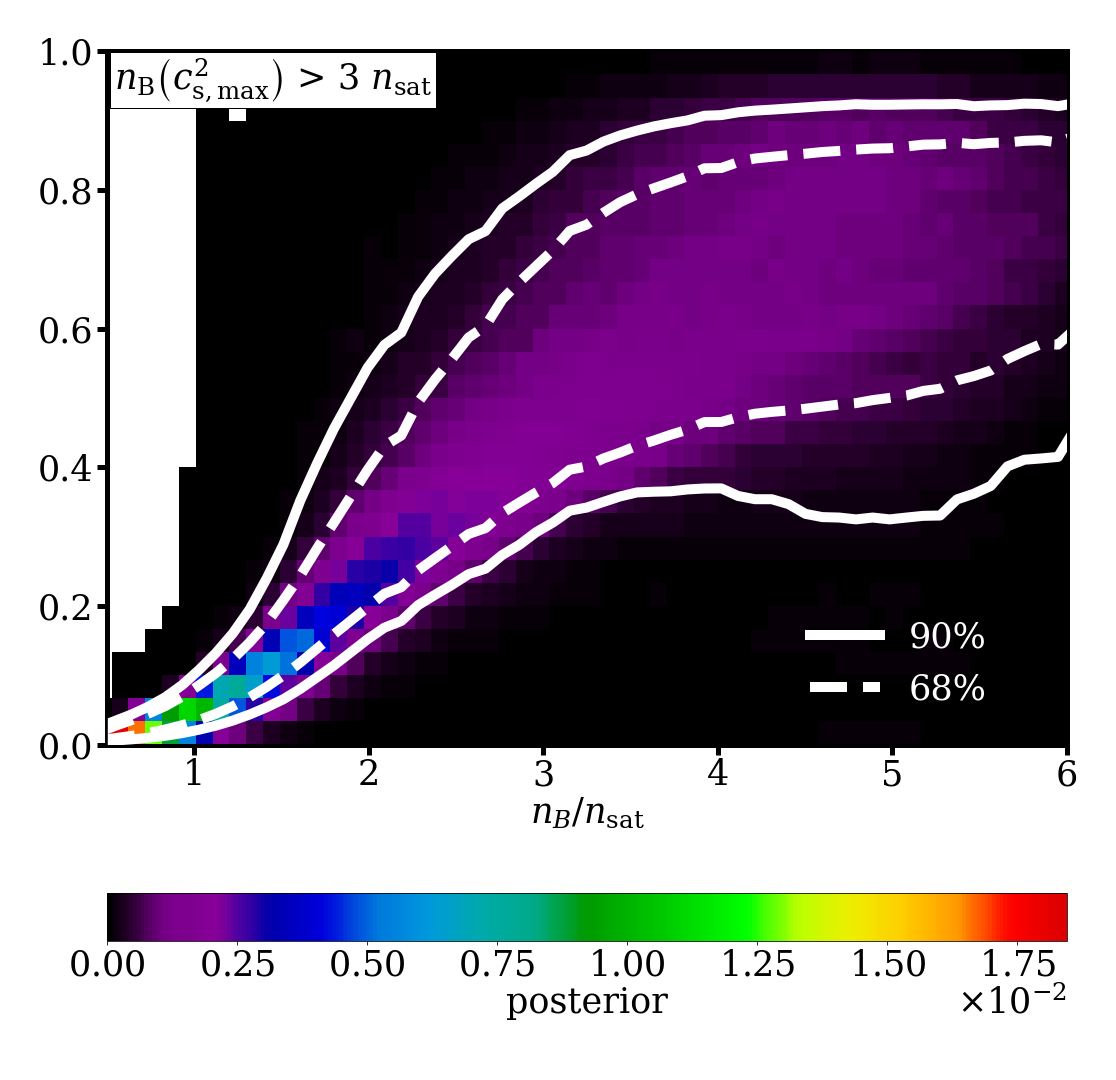}
   \end{tabular}
  \caption{EoS posteriors for the case when a global maximum in the speed of sound is present below (left) and above (right) 3 $n_{\rm sat}$ for mGP EoS. The EoS are represented by the speed of sound squared in units of $c^2$ as a function of baryon number density in units of $n_\textrm{sat}$. The posterior probability distributions are produced by binning the EoS by the speed of sound and number density, weighing each EoS by the corresponding likelihood, then normalizing the heights of the bins such that the sum of all bin heights is equal to one. Also shown in the posterior plots are the 90\% and 68\% credible regions for the speed of sound squared at a given density for $0.5 \leq n_B \leq 6.0 \ n_{\rm sat}$. The posterior probability that the central density for a maximally massive star is greater than $\sim6.0 \ n_{\rm sat}$ is negligible in both cases.}
  \label{fig:cs2twocases}
\end{figure*} 

The left panel of Fig.~\ref{fig:posteriornBatglobalmax} shows the prior for $n_B(c^2_{s,{\rm max}})$ for both the benchmark GP and mGP models. The maximum value of $c_s^2$ can occur at any density up to $n_B\leq 10$ $ n_{\rm sat}$, as seen for both priors.  However, already at the level of the prior, we do see differences between the two models. The benchmark GP model prior has a peak at approximately the same density as the $n_B^{\rm max}$ posterior shown in Fig.~\ref{fig:nBmax}. This feature indicates that benchmark GP samples mostly reach a global maximum in $c_s^2$ at or near their maximal baryon density, i.e., $n_B(c^2_{s,{\rm max}}) \approx n_B^{\rm max}$. This result is compatible with the benchmark GP assumption that the low density (below $\sim 1.1 n_{\rm sat}$) and high density (above $\sim 1.1 n_{\rm sat}$) regions display correlations of length $\ell = 1$ in units of $\log_{10}$ erg cm$^{-3}$. The EoS in the low density regime must be smooth and soft to be in agreement with symmetry energy estimates, while astronomical observations require an EoS stiff enough to support 2 $M_\odot$ stars but not too stiff in the regime below 3 n$_{\rm sat}$ because of tidal deformability constraints around 1.4 $M_\odot$ stars. This transition from soft to stiff, by construction, happens over a range in pressure corresponding to the hyperparameter $\ell$. Since we imposed a reasonably large value for $\ell$, bumps are less likely in the benchmark GP model. In contrast, the mGP model has a prior that is relatively uniform in the range between $1.1-8$ $n_{\rm sat}$, rather than peaked at densities above 3 $n_{\rm sat}$, as in the benchmark GP case. That implies that some of these EoS have a low-density bump in $c_s^2$.  In this case, the assumption that the low and high density regimes are correlated over a long range in pressure is relaxed, allowing for low-density bumps to appear. 

In the right panel of Fig.~\ref{fig:posteriornBatglobalmax}, we show the posterior for the density at which the speed of sound is maximal (computed from a kernel density estimate), together with the prior, for both the benchmark GP (blue) and mGP models (green). The posteriors present interesting features and  striking differences between both models. The benchmark GP model leads to a posterior with a maximum consistent with its prior, and thus, consistent with the maximum density $n_B^{\rm max}$. However, the mGP posterior distribution for $n_B$ at the global $c_s^2$ maximum is \textit{bimodal}, with peaks at $\sim 2$ $n_{\rm sat}$ and $\sim 5$ $n_{\rm sat}$. The peak of the mGP posterior distribution centered at $n_B\sim 5$ $n_{\rm sat}$ is somewhat larger than the peak at $n_B\sim 2$ $n_{\rm sat}$. Nonetheless, the peak at $n_B\sim 2$ $n_{\rm sat}$ is still quite significant and it is clearly a result of the extra structure built into the mGP model. We note that this low-density bump is consistent with recent preliminary findings from heavy-ion collisions \cite{Oliinychenko:2022uvy} and may be an indication of a crossover phase transition (a possible explanation is quarkyonic matter, see, e.g., Ref.~\cite{McLerran:2018hbz}). We should be careful in our interpretation of these results, however, since the posterior is bimodal, and a monotonically increasing $c_s^2$ cannot be ruled out. It remains to be seen if the low-density peak in the posterior for $n_B(c^2_{s,{\rm max}})$ will be further enhanced or suppressed by future astrophysical observations of neutron stars. 

We now investigate the differences between the two distinct peaks in the mGP posterior shown in Fig.~\ref{fig:posteriornBatglobalmax} by dividing the mGP EoS into two groups: one with $n_B(c^2_{s,{\rm max}})< 3$ $n_{\rm sat}$ and one with $n_B(c^2_{s,{\rm max}})>3$ $n_{\rm sat}$. In Fig.~\ref{fig:cs2twocases}, we show the mGP posterior for $c_s^2(n_B)$ in the first group (left panel) and in the second group (right panel), using the binned probability densities from Eqs.~(\ref{eqn:Hprior}-\ref{eqn:Htot}), together with the constant density 90\% and 68\% credible regions. The resulting posteriors are qualitatively different between the two groups. We see a much sharper increase in $c_s^2$ at low densities for the $n_B(c^2_{s,{\rm max}})<3$ $n_{\rm sat}$ group in the left panel. In that case, we see that $c_s^2$ may have a peak, followed by a decrease, or it may plateau, at larger $n_B$. Interestingly, the $n_B(c^2_{s,{\rm max}})<$ $3n_{\rm sat}$ group allows for the possibility of a softening in the EoS at large densities that is not seen in the other group. The $n_B(c^2_{s,{\rm max}})>3$ $n_{\rm sat}$ group on the right panel appears to have a monotonically increasing posterior that ends at a large $c_s^2$ at large $n_B$. This group more closely resembles nucleonic-only EoS.  Another interesting difference is that, unlike the $n_B(c^2_{s,{\rm max}})<3$ $n_{\rm sat}$ group, the $n_B(c^2_{s,{\rm max}})>3$ $n_{\rm sat}$ group has a tight $c_s^2$ distribution at large $n_B$, which drives $c_s^2$ to large values.

We can also analyze quantitative differences between these posteriors.
In the left panel, at 2 $n_{\rm sat}$,  the  $n_B(c^2_{s,{\rm max}})<3$ $n_{\rm sat}$ group allows for $c_s^2$ as high as $\sim 0.80$ at 90\% credibility and $\sim 0.55$ at 68\% credibility. In contrast, the second group predicts much smaller $c_s^2$ for $n_B=2$ $n_{\rm sat}$, $\sim 0.45$ and $\sim 0.35$ at 90\% and 68\% credibility, respectively. At densities above 3 $n_{\rm sat}$, the $n_B(c^2_{s,{\rm max}})<3$ $n_{\rm sat}$ group continues to allow for a wide range of $c_s^2$, displaying $c_s^2$ values as low as $\sim 0.2$ at 90\% credibility, and as high as $\sim 0.8$ at 90\% credibility at 5 $n_{\rm sat}$. At 68\% credibility, the lower and upper bands are at roughly $c_s^2 \approx 0.3$ and $c_s^2 \approx 0.7$, respectively, at 5 $n_{\rm sat}$. In contrast, at 5 $n_{\rm sat}$, the ranges for the $n_B(c^2_{s,{\rm max}})>3$ $n_{\rm sat}$ group are approximately $[0.5,0.8]$ at 68\% credibility and $[0.3,0.9]$ at 90\% credibility. Overall, if a global maximum occurs below 3 $n_{\rm sat}$, our results indicate that we can expect an EoS that is stiffer at low densities and softer at high densities. On the other hand, if a global maximum occurs above 3 $n_{\rm sat}$, the $c_s^2$ posterior suggests that the EoS is stiffer and above the conformal value of 1/3 for all $n_B>3\ n_{\rm sat}$.

Recalling an earlier discussion about what the absence of a clear peak-like structure in $c_s^2$ means, we emphasize that the EoS in both posteriors shown in Fig.~\ref{fig:cs2twocases} were generated using the mGP framework. Therefore, all these samples contain nontrivial features. Thus, it is possible that $n_B(c^2_{s,{\rm max}})>3$ $n_{\rm sat}$ group may have a small bump in $c_s^2$ at low densities but then the EoS continues to become stiffer, ending at an even larger $c_s^2$ near $n_B^{\rm max}$.

\subsection{Are there nontrivial features in $c_s^2(n_B)$?}

We have established that different assumptions about the scale of correlations across densities in the speed of sound functional does not lead to significantly different predictions for the mass-radius relation or $c_s^2(n_B)$ given current constraints. On the other hand, introducing multi-scale correlations via the mGP model had a significant impact on the posterior for the value of $n_B$ at the maximum $c_s^2$. What we learn from this is that both the benchmark GP and the mGP models can describe astronomical observations, while respecting symmetry energy and pQCD constraints. We can now ask if the data \textit{prefers} one of the two models.

As discussed in Sec.~\ref{subsec:primer}, the model evidence quantifies the level of support of the data for a given model, and the ratio between the evidence for two different models, the Bayes factor, quantifies if one of the models is preferred over the other by the data. Using Eq.~(\ref{eq:evidencefinite}), we separate the benchmark GP and the mGP samples and compute the evidence for each. We find a Bayes factor of \footnote{In a previous note \cite{Mroczek:2023eff}, we reported a Bayes factor of K=1.126, which was obtained using a normalization factor ($N_{m}$ in Eq.~\ref{eq:evidencefinite}) that reflected the size of the subset of the priors for each model that passed the checks (using the notation introduced in Secs.~\ref{subsec:hyperprior} and \ref{subsec:model_evidence}, these sets are $\Phi_\textrm{benchmark GP}\cap\Phi_\checkmark$ and $\Phi_\textrm{mGP}\cap\Phi_\checkmark$). This choice essentially ensured a Bayes factor of $\sim 1$, because both priors have information about astrophysical constraints (Ref.~\cite{Essick:2023fso} also pointed this out). Here, $N_{m}$ corresponds to the full prior sample size (using the notation introduced in Secs.~\ref{subsec:hyperprior} and \ref{subsec:model_evidence}, these sets are $\Phi_\textrm{benchmark GP}$ and $\Phi_\textrm{mGP}$).}

\begin{equation}\label{eq:bayesfac}
    K = \dfrac{\mathcal{E}_{\rm benchmark GP}}{\mathcal{E}_{\rm mGP}} = 1.480.
\end{equation}
This value is not a significant deviation from unity, which means that current constraints do not favor either model. The physical interpretation is that multi-scale correlations and nontrivial features in $c_s^2(n_B)$ are not ruled out by current constraints, but neither are they required.  

\section{Conclusions and Discussion}\label{sec:conclu}

Nuclear physics models with phase transitions and exotic degrees of freedom contain multiscale features that present as non-trivial structure in $c_s^2$.  In this work, we developed a new framework, which we named modified Gaussian processes (mGP), as a novel approach to producing functional forms of the EoS for the cold, catalyzed nuclear matter in neutron stars that each contain long-, medium-, and short-range correlations. These EoS can be generated with high computational efficiency and contain features that are indicative of the emergence of exotic degrees of freedom. 

We compared our new mGP EoS model to a benchmark GP model that contains only long-range correlations (i.e.,~does not contain any short-  or medium-range correlations), using a Bayesian analysis that incorporated astrophysical data, low-energy nuclear physics constraints, and input from pQCD calculations. From our Bayesian analysis, we found that both the benchmark GP and mGP models provide nearly equivalent results for the mass-radius, $c_s^2(n_B)$, and maximum central density posteriors. In fact, the Bayesian evidence for both models is nearly the same, leading to a Bayes factor of 1.5 between them. Thus, it is clear that we cannot rule out nontrivial features in the speed of sound from the data, and also that these EoS are as valid as a smooth EoS, given current data. 

Given the very similar posteriors for both the benchmark GP and the mGP models, one may wonder if there are any differences between the two.  We found that the main difference between the two models is that mGP model allows for EoS that have bump in $c_s^2$ at low densities.  In fact, the posterior for the baryon density at which the maximum of $c_s^2$ occurs leads to a binomial distribution with peaks at $n_B\sim 2$ $n_{\rm sat}$ and $n_B\sim 5 n_{\rm sat}$.  In contrast, the benchmark model only produces a definite peak at $n_B\sim 5$ $n_{\rm sat}$.  Thus, we must conclude that the benchmark model is not adequately exploring the possibility of a bump in $c_s^2$ at low densities, due to long correlation lengths. We argue that it is important to explore the possibility of peaks around $n_B\sim 2$ $n_{\rm sat}$ because a global maximum at such densities is compatible with the onset of exotic phases in the core of neutron stars and, in this analysis, its existence is completely driven by astrophysical data.  

Another question we explored is if the presence of sharp features in $c_s^2$ is potentially excluded due to pQCD constraints at high $n_B$. Similar to what was done in \cite{Somasundaram:2022ztm}, we applied the constraints at the maximum central density that is peaked between $5 - 6 \ n_{\rm sat}$ for both models.  We find that these pQCD constraints only have a small effect on our results, with no visible effect on our posteriors. Only $0.0083\%$ of all EoS in our study were inconsistent with the pQCD constraints entirely, and only $3.5\%$ were in some degree of tension. Tension can occur because of uncertainty in an undetermined scale that arises from the contribution of missing higher order terms. One sets a range of values to that scale and some of those values may exclude an EoS whereas others may not. We plan to further explore the consequences of these pQCD constraints in a follow-up analysis.

Other approaches have been used to tackle similar questions as studied here, such as a deep neural network \cite{Fujimoto:2019hxv} or linear segments in $c_s^2$ \cite{Somasundaram:2021clp}. It would be relevant to directly compare these different methods to our mGP model in future work to study their ability to reproduce specific features in $c_s^2$ from nuclear physics models. 
Additionally, Essick \textit{et al.}~\cite{Essick:2023fso} developed a new technique to extract features indicative of phase transitions from functional forms of the EoS.  Rather than modifying samples from a benchmark model, Essick \textit{et al.} sample from a mixture of GPs that contain a wide range of correlation lengths that are fixed for a given EoS, including short correlation lengths that lead to behavior in $c_s^2$ that mimics arbitrary phase transitions. Using this method, which we emphasize does not include multi-scale correlations in the EoS, Essick \textit{et al.} also find that current data is not yet constraining enough to rule in favor or against nontrivial features in $c_s^2$, with the exception of very strong first-order phase transitions (latent energy per particle $\gtrsim$ 100 MeV). Interestingly, they also conclude from realistic simulations of future data that a catalog of around one hundred events would at best lead to a Bayes factor of $\sim 10:1$ in favor of a phase transition, even when the true EoS contains a strong phase transition. Their conclusions support the idea that astrophysical observations of neutron stars, currently and in the near future, will not be able to constrain short-range correlations in the EoS.

Overall, the results presented here suggest that current constraints are not enough to rule definitively in favor of or against phase transitions to exotic degrees of freedom in the core of neutron stars, and that unambiguous signatures of structure in the EoS still require investigation. A clear ruling regarding the existence of exotic matter in the core of neutron stars will require more precise input from astronomical observations, laboratory measurements, and input from effective theories and QCD at high densities \cite{Lovato:2022vgq}. Fortunately, more data is anticipated from the NICER collaboration both in terms of better statistics on existing measurements, and radii from new neutron stars. Additionally, LIGO/Virgo/KAGRA's fourth observing run started in May 2023 with better sensitivity than during the third observing run, such that more neutron star mergers that will provide $\Lambda$ constraints are anticipated \cite{Carson:2019rjx} and the binary love relation may provide further insight into structure in $c_s^2$ \cite{Tan:2021nat}.  Finally, more nuclear physics data is anticipated from the Facility for Rare Isotope Beams that will help constrain the low density regime of the EoS, and from low-energy heavy-ion collisions that will probe the large density, low-temperature region of the QCD phase diagram \cite{Sorensen:2023zkk}.

\section*{Acknowledgments} 
We thank Reed Essick for a detailed reading of the original version of the manuscript and thorough feedback and Katerina Chatziioannou for insightful discussions on Gaussian processes. D.M is supported by the National Science Foundation Graduate Research Fellowship Program under Grant No. DGE – 1746047 and the Illinois Center for Advanced Studies of the Universe Graduate Fellowship. 
J.N.H, D.M., and N.Y.\ were supported in part by the National Science Foundation (NSF) within the framework
of the MUSES collaboration, under grant number OAC2103680. J.N.H. acknowledges financial support from the US-DOE Nuclear Science Grant
No. DESC0020633 and DE-SC0023861.  M.C.M. was supported in part by NASA ADAP grant 80NSSC21K0649.
N.Y. was supported in part by NSF Award PHYS-2207650.
The authors also acknowledge support
from the Illinois Campus Cluster, a computing resource that
is operated by the Illinois Campus Cluster Program (ICCP)
in conjunction with the National Center for Supercomputing
Applications (NCSA), which is supported by funds from
the University of Illinois at Urbana-Champaign.

\bibliography{inspire,notinspire}

\end{document}